\documentclass[a4paper,fleqn,usenatbib]{mnras}
\usepackage{mathptmx}

\usepackage[T1]{fontenc}
\usepackage{ae,aecompl}

\usepackage{graphicx}	
\usepackage{amsmath}	
\usepackage{amssymb}	

\usepackage{xcolor}
\usepackage[normalem]{ulem}

\definecolor{darkred}{rgb}{0.79,0,0}

\usepackage{amsmath}

\title[Triple star candidates in the CoRoT fields]{A search for tight hierarchical triple systems amongst the eclipsing binaries in the CoRoT fields}

\author[T. Hajdu et al.]{
T. Hajdu,$^{1,2}$\thanks{E-mail: t.hajdu@astro.elte.hu}
T. Borkovits,$^{3,4}$
E. Forg\'{a}cs-Dajka,$^{1}$
J. Sztakovics,$^{1,2}$
G. Marschalk\'{o},$^{2,3}$\newauthor
J. M. Benk\H o,$^{4}$
P. Klagyivik,$^{4}$
and M. J. Sallai$^{1}$
\\
$^{1}$E\"{o}tv\"{o}s University, Department of Astronomy, H-1118 P\'{a}zm\'{a}ny P\'{e}ter stny. 1/A, Budapest, Hungary  \\
$^{2}$Wigner Research Centre for Physics of HAS, PO Box 49, H-1525, Budapest, Hungary \\
$^{3}$Baja Astronomical Observatory of Szeged University, H-6500 Baja, Szegedi \'ut, Kt. 766, Hungary \\
$^{4}$Konkoly Observatory, Research Centre for Astronomy and Earth Sciences, Hungarian Academy of Sciences, H-1121 Budapest,\\ Konkoly Thege Mikl\'os \'ut 15-17, Hungay
}

\date{Accepted XXX. Received YYY; in original form ZZZ}

\pubyear{2017}

\begin{document}
\label{firstpage}
\pagerange{\pageref{firstpage}--\pageref{lastpage}}
\maketitle


\begin{abstract}
We report a comprehensive search for hierarchical triple stellar system candidates amongst eclipsing binaries (EB) observed by the CoRoT spacecraft. We calculate and check eclipse timing variation (ETV) diagrams for almost 1500 EBs in an automated manner. We identify five relatively short-period Algol systems for which our combined light curve and complex ETV analyses (including both the light-travel time effect and short-term dynamical third-body perturbations)  resulted in consistent third-body solutions. The computed periods of the outer bodies are between 82 and 272~days, (with an alternative solution of 831~days for one of the targets).
We find that the inner and outer orbits are near coplanar in all but one case. The dynamical masses of the outer subsystems determined from the ETV analyses are consistent with both the results of our light curve analyses and the spectroscopical information available in the literature. One of our candidate systems exhibits outer eclipsing events as well, the locations of which are in good agreement with the ETV solution. We also report  another certain triply eclipsing triple system which, however, is lacking a reliable ETV solution due to the very short time range of the data, and four new blended systems (composite light curves of 2 eclipsing binaries each), where we cannot decide whether the components are gravitationally binded or not. Amongst these blended systems we identify the longest period and highest eccentricity eclipsing binary in the entire CoRoT sample. 
\end{abstract}


\section{Introduction}

Multiplicity is a common feature amongst binary star systems. For example, according to an investigation of \citet{Tokovininetal06} almost two thirds of their surveyed 165 solar-type spectroscopic binary systems have at least one more stellar companion. Furthermore, in the same sample, amongst the shortest period binaries ($P\leq2\fd9$) this ratio practically reaches 100\%. These findings are in good agreement with the recently most commonly accepted formation theory of the closest binary systems (typically with periods of a few days), the so called Kozai Cycles with Tidal Friction (KCTF) mechanism, which first was proposed by \citet{kiselevaetal98} and later was quantitatively investigated in details by e.g., \citet{fabryckytremaine07,naozfabrycky14}. Multiplicity, however, may have fundamental influence on binary star evolution (and, of course, directly or indirectly on the stellar evolution of binary members) not only in the formation period of a binary system, but at every stage of its evolution from the birth to the death of the binary members. Some examples are the hypothetised importance of multiplicity in the formation of blue stragglers \citep{peretzfabrycky09,naozfabrycky14}, and different kinds of binaries formed by degenerate components \citep{shappeethompson13,naozetal16}. Furthermore, \citet{taurisvandenheuvel14} amongst others have shown that the presence of a third stellar component may prevent a close binary system from disintegration even when on of its components undergoes a supernova explosion. Therefore, the identification of third (or even more) additional stellar companions to binary star systems has great astrophysical importance both from a general theoretical perspective (as probing the current theories) and on the other hand to understand the evolution of any given, individual system.
  
In the case of eclipsing binaries (EB) one long-lasting, traditionally used method for the identification of third, more distant companions is based on the detection and analysis of the eclipse timing variations (ETV) of the binary star which occur due to the light-travel time effect (LTTE) as the EBs distance to the observer periodically varies revolving on its orbit around the common centre of mass of the triple (or multiple) system.

According to our knowledge, \citet{Chandler1888} was the first who mentioned LTTE as a possible origin of the observed ETVs of Algol. After the preliminary analytical works of \citet{Woltjer1922}, the widespreadly used mathematical description of an LTTE forced ETV was given by \citet{Irwin1952,irwin59} who also gave a graphical fitting procedure for determining the elements of the light-time orbit from the ETVs that had been traditionally investigated by the use of  eclipse timing diagrams, which in the century-long history of the variable star research traditionally was called as $O-C$ (observed minus calculated) diagram \citep[see e.g.][for a short review on the advantages and obstacles of the application of $O-C$ diagrams in the analysis of period variations of different kinds of variable stars]{sterken05}.  

There are, however, various other mechanisms capable to produce ETVs in EBs, some of them may even strongly mimic LTTE-like behaviour. Therefore, certain detection of third components in such a manner is far from being an easy matter. In this regard \citet{Frieboes-Conde1973} listed four criteria that an ETV curve should fulfil for an LTTE solution which can be taken seriously. These criteria can be summarized as follows. (1) The shape of the ETV curve must follow the analytical form of an LTTE solution. (2) The ETVs of the primary and secondary minima must be consistent in both phase and amplitude with each other. (3) The estimated mass or lower limit to the mass of the third component, derived from the amplitude of the LTTE solution must be in accord with photometric measurements or limits on third light in the system. (4) Variation of the systemic radial velocity (if it is available) should be in accord with the LTTE solution. Recently this list was complemented with two subsequent criteria by \citet{Borkovits2016}, as follows. (5) The times of the maxima of the ellipsoidal variations (if they are detectable with sufficient accuray), at least in EBs that have circular orbits, should be in accord both in phase and amplitude with the ETVs. Furthermore, for triples exhibiting outer eclipses, an additional natural criterion for identifying the outer eclipsing body with the source of the observed LTTE is that (6) the LTTE should exhibit the same period as the extra eclipses, and these latter should occur around the inferior and/or superior conjunction points of the LTTE-orbit.\footnote{Thanks to some remarks of the referee of the present paper we realized that this last criterium was set erroneously in \citet{Borkovits2016}, where, originally, it was stated that the outer eclipses ``should occur around the extrema of the LTTE''. This latter statement is strictly valid only in the case of a circular outer orbit or an orbit seen from the direction of its major axis.}

Before the era of small, but ultraprecise photometric space telescopes (e.g. as CoRoT and {\em Kepler}) the vast majority of the known third companions (or, more strictly, candidates) had orbital periods of several years or, even decades, and only a very limited number of hierarchical triple systems with outer periods less than a year were known  \citep[see~e.g.][]{tokovinin04}. As dynamical stability criteria \citep[see~e.g.][]{mardlingaarseth01} would allow the presence of 1-2-month-long outer period stellar companion to a typical eclipsing binary with period from a few hours to few days, it was not clear whether the small known number of such systems was a consequence of some selection effects, or of yet-unknown evolutionary origin(s) \citep[see e.g.][]{tokovinin14}. In this regard the four-year-long almost continuous measurements of {\em Kepler} spacecraft \citep{boruckietal10} resulted in a significant improvement, allowing to explore regions of the parameter space previously out of the reach of ground-based ETV studies due to the small LTTE amplitude involved, like triple stellar systems in the shortest theoretically possible outer period regime. Systematic analyses of the ETVs of more than 2700 EBs (and ellipsoidal variables), observed continuously during the prime {\em Kepler} mission have led to the discovery of more than 200 hierarchical triple system candidates \citep{Rappaport2013,conroyetal14,Borkovits2015,giesetal15,Borkovits2016,kirketal16}. Considering e.g. the survey of \citet{Borkovits2016}, more than a hundred of the 222 triple system candidates investigated by them have outer periods less than 1000\,days.

The observing strategy, i.e. the short, four-five month-long observing sessions of the complementer and contemporary mission of the European CoRoT spacecraft \citep{auvergneetal09}, however, unfortunately, was less favourable from the point of view of searching for additional, distant companions around EBs by the use of ETV analysis. This particularly holds when we restrict ourselves to the most certain detections which, according to the criteria of \citet{conroyetal14} and similarly, of \citet{Borkovits2016} require at least two fully covered orbital periods of the outer orbit to be observed. On the other hand, in a less restrictive sense the 100-150-day long CoRoT ETV data series may allow us to identify some of the tightest triple star {\em candidates}. For example,  the work of \citet{Borkovits2016} mentioned above reported 13 triples (i.e. approximately 0.5\% of the total number of the {\em Kepler} EBs) with outer period less than 150\,days. The characteristic shape of the ETVs of the majority of these tightest triple systems would allow us to identify them as probable triple system candidates even from the short ETV sections available from CoRoT observations.

A further glance at the results of \citet{Borkovits2016} reveals that in the ETVs of these tight systems the dynamical, hierarchical third-body perturbations play a dominant role, or are at least as important in magnitude as the classical LTTE, therefore their contribution should also be taken into account. This latter, dynamical ETV contribution was analytically described in a series of papers by \citet{Borkovits2003, Borkovits2011, Borkovits2015}. (Some preliminary works on this field had also been carried out by \citealt{Soderhjelm1975} and \citealt{Mayer1990}.)

In this paper we are searching for hierarchical triple star candidates amongst the EBs observed by the CoRoT spacecraft, primarily with the analysis of their ETVs. For the analysis we use publicly available CoRoT photometric data. In Section \ref{data processing} we outline the steps of our investigation, starting with the methods used for data acquisition and automatic ETV curve generation, then continuing with the system selection procedures and, finally, closing with a short description of some details of the applied ETV and the auxiliary light curve analyses as well.

The results of the detailed analysis of the ETV and light curves of the five newly identified tight hierarchical triple candidates as well as some other interesting by-products of our research, are discussed in Sections \ref{third bodies} and \ref{Discussion}. 

Finally, a short summary is given in Section \ref{summary}.
 
\section{Basic steps of the analysis}\label{data processing}

\subsection{Data acquisition and preparation}

The space mission CoRoT performed wide-field stellar photometry at ultra-high precision \citep{1998EM&P...81...79R,2006cosp...36.3749B}. The mission took 6 years from the end of 2006 to the November of 2012. During an observation, up to 12\,000 stars were monitored  simultaneously  and continuously  over  150  days  of  observation. All observations of CoRoT spacecraft, the so-called LEGACY data (version 4, see \citealp{chaintreuiletal16}) are publicly available now, e.g. through the VizieR archive service\footnote{http://vizier.u-strasbg.fr/viz-bin/VizieR?-source=B/corot}.

We downloaded all the corrected \citep[LCC, see][Sect.\,1.2]{chaintreuiletal16} light curves of the CoRoT Bright and Faint Star Catalogs. Then we performed a visual inspection of all the over 177\,700 light curves individually, and identified about 1500 EBs (including binaries with ellipsoidal variations, but without real eclipses, too). This number is nearly the same as the one reported in \citet{baudinetal16}. Unfortunately, these authors do not give the complete list of these EBs, therefore we were unable to cross-check our findings with theirs. We made comparison, however, with the so-called ``Unofficial CoRoT Eclipsing Binary Catalog'' of Jonathan Devor\footnote{ http://www.astro.tau.ac.il/$\sim$jdevor/CoRoT\_catalog/catalog.html}, which contains also 1479 items. Surprisingly, there is a remarkable ($\approx40\%$) amount of mismatch between our findings and this latter catalog. In our opinion, several very short period ($P\lesssim0\fd15$), low amplitude ($\lesssim1\%$ in normalized flux) light curves listed in Devor's catalog should belong to pulsating stars rather than eclipsing or ellipsoidal variables and, therefore, we do not count them amongst the CoRoT EBs. On the other hand, a significant amount of certain CoRoT EBs are not included into this unofficial catalog, the triply eclipsing CoRoT\,104079133 (i.e. one of the five EBs which we study in this paper in details) being a notable example.

We then prepared the light curves of these EBs for the forthcoming analysis in an iterative and automatic manner, using our own GPU-based code, written in CUDA language.

As a first step, the code determines preliminary eclipsing periods for each EB using the Lomb-Scargle periodogram method. Then, using these periods, it creates folded, binned, averaged light curves for each EBs in the following manner. The light curves are binned into 500-1000 equally spaced phase-cells, according to the orbital phases of each measured point. Then the average fluxes are calculated cell by cell, and associated to the phase of the cell midpoints. In the next step the code identifies the locations (phase-domains) of the eclipses (primary and secondary) in the folded light curves, and then calculates template minima, fitting sixth-order polynomials on the previously identified phase ranges. 

In the next part of the analysis the code scans the original light curves of each EB, identifies all the individual primary and secondary eclipses, and applies the appropriate template minimum for the determination of the individual mid-minimum moments. For this purpose the code fits a three-parameter model to the data with the Levenberg\,--\,Marquardt method. The fitted template curve is the following:
\begin{equation}
f'(x)=a_0+a_1\cdot f(x-a_2)
\end{equation}
where $f(x)$ is the previously determined polynomial template function, and the most important parameter is $a_2$, which gives the phase-lag between the template and the current eclipse \citep[see][Sect.\,4, for further details]{Borkovits2015}.

In the last step the code calculates ETV diagrams using the preliminary period obtained in the very first step, and determines the average slope of each ETV curve with linear regression. The period is then corrected with this average slope (and the epoch, i.e. the moment of the zero phase is corrected, too), and the whole procedure is reiterated until convergence of the period. We found that two iteration steps were enough for all light curves. 

Extra care was required for some of the light curves, due to the presence of extra-eclipsing events. The data points affected by these extra events were excluded from both the light curve folding and the ETV forming procedures.

\subsection{System selection}

For further analysis we selected (by visual inspection) systems having sine-like variations or at least significant curveture(s) in their  ETV curve(s). We dropped out, however, some short period (most probably overcontact) binaries where the primary and secondary ETVs exhibited quasi-sinusoidal variations in opposite phase to each other, which might be the product of light curve distortions originating from stellar spot rotation \citep{Tran2013,balajietal15}. We investigated the ETVs of EBs showing clearly visible extra eclipsing events with extra care. Finally, we found five EBs in the investigated CoRoT sample for which we were able to obtain preliminary ETV solutions. Then in the last step we carried out a more complex study of these EBs, including light curve analyses, too.

\subsection{Analysis of the folded light curves}

The supplementary light curve analyses of the selected systems were carried as follows. Rather than conducting a more sophisticated investigation, our primary aim was to obtain the values of those parameters from the light curve that provide significant auxiliary information to the ETV analysis. These parameters are the eccentricity ($e_1$) and argument of periastron ($\omega_1$) of the eclipsing pairs, and the amount of the third light ($l_3$) in the light curve. The photometrically obtained values of $e_1$ and $\omega_1$ can be used directly for the ETV analysis (see below), while the presence (or absence) of an extra light source (and its light ratio) may be a good additional indicator of the reliability of our third-body solutions. 

For this study we used the folded and binned, averaged CoRoT light curves. The analysis of these light curves was carried out with an MCMC-based parameter search which was recently incorporated into the newest version of our own {\sc lightcurvefactory} light curve synthesis program \citep{borkovitsetal13,borkovitsetal14}. We used an own implementation of the generic Metropolis-Hastings algorithm, and uniform priors. For the sake of a quick convergence we used in general a special set of nine adjusted parameters. These were as follows: (i) length of the primary eclipse ($\Delta t$); (ii) ratio of the stellar radii ($R_\mathrm{B}/R_\mathrm{A}$)\footnote{In order to avoid confusion we use numerical indices, (i.e. 1 and 2) only for the quantities referring to the inner and outer orbits, while the parameters associated with the stellar components are indexed with capitals, with A, B and C marking the primary and secondary components of the inner (eclipsing) binary, and the the third, more distant component, respectively.}; (iii) mass ratio ($q_1=m_\mathrm{B}/m_\mathrm{A}$); (iv) temperature ratio ($T_\mathrm{eff,B}/T_\mathrm{eff,A}$); (v) eccentricity ($e_1$); (vi) mid-phase of the secondary eclipse ($\phi_\mathrm{II}$); (vii) mid-time of the primary eclipse; (viii) inclination ($i_1$); (ix) third light ($l_3$)\footnote{We emphasize again that the light curve analyses were carried out on phase-folded light curves. Thus, the orbital period ($P_1$) was not an adjustable parameter.}. Then the additional system parameters like the fractional radii of the two stars ($r_\mathrm{A,B}=R_\mathrm{A,B}/a_1$) and the remaining orbital parameters (argument of periastron -- $\omega_1$ and periastron passage time -- $\tau_1$, or its some equivalents) were calculated by the use of the relations given in \citet[][Eq.\,6--10]{rappaportetal17}. The advantage of this set of parameters is that three of them (i, vi and vii) are direct observables and therefore their initial values can be determined very easily from the folded light curve. According to our experiences, if the initial values of these three parameters are set properly, then our Markov Chains converged quickly for any arbitrary (but physically realistic) initial values of the other parameters. Regarding the mass ratio ($q_1$), however, some caution has to be taken. For detached EBs with almost spherical stellar components, the mass ratio has only minor influence on the light curve, therefore pure light curve analysis can derive its actual value only with a large uncertainty. This is especially true when other complicating effects (like chromospheric activity, pulsation, etc.) distort the light curves with magnitudes similar to (or greater than) the mass ratio dependent effects. This was exactly the situation for CoRoT\,110830711. Therefore in this case we used some astrophysical contraints for $q_1$, as discussed in Sect.\,\ref{Subsect:ETVsolution}.

Regarding other, higher order effects influencing a light curve solution, ellipsodial variation, Doppler-boosting and reflection/irradiation effects were also taken into account in our analysis. Limb darkening was considered according to the logarithmic law \citep{klinglesmithsobieski70}. The corresponding coefficients were computed internally by the use of the ``in-house tables'' of the Phoebe software\footnote{http://phoebe-project.org/1.0} \citep{prsazwitter05}. In case of CoRoT\,102698865, however, the use of these precomputed coefficients resulted in systematically biased residuals during the eclipses with a magnitude of about 3-4000 ppm, which diminished remarkably when the adjustment of these parameters was switched on.

The results of our analysis on the selected five EBs will be discussed in Sect.\,\ref{third bodies}. Here we only briefly mention that three of the five EBs were found to be dominated by the flux(es) of extra source(s). As a consequence, we may assume that for these three systems the spectroscopic information e.~g. temperature, spectral type given either in the ExoDat Information System\footnote{http://cesam.oamp.fr/exodat/} \citep{ExoDAT}, or in \citet{sarroetal13} may refer to the extra source(s) and therefore, unfortunately, we cannot use them with full confidence to convert the direct outputs of the light curve analysis, which are dimensionless, relative quantities (e.~g. temperature and mass ratios, and the dimensions of the stars relative to the semimajor axis) to physical units. There are, however, two possibilities for getting some information, or at least reasonable estimations, for the real nature of the binary components. 

First, the combination of the relative stellar radii and the mass ratio offer an indirect possibility to infer at least a probable luminosity class for the binary components via a reliable estimation of the local surface gravity ($g$). In order to show this, we approximate the local surface gravity of (let's say), the primary component as:
\begin{equation}
g_\mathrm{A}=\frac{Gm_\mathrm{A}}{r_\mathrm{A}^2a_1^2},
\end{equation}
which, by the use of Kepler's third law can be written as
\begin{eqnarray}
g_\mathrm{A}&=&r_\mathrm{A}^{-2}\left(\frac{2\upi}{P_1}\right)^{4/3}\frac{(Gm_\mathrm{A})^{1/3}}{(1+q_1)^{2/3}} \nonumber \\
&=&g^*_\mathrm{A}m_\mathrm{A}^{1/3},
\label{Eq:g*def}
\end{eqnarray}
where $g^*_\mathrm{A}$ can be calculated directly from the light curve solution. (For the secondary component $q_1$ should be replaced with $q_1^{-1}$.) Expressing $g$ in its usual logarithmic form (and using the usual astrophysical units)
\begin{equation}
\log g_\mathrm{A}=\log g^*_\mathrm{A}+\frac{1}{3}\log m_\mathrm{A},
\end{equation}
it can be seen that the mass-dependent, unknown last term gives only a minor contribution to the sum for a wide range of the physically reasonable stellar masses \citep[cf.][]{southworthetal04}. Therefore $\log g^*_\mathrm{A}$ can be used as a good estimate of $\log g_\mathrm{A}$ and thus of the probable luminosity class of the given star.\footnote{On the other hand, however, one should again keep in mind that, as discussed above, the photometric mass ratio $q_1$ is a weakly determined quantity for our detached EBs. This results in an uncertainty in $\log g^*$ and, therefore, in the estimated local surface gravity. From Eq.\,(\ref{Eq:g*def}) one can see that for the more massive component (i.e. for which $q\leq1$) this uncertainty has an upper limit of $\Delta{\log g^*_{(q\leq1)}}\leq\frac{2}{3}\log2\approx0.2$. For the less massive component there is no such upper limit. Conversely, for extreme mass ratios, the uncertainty, in theory, may tend to infinity. In practice it can be used an additional indicator of inappropriate light curve solutions with unphysical mass-ratios.}

Second, as it will be discussed in the next subsection, by combining the outputs of the light curve analysis with the results of a joint LTTE+dynamical ETV analysis, we can infer the masses of the EB components in a dynamical way.

\subsection{Overview of the ETV analysis}

The ETV analysis of each system was carried out with the newest version of the {\sc omincfit} code of T. Borkovits. This version differs from the previous ones only by the inclusion of an MCMC-based parameter search. The theoretical base and the applied analytical formalism of the analysis remained unchanged, however, and was described in detail in \citet{Borkovits2015,Borkovits2016}. Therefore here we give only a brief summary. 

We define ETV as the time difference of the observed and calculated mid-minima times of each individual eclipses:
\begin{equation}
\Delta = T(E)-T_0-P_\mathrm{s}E,
\end{equation}
where $T(E)$ stands for the observed mid-minimum time of the $E$th eclipse (cycle number $E$ is integer for primary and half-integer for secondary eclipses, respectively), $T_0$ indicates the reference epoch, i.e. the observed mid-eclipse time of the `zeroth' event, and $P_\mathrm{s}$ is the constant sidereal (eclipse) period. The ETV is then basically modelled in the following form:
\begin{equation}
\Delta=c_0+c_1E+\left[\Delta_\mathrm{LTTE}+\Delta_\mathrm{dyn}+\Delta_\mathrm{apse}\right]_0^E,
\end{equation}
where $c_{0,1}$ give corrections to the reference epoch and the eclipse period, respectively (independent on their origins), while $\Delta_\mathrm{LTTE}$, $\Delta_\mathrm{dyn}$, and $\Delta_\mathrm{apse}$ refer to the contributions of LTTE, short-period dynamical third-body perturbations (i.e. those with periods equal to, or related to the orbital period $P_2$ of the third, outer component), and apsidal motion effect to the ETVs, respectively.

The LTTE contribution takes the following form \citep{Irwin1952}:
\begin{equation}
\Delta_\mathrm{LTTE}=-\frac{a_\mathrm{AB}\sin i_2}{c}\frac{\left(1-e_2^2\right)\sin\left(v_2+\omega_2\right)}{1+e_2\cos v_2},
\end{equation}
or, changing to eccentric anomaly:
\begin{equation}
\Delta_\mathrm{LTTE}=-\mathcal{A}_\mathrm{LTTE}\sin\left(\mathcal{E}_2+\phi\right)+\frac{\mathcal{A}_\mathrm{LTTE}}{\sqrt{1-e_2^2\cos^2\omega_2}}e_2\sin\omega_2,
\end{equation}
where $a_\mathrm{AB}$ denotes the semi-major axis of the EB's center of mass around the center of mass of the triple system, while $i_2$, $e_2$, $\omega_2$ stand for the inclination, eccentricity, and argument of periastron of the relative outer orbit (i.e. the orbit of the third component relative to the center of mass of the EB), respectively. Furthermore, $v_2$ and $\mathcal{E}_2$ are the true and eccentric anomalies of the third component, respectively, and $c$ is the speed of light. Note the negative sign on the right hand sides, which arises from the use of the {\em companion's} argument of periastron $\omega_2$, instead of the argument of periastron of the light time orbit of the EB ($\omega_\mathrm{AB}=\omega_2+\upi$). This modification was necessary for the use of the dynamical perturbation terms which are expressed in the orbital elements of the third component's relative outer orbit (see below). Moreover, the amplitude of the LTTE curve is
\begin{equation}
\mathcal{A}_\mathrm{LTTE}=\frac{a_\mathrm{AB}\sin i_2}{c}\sqrt{1-e_2^2\cos^2\omega_2},
\end{equation} 
while its phase $\phi$ can be calculated as:
\begin{equation}
\phi=\arctan\left(\frac{\sin\omega_2}{\sqrt{1-e_2^2}\cos\omega_2}\right).
\end{equation}

The ETV contribution of the short-period dynamical perturbations ($\Delta_\mathrm{dyn}$) has a very complicated dependence on the orbital elements of the inner and outer orbits, and their relative configurations as well. Furthermore, for eccentric inner orbits even the orbits' relative orientation to the observer becomes an additional important factor. The most thorough discussion of these effects can be found in \cite{Borkovits2015}. Here, for simplicity, we give only the most dominant terms of the analytic description:
\begin{eqnarray}
\Delta^\mathrm{lead}_\mathrm{dyn}&=&\mathcal{A}_\mathrm{dyn}\left(1-e_1^2\right)^{1/2}\left\{\left(1\mp\frac{3}{2}e_1\sin\omega_1\right)\right. \nonumber \\
&&\times\left[\left(1-\frac{3}{2}\sin^2i_\mathrm{m}\right)\mathcal{M}+\frac{3}{4}\sin^2i_\mathrm{m}\mathcal{S}\right] \nonumber \\
&&\mp\frac{15}{4}e_1\sin(\omega_1-2g_1)\nonumber \\
&&\times\left[\sin^2i_\mathrm{m}\mathcal{M}+\frac{1}{2}\left(1+\sin^2i_\mathrm{m}\right)\mathcal{S}\right] \nonumber \\
&&\left.\pm\frac{15}{4}e_1\cos(\omega_1-2g_1)(1+\cos i_\mathrm{m})\mathcal{C}\right\} +\mathcal{O}(e_1^2),
\label{Eq:Deltadynlead}
\end{eqnarray}
where
\begin{equation}
\mathcal{A}_\mathrm{dyn}=\frac{1}{2\upi}\frac{m_\mathrm{C}}{m_\mathrm{ABC}}\frac{P_1^2}{P_2}\left(1-e_2^2\right)^{-3/2},
\label{Eq:Adyn}
\end{equation}
which, as was found in \citet{Borkovits2016}, in most cases, gives a reasonable estimation for the ETV-amplitude of the short-term dynamical contribution. The time-dependence is buried within the trigonometric expressions:
\begin{eqnarray}
\mathcal{M}&=&v_2-l_2+e_2\sin v_2, \\
\mathcal{S}&=&\sin(2v_2+2g_2)+e_2\left[\sin(v_2+2g_2)+\frac{1}{3}\sin(3v_2+2g_2)\right],  \\
\mathcal{C}&=&\cos(2v_2+2g_2)+e_2\left[\cos(v_2+2g_2)+\frac{1}{3}\cos(3v_2+2g_2)\right].
\end{eqnarray}
Moreover, $i_\mathrm{m}$ means the mutual (relative) inclination of the inner and outer orbits, while $l_2$ stands for the mean anomaly of the tertiary and $g_{1,2}$ denote the arguments of periastron of the inner and outer orbits, measured from the intersections of the respective orbital planes and the invariable plane of the triple. Note, in Eq.\,(\ref{Eq:Deltadynlead}) the upper and lower signs refer to the primary and secondary eclipses, respectively.

Comparing the amplitudes of the LTTE and dynamical terms, \citet{Borkovits2016} showed that they fulfill the inequality:
\begin{equation}
\frac{\mathcal{A}_\mathrm{dyn}}{\mathcal{A}_\mathrm{LTTE}}\geq 1.45\times10^3m_\mathrm{ABC}^{-1/3}\frac{P_1^2}{P_2^{5/3}},
\label{Eq:Adyn/ALTTE}
\end{equation}
where $P$'s should be expressed in days, and $m_\mathrm{ABC}$ in solar mass. As all of our five triple system member candidate EBs eclipsing periods $P_1<4$\,days, we can substitute this upper limit into the above equation. Then one can obtain
\begin{equation}
\frac{\mathcal{A}_\mathrm{dyn}}{\mathcal{A}_\mathrm{LTTE}}\gtrsim m_\mathrm{ABC}^{-1/3}\left(\frac{P_1}{4}\right)^2\left(\frac{416}{P_2}\right)^{5/3},
\label{Eq:Adyn/ALTTE2}
\end{equation}  
which illustrates, that the dynamical contribution should be likely larger than, or at least comparable to the LTTE contribution for such short outer period third bodies found in our sample.

The combination of the LTTE and dynamical contributions allows us to calculate both the total mass of the inner EB and the individual mass of the third component in a dynamical way. This is so, because, similar to the radial velocity solution of a single line spectroscopic binary, the mass function of the distant component $C$ can be calculated from the LTTE component as
\begin{equation}
f(m_\mathrm{C})=m_\mathrm{C}\sin^3 i_2\left(\frac{m_\mathrm{C}}{m_\mathrm{ABC}}\right)^2=\frac{4\upi^2 a_\mathrm{AB}^3\sin^3 i_2}{GP_2^2}.
\end{equation}
This shows that if the outer mass ratio ($m_\mathrm{C}/m_\mathrm{ABC}$) and the inclination of the outer orbit ($i_2$) were known, the mass of the third companion ($m_\mathrm{C}$) and the total mass of the inner EB ($m_\mathrm{AB}$) could be calculated. Now the outer mass ratio $m_\mathrm{C}/m_\mathrm{ABC}$ is a direct output of the dynamical contribution, therefore the projected masses $m_\mathrm{AB}\sin^3i_2$ and $m_\mathrm{C}\sin^3i_2$ can be immediately calculated from a combined ETV solution. Regarding the outer inclination $i_2$, it can be derived e.g. from the expression
\begin{equation}
\sin i_2=\left|\frac{\sin(\omega_1-g_1)}{\sin(\omega_2-g_2)}\right|\sin i_1,
\end{equation}
where $\omega_{1,2}$, $g_{1,2}$ are direct outputs of the ETV solution, while $\sin{i_1}$ can be taken from the light curve solution. The verification and detailed discussion of this relationship is given in \citet[][Appendix D]{Borkovits2015}. Note, however that, as it will be discussed in the next section, five of our six ETV solutions resulted in almost coplanar orbits ($i_\mathrm{m}<5\degr$) and, therefore, for these cases the $\sin i_2\approx\sin i_1$ approximation would be just as adequate.

As mentioned before, three of the five EBs were found to have eccentric inner orbits. Moreover, for two of them we detected evidences of apsidal motion \citep[see, e.g.,][]{cowling38,sterne39,gimenezgarcia83}, too. Therefore these three systems required the inclusion of the apsidal motion related terms, too, into their ETV analysis, as follows:
\begin{eqnarray}
\Delta_\mathrm{apse}&=&\pm\frac{P_1}{2\upi}\left[2\arctan\left(\frac{e_1\cos\omega_1}{1+\sqrt{1-e_1^2}\mp e_1\sin\omega_1}\right)\right. \nonumber \\
&&\left.+\sqrt{1-e_1^2}\frac{e_1\cos\omega_1}{1\mp e_1\sin\omega_1}\right],
\label{Eq:Delta_apse}
\end{eqnarray}
where, as before, the alternate signs refer again to the primary and secondary eclipses, respectively. Since this expression gives the displacement of the secondary eclipse from photometric phase $0\fp5$, it carries important information on the eccentricity, or, strictly speaking, on the $e_1\cos\omega_1$ parameter of the EB, even in the absence of any detectable apsidal motion. The apsidal motion of the EB's orbit then is included into Eq.\,(\ref{Eq:Delta_apse}) through the time dependence of $\omega_1$. Our code allows different modes for modelling the apsidal motion. In the present work two of them were applied. In mode AP1 the apsidal motion is considered to be linear in time, i.e. the apsidal advance rate is an additional constant parameter, which is unconstrained, i.e. can be adjusted freely. On the contrary, in mode AP2 it is treated as a fixed quantity calculated internally from the third-body perturbation equations, as described in \citet[][Appendix C]{Borkovits2015}.

\section{Results}\label{third bodies}

We found five EBs in the whole CoRoT sample for which we were able to establish reliable, physically consistent results. We list the basic parameters of these five EBs in Table\,\ref{Binaries}. (Note, for simplicity, in all the Tables we use reduced BJDs -- hereafter RBJD --, i.\,e.\, BJD\,--\,2\,400\,000.) As it can be seen, all of them are relatively short-period, detached binaries. In this section we discuss the complex analysis of each of these ternary system candidates separately, though the numerical results are tabulated collectively in Tables\,\ref{Tab:lcfoldfit} (light curves), \ref{LTTEdyn} (ETVs) and \ref{Tab:EBfullcomb} (derived parameters obtained by the combination of the two kinds of analysis).\footnote{The times of minima of the five EBs are tabulated in Appendix\,\ref{App:ToM}.} We then give a short description of some additional EBs for which, although an ETV solution was not possible, nevertheless, some of their light curve features may imply a multiple nature with various probabilities.

\subsection{Hierarchical triple system candidates with consistent ETV and light curve solutions.}
\label{Subsect:ETVsolution}

\begin{table*}
\caption{Properties of the investigated systems. Most of the data, with the exception of the binary ephemerides (i.e. zero epoch -- $T_0$; and period -- $P_1$), were taken from the ExoDat Information System.  $T_0$ and $P_1$ are obtained from our ETV analysis, and may serve as ephemerides for future follow-up observations (see Sect.\,\ref{Subsect:follow-up} for details).}
\label{Binaries}
\centering
\begin{tabular}{c c c c c @{\hspace{5pt}}c}
\hline
CoRoT Id & Run(s) & Mag. & $T_0$ & $P_1$ & SpT.\\
	 &	  & in R & RBJD &[days] & \\
\hline
100805120& LRc01 & 12.83& 54238.2824(3) & 2.271722(8)   & K0III\\
101290947& LRc01 & 13.74& 54237.66574(1) & 2.048813(1)  & G0III\\
102698865& LRa01, LRa06&13.99&54398.5431(1)&3.7735657(2)& A0V\\
104079133& LRc04 & 15.01& 55022.7375(1) & 2.764624(5)   & G2V\\
110830711& LRa02 & 12.85& 54789.29970(4)& 2.545875(2)   & F5V\\
\hline
\end{tabular}
\end{table*}

\begin{table*}
\caption{Parameters obtained from the light curve solutions, together with the epoch ($T_0$) and period ($P_1$) used for creating the folded light curves. Numbers in parentheses denote 1-$\sigma$ uncertainties in the last digits. (Parameters without uncertainties were kept on fixed values.)}
\label{Tab:lcfoldfit}
\centering 
\begin{tabular}{lccccc}
\hline
Parameters & 100805120 & 101290947 & 102698865$^b$ & 104079133 & 110830711 \\
\hline
$T_0$ [RBJD]& 54238.2790 & 54237.6683 & 54398.6000 & 55022.73835 & 54789.2988\\
$P_1$ [d] & 2.27175 & 2.04882 & 3.773576 & 2.76476830 & 2.54588\\
$e_1$  & 0.020(7) & 0 & 0.080(3) & 0.005(2) & 0 \\
$\omega_1$ [deg] & 101(2) & $-$ & 43(2) & 292(9) & $-$ \\
$i_1$ [deg] & 81.81(9) & 88.5(5) & 89.0(3) & 88.4(1) & 88.1(1)\\
$(\lambda_0)_1^a$ [deg] & 269.7(1) & 269.98(3) & 276.9(3) & 270.2(1) & 269.993(6)\\
$q_1$ & 0.20(14) &$0.93^{+0.13}_{-0.55}$& 0.64(6) & 0.34(3) & 0.491(1)$^c$\\
$r_\mathrm{A}$ & 0.205(13) & 0.0783(5) & 0.1502(7) & 0.0840(7) & 0.1102(3)\\
$r_\mathrm{B}$ & 0.080(13) & 0.0763(5) & 0.0968(6) & 0.0650(7) & 0.0547(3)\\
$T_\mathrm{B}/T_\mathrm{A}$ & 0.784(7) & 0.982(3)& 0.914(4) & 0.800(1) & 0.625(2)\\
$L_\mathrm{A}/(L_\mathrm{A}+L_\mathrm{B})$ & 0.95(2) &$0.53^{+0.02}_{-0.09}$& 0.775(4) & 0.802(4) & 0.9637(6)\\
$l_3$ & 0.907(12)& 0.907(9)& 0.151(8) & 0.726(4) & 0.06$^{+0.01}_{-0.03}$\\
$\log g^*_\mathrm{A}$&4.04(6)&4.75$^{+0.08}_{-0.02}$&3.93(1)&4.67(1)&4.438(2)\\
$\log g^*_\mathrm{B}$&4.40(1.02)&4.73$^{+0.04}_{-0.18}$&4.17(4)&4.57(8)&4.840(5)\\
\hline
\end{tabular}

Notes. $a$: True longitude (i.e. ${l_0}_1+\omega_1$) of the secondary component at epoch $T_0$; $b$: monochromatic logarithmic limb-darkening coefficients were adjusted. Results: $x_\mathrm{A}=0.376^{+0.049}_{-0.027}$; $x_\mathrm{B}=0.496^{+0.020}_{-0.013}$; $y_\mathrm{A}=0.277^{+0.058}_{-0.034}$; $y_\mathrm{B}=0.227^{+0.047}_{-0.033}$; $c$: constrained by the formulae of \citet{toutetal96}, see text for details.
\end{table*}

\begin{table*}
\caption{Orbital elements from combined dynamical and LTTE solutions. For eccentric inner EBs the ETV-derived values of the EB's eccentricity ($e_1$), argument of periastron ($\omega_1$), and apsidal motion period ($U$) are also given in the notes. (Note $U_\mathrm{fit}$ and $U_\mathrm{calc}$ refer to freely adjusted -- AP1 -- and constrained -- AP2 -- mode apsidal motion solutions.) Furthermore, for the two non-coplanar solutions (i.e. $i_\mathrm{m}>1\degr$) the calculated observable inclination ($i_2$) of the outer orbit, and the sky-projected angular distance of the ascending nodes ($\Delta\Omega=\Omega_2-\Omega_1$) are also listed in the notes.} 
\label{LTTEdyn}
\centering 
\begin{tabular}{lcccccccccc}
\hline
CoRoT Id & $P_2$ & $a_2$ & $e_2$ & $\omega_2$ & $\tau_2$ & $i_\mathrm{m}$ & $f(m_\mathrm{C})$ & $\frac{m_\mathrm{C}}{m_\mathrm{ABC}}$ & $m_\mathrm{AB}$ & $m_\mathrm{C}$  \\
& (d) & (R$_{\sun}$) & & ($\degr$) & (RBJD) & ($\degr$) & (M$_{\sun}$) & & (M$_{\sun}$) & (M$_{\sun}$) \\
\hline
100805120$^a$& 104(1) & 141(10)& 0.16(1) &  49(4)   & 54259(2)  &$0.3^{+1.8}_{-0.3}$ &0.71(10) & 0.60(3) & 1.38(31) & 2.06(44)  \\
101290947    &110.2(1)& 139(5) &0.350(5) & 106(4)   & 54254.7(1)& 0.0($-$) &0.109(6) &0.33(1)& 1.96(24)& 0.98(12) \\
102698865$^b$& 272(1) & 315(8) & 0.32(4) & 103(5)   & 54272(5)  &$0.3^{+2.4}_{-0.3}$ &0.04(1)& 0.19(2) & 4.58(44) & 1.09(10) \\
102698865$^c$& 831(34)& 679(38)& 0.43(14)& 343(73)  & 54456(170)& 39(3)    & 0.11(4) &0.31(5)& 4.20(84)& 1.88(30) \\
104079133$^d$&  90(2) & 108(5) & 0.20(2) & 349(2)   & 55047.4(6)& 0.9(6)   & 0.12(5) &0.39(7)& 1.28(21)& 0.81(20) \\
110830711$^e$& 82(2)  & 108(6) &0.119(8) &  14(2)   & 54754(2)  &$4.9^{+5.2}_{-1.3}$ & 0.04(2) & 0.25(5) & 1.87(36) & 0.62(13) \\
\hline
\end{tabular}

Notes. $a$:  $e_1=0.026(3)$; $\omega_1=91.2(6)\degr$; $U_\mathrm{calc}=28$\,yr; $b$: $e_1=0.078(6)$; $\omega_1=41(3)\degr$; $U_\mathrm{fit}=279(39)$\,yr; $c$: $e_1=0.100(3)$; $\omega_1=54(2)\degr$; $U_\mathrm{fit}=415(22)$\,yr; $\Delta\Omega=-24(7)\degr$; $i_2=58(9)\degr$; $d$: $e_1=0.0040(6)$; $\omega_1=300(4)\degr$; $U_\mathrm{fit}=2.4(7)$\,yr; Conjunctions of the outer orbit (in RBJD): $t_\mathrm{inf}=55\,033(1)$, $t_\mathrm{sup}=55\,067(1)$; $e$: $\Delta\Omega=-2(7)\degr$; $i_2=92(4)\degr$.
\end{table*}

\begin{table}
\caption{Physical parameters of the eclipsing binaries from the combination of the light curve and ETV solutions.
\label{Tab:EBfullcomb}}
\centering 
\begin{tabular}{lccccc}
\hline
CoRoT Id & $a_1$ & $m_\mathrm{A}$ & $m_\mathrm{B}$ & $R_\mathrm{A}$ & $R_\mathrm{B}$ \\ 
& (R$_{\sun}$) & (M$_{\sun}$) & (M$_{\sun}$) & (R$_{\sun}$) & (R$_{\sun}$)\\
\hline
100805120 & 8.1(6) & 1.15(29) & 0.23(14) & 1.66(16)& 0.65(12)\\
101290947 & 8.5(3) & 1.01$^{+0.31}_{-0.14}$ & $0.94^{+0.13}_{-0.31}$ & 0.67(2)& 0.65(2)\\
102698865 &16.9(5) & 2.79(29) & 1.79(20) & 2.54(8)& 1.64(5)\\
102698865 &16.4(1.1)& 2.56(52) & 1.64(34) & 2.47(17)& 1.59(11)\\
104079133 & 9.1(5) & 0.95(16) & 0.32(6) & 0.76(4)& 0.59(3)\\
110830711 & 9.7(6) & 1.25(24) & 0.61(12) & 1.07(7)& 0.53(3) \\
\hline
\end{tabular}
\end{table}

\begin{figure}
 \includegraphics[width=\columnwidth]{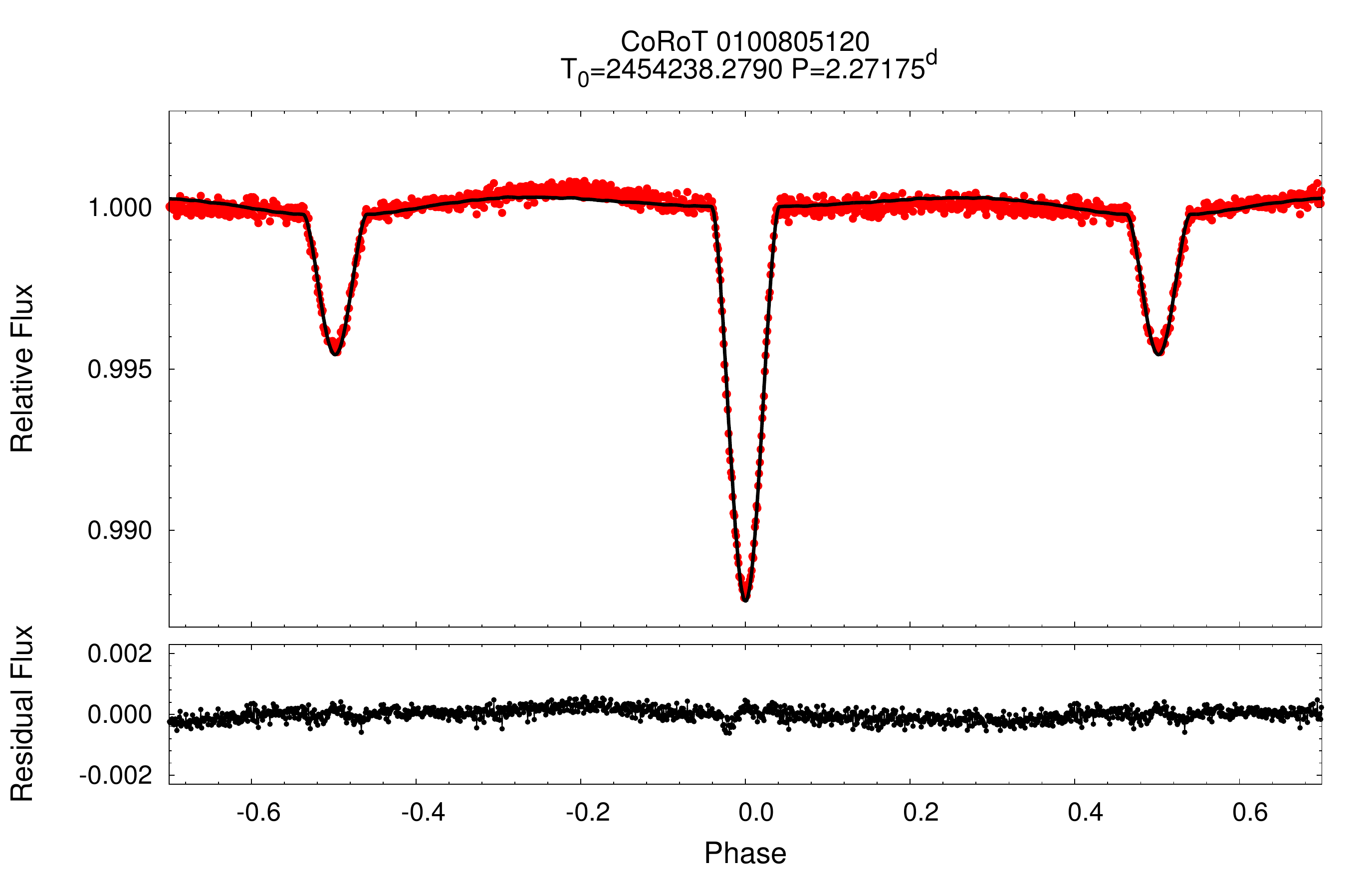}
  \caption{Folded, binned, averaged light curve of CoRoT 100805120 (red) together with the synthesized light curve solution (black) and the residual curve (below).}
  \label{Fig:C0100805120lcfit}
\end{figure}

\begin{figure}
\includegraphics[width=\columnwidth]{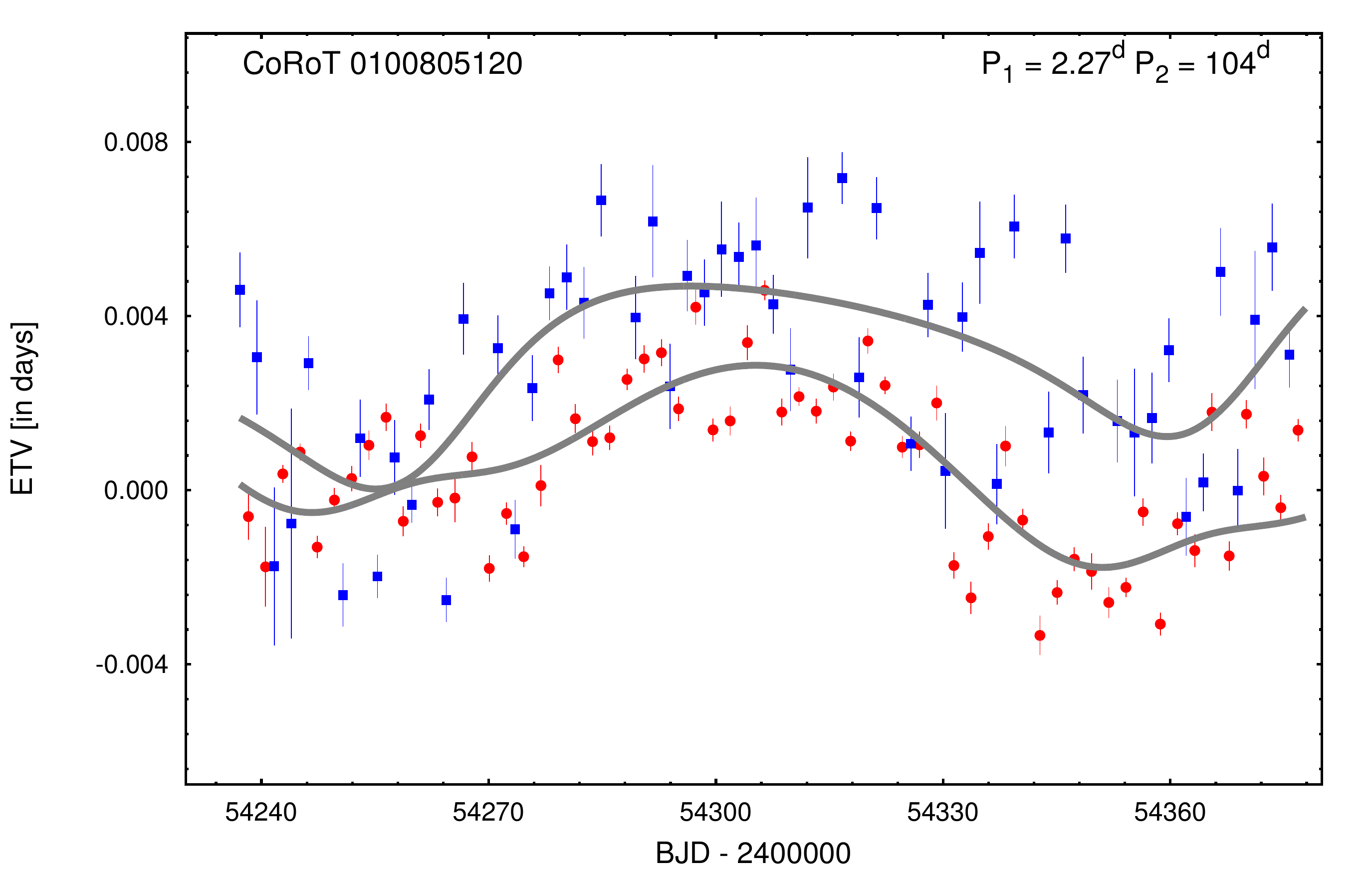}
  \caption{Primary (red points) and secondary (blue rectangles) ETV curves of CoRoT 100805120 together with the combined LTTE+dynamical ETV solution (grey).}
  \label{100805120fit}
\end{figure}

{\it CoRoT 100805120}. This is an Algol-type EB with $P\approx2\fd27$ days period in the CoRoT-LRc01 field \citep{cabreraetal09}. The eclipse depths are only $\sim1.1\%$ and $\sim0.4\%$ for the primary and secondary minima, respectively. Such small amplitude eclipses with relatively high secondary to primary eclipse depth ratio and long eclipse durations ($\sim0\fp075$ in phase) are good indicators of a significant amount of third light in the light curve. Our light curve analysis (see the first column of Table\,\ref{Tab:lcfoldfit}, and also Fig.\,\ref{Fig:C0100805120lcfit}) reveals that only less than 8\% of the total flux of the CoRoT light curve comes from the eclipsing pair. (Note, the ExoDat catalog gives only 1.6\% contamination rate, therefore the source of the extra light should really be the primary, unresolved CoRoT target.) According to our analysis the EB has a slightly eccentric orbit seen almost along the direction of the major axis. The asymmetric out-of-eclipse section of the folded light curve may also exhibit some rotational modulations.

The ETVs of both the primary and secondary minima show sinusoidal features and furthermore, the slight divergence between the two curves might be indicative of apsidal motion. Therefore, we were looking for combined LTTE+dynamical ETV solution, allowing apsidal motion, too. We made runs both with freely adjusted (i.e unconstrained -- mode AP1, see \citealt{Borkovits2015}, Sect. 2.2) and dynamically constrained (mode AP2) apsidal advance rates. As the unconstrained solution resulted in an apsidal advance rate close to the constrained one, we kept the latter, constrained solution. The model ETV curves (together with the observed ETVs) are plotted in Fig.\,\ref{100805120fit}, while the main parameters of our solutions are listed in the first row of Table\,\ref{LTTEdyn}. Note that, in addition of the direct output parameters of the ETV solution, we also listed the the derived masses of the outer binary, i.e. $m_\mathrm{AB}$ and $m_\mathrm{C}$, in the last two columns.

According to our solution the inner and outer orbits are almost coplanar, which is in good agreement with the fact that no eclipse depth variations were detected during the $\sim141$-day-long CoRoT observations.

Considering the masses, we found $m_\mathrm{C}>m_\mathrm{AB}$; that is, the third component seems to be the most massive object in the triple. Therefore, it should probably be the brightest star in the system, unless it was a degenerate object. Its derived mass $m_\mathrm{C}=2.1\pm0.4\,\mathrm{M}_{\sun}$ is in good agreement with the spectral classification of K0III which is given in the ExoDat catalog. Although our solution gives only the total mass ($m_\mathrm{AB}=1.4\pm0.3\,\mathrm{M}_{\sun}$) of the inner, eclipsing binary, with the use the photometric mass ratio\footnote{Note, that \citet{Borkovits2015} showed that the mass ratio $q_1$ of the EB can also be determined from the dynamical ETV analysis when the terms higher order in the period ratio $(P_1/P_2)$ were considered, too. In the present study, however, for the shortness of the data series and their limited accuracy (at least relative to the measurements of the prime {\em Kepler} mission) we decided not to include these terms.} $q_1=0.2\pm0.1$ obtained from the light curve solution above, one can get the individual masses of the two stars forming the inner EB. Furthermore, combining the physical dimensions of the semi-major axis of the EB's orbit deduced from the ETV solution with the fractional radii, one can calculate the stellar radii in physical units, too. We tabulate these derived values in Table\,\ref{Tab:EBfullcomb}. This way we can give another rough estimation for the expected amount of the photometric third light ($l_3$). According to our results, the primary component of the EB might be a moderately evolved solar-like star. Taking therefore the crude estimation  $T_\mathrm{eff,A}=6\,000\pm500$\,K, and using $R_\mathrm{A}=1.7\pm0.2\,\mathrm{R}_{\sun}$ (see Table\,\ref{Tab:EBfullcomb}), we get $L_\mathrm{A}=3.4\pm1.4\,\mathrm{L}_{\sun}$. Assuming that the third component is really a K0III star, its luminosity is expected to be in the range $25\lesssim L_\mathrm{C}\lesssim 100\,\mathrm{L}_{\sun}$ \citep[see, e.g.][]{kumaretal11} and therefore, one can get $0.83\lesssim l_3\approx L_\mathrm{C}/(L_\mathrm{A}+L_\mathrm{B}+L_\mathrm{C})\lesssim0.98$. This result is in good agreement with the amount of the third light found in our light curve analysis. Thus, we conclude, that this triple candidate could join the still few-membered club of {\em Kepler}-spacecraft discovered compact hierarchical triple systems, in which the distant tertiary component is a red giant star (e.g. HD\,181068 -- \citealp{Derekas2011,borkovitsetal13},  KIC\,07690843  -- \citealp{gaulmeetal13,Borkovits2016},   KIC\,07955301   -- \citealp{Rappaport2013,gaulmeetal13}).

\begin{figure}
\includegraphics[width=\columnwidth]{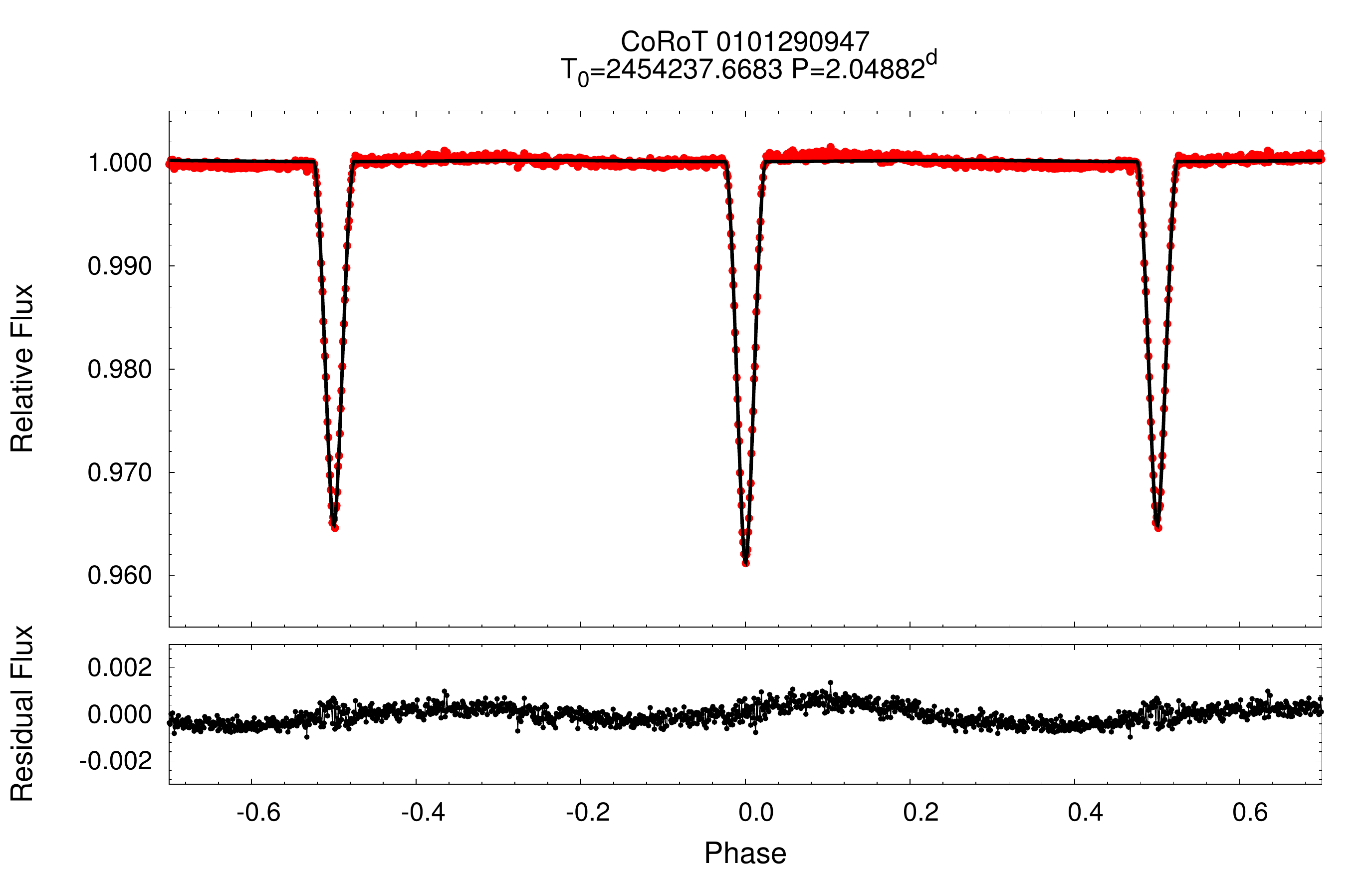}
  \caption{Folded, binned, averaged light curve of CoRoT 101290947 (red) together with the synthesized light curve solution (black) and the residual curve (below). The small systematic residuals may be the manifestation of rotational modulations due to chromospheric activity.}
  \label{C0101290947lcfit}
\end{figure}

{\it CoRoT 101290947} is another, $\sim2$ day-period detached binary in the LRc-01 field, with almost equally shallow primary and secondary eclipses (with a depth of $~3.5\%$, see Fig.\,\ref{C0101290947lcfit}). Its eclipsing nature was first reported by \citep{cabreraetal09}. Moreover, the large amplitude, sine-like ETV has also been noticed and interpreted as LTTE by the same group but, apart from a conference poster, their findings have remained unpublished (Cabrera, 2016, private communication). Our light curve solution (second column in Table\,\ref{Tab:lcfoldfit}) has resulted in an eclipsing pair seen almost edge-on, formed by two very similar stars, and a huge $l_3\approx90\%$ third flux contribution, $\sim 10-11\%$ of which -- according to ExoDat -- may arise from resolved contaminating sources. Therefore, similar to the previous system, the spectral information given in previous works cannot be used for discussing the fundamental physical properties of the binary.\footnote{For this system ExoDat and \citet{sarroetal13} contradict each other. The former gives spectral classification G0III, while the latter gives $\log g=4.6$, which suggest luminosity class V (i.e. main-sequence star).} Considering, however, the obtained surface gravity indicators $(\log g^*_\mathrm{A,B}=4.7)$, we may assume that the binary is composed of two low-mass main-sequence stars.\footnote{A little caution is needed here on the accurate values of the two strongly correlated quantities of the third light ($l_3$) and the inclination ($i_1$). Our $\log g^*_{A,B}$ values would suggest either very low-mass or undersized main-sequence components. Despite the fact that the obtained $i_1$ and $l_3$ values were found to be very similar and robust in all Markov Chains, we cannot exclude the possibility that this result might have been affected by the evident out-of-eclipse distortions, and the true inclination and third flux may be somewhat lower and therefore, the radii of the stars a bit larger.} Finally, we note that the residual light curve shows a systematic sine-like structure of the order $\sim1000$\,ppm. This feature may come from rotational modulation (see bottom panel of Fig.\,\ref{C0101290947lcfit}).

\begin{figure}
\includegraphics[width=\columnwidth]{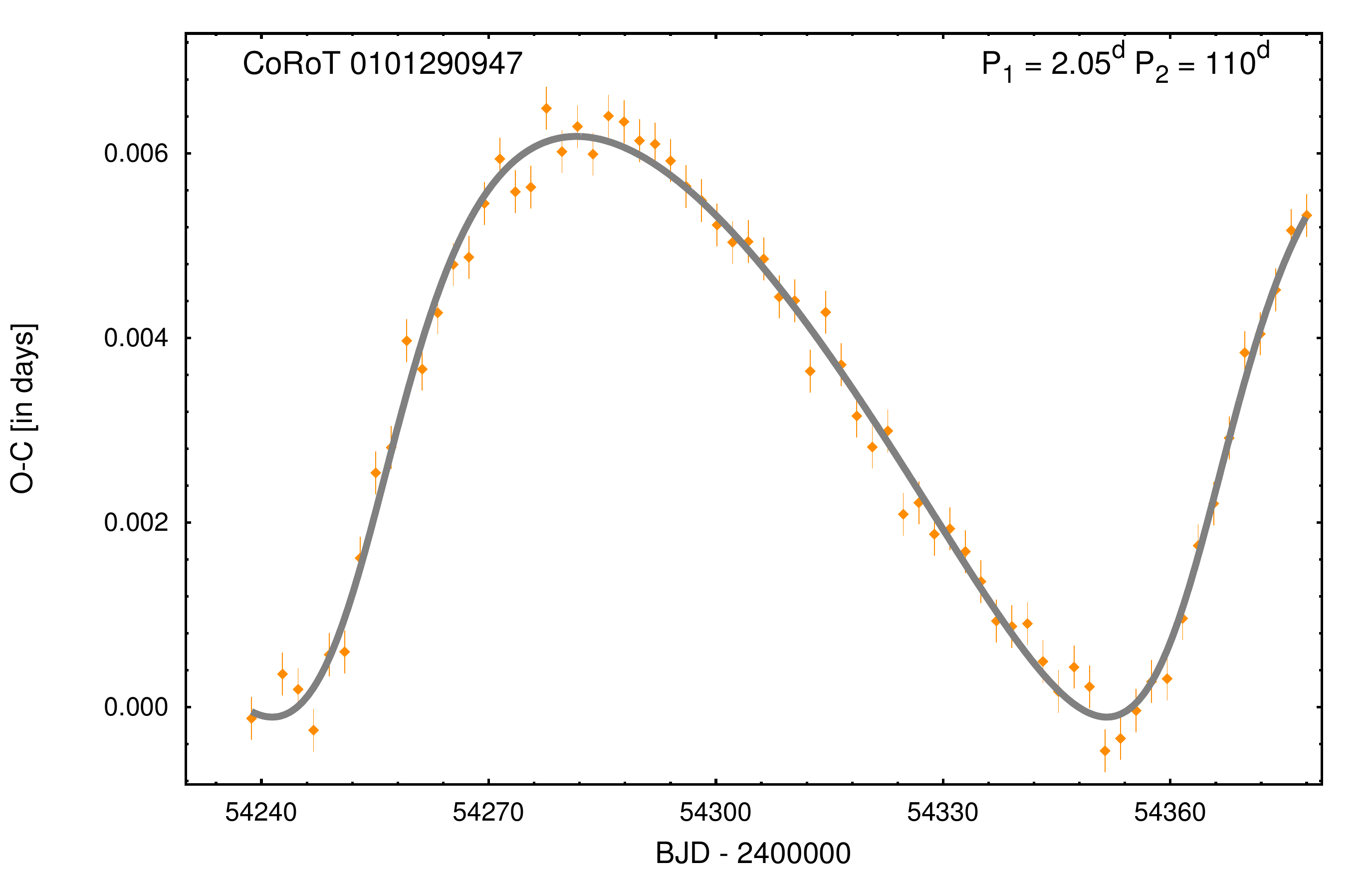}
  \caption{Timing curve for the averaged ETV data of CoRoT 101290947 together with the combined LTTE+dynamical solution for coplanar configuration.}
  \label{101290947fit}
\end{figure}

Turning to the ETV analysis, for the circular inner orbit we used the average of equally good quality primary and secondary ETV curves for our analysis.(The advantages of the use of averaged ETVs were discussed in \citealt{Borkovits2016}.) Our first runs with freely adjusted mutual inclinations resulted in a solution with a mutual inclination $i_\mathrm{m}=21\fdg5^{+3.6}_{-7.8}$. We found, however, that all the physically realistic configurations in this mutual inclination range would have resulted in fast inclination variations ($\Delta i_1\sim\pm1-2\degr$) during the four months of CoRoT observations. Such a large inclination variation would have given rise to remarkable eclipse depth variations, which was not observed. Therefore we omitted these models, and fixed $i_\mathrm{m}=0\degr$. Our coplanar solution is tabulated in the second row of Table\,\ref{LTTEdyn} (see also Fig.\,\ref{101290947fit}), while the derived individual masses and physical radii of the EB components are given in Table\,\ref{Tab:EBfullcomb}. According to our solution the three stars would have similar masses around $\sim 0.9-1.0\,\mathrm{M}_{\sun}$. Some caution, however, is necessary, as in this case the EB members would be remarkably undersized. On the other hand, the similar mass of the tertiary did not necessarily contradict to the large amount of the third light. This question may be resolved by assuming either that the third component -- again -- is a red giant star, or that part of the extra light comes from a fourth, unresolved source, not necessarily binded to the triple.

\begin{figure}
\includegraphics[width=\columnwidth]{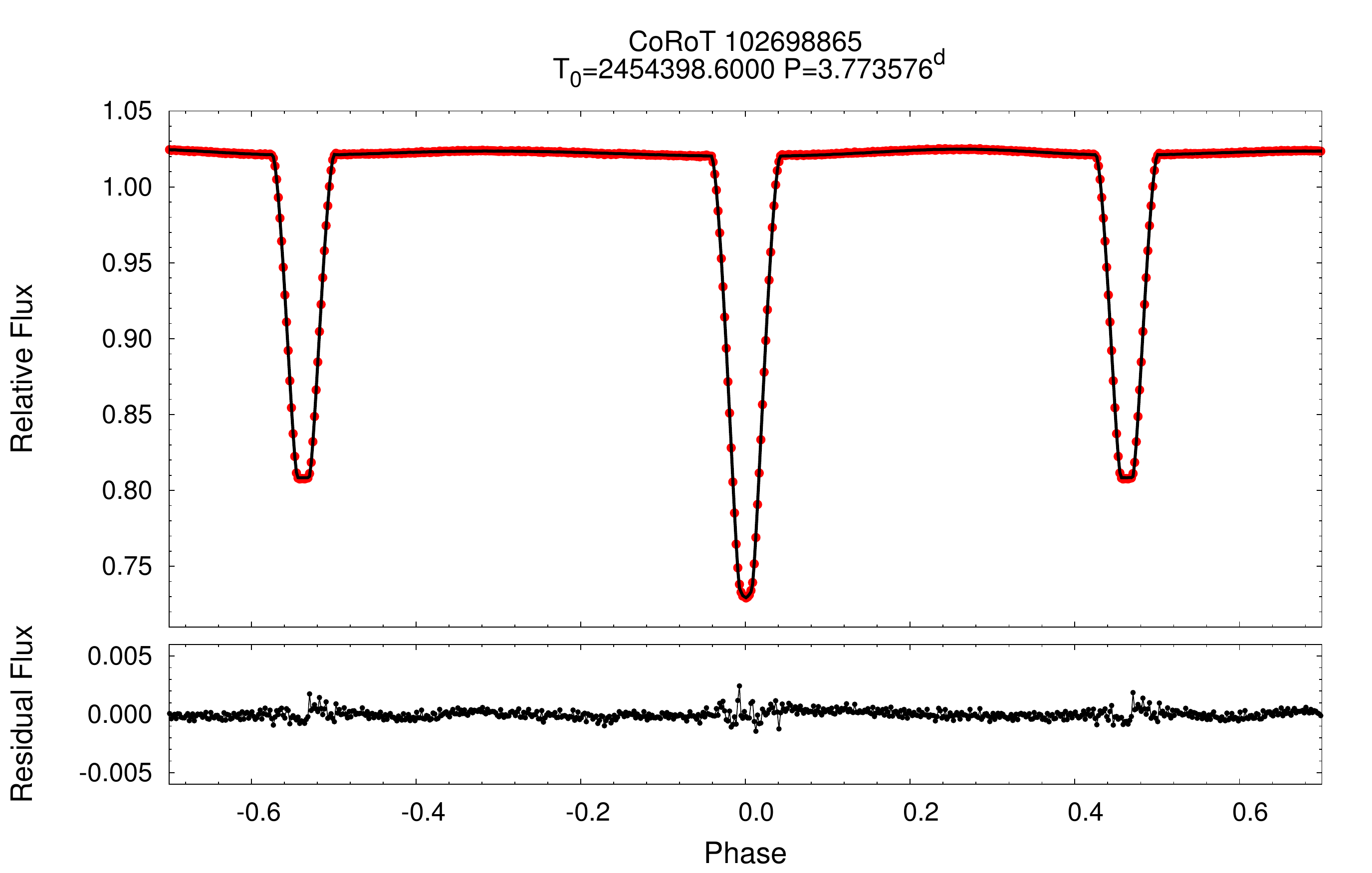}
  \caption{Folded, binned, averaged light curve of CoRoT 102698865 (red) together with the synthesized light curve solution (black) and the residual curve (below).}
  \label{C0102698865lcfit}
\end{figure}

{\it CoRoT 102698865}. This is the longest period ($P_1\sim3\fd77$) EB in our sample, and the only one observed by the spacecraft during two different runs. The datasets LRa1 and LRa6 cover $\sim131$ and $\sim77$ days respectively, with a gap of cca. $\sim1410$ days between the two. The light curve exhibits total eclipses with primary transits and secondary occultations, the latter being slightly displaced from phase $0\fp5$ (Fig.\,\ref{C0102698865lcfit}).
 
As before, first we carried out the analysis of the phase-folded, averaged light curve. The analysis resulted in a significant, but nevertheless not dominant third light ($l_3\approx15-20\%$). (The contamination rate tabulated in ExoDat catalog is $~0.5\%$.) Therefore in this case we assumed that the spectral classification (A0V) given in the ExoDat catalog indeed refers to the primary of the eclipsing pair, thus its temperature, bolometric albedo and surface gravity exponent were set accordingly. Despite the correctly set limb-darkening and other atmospheric parameters our, solutions failed in the sense that we were not able to modell the eclipses better than with $\sim5000$\,ppm residuals. Such, relatively higher light curve residuals were found by other authors too for the high-accuracy {\em Kepler}- and CoRoT light curves. The possible reasons, including the not fully adequate physical models of the stellar atmospheres, were discussed briefly by \citet{hambletonetal13}. Therefore, similar to, e.g. \citet{southworthetal11}, we decided to adjust the (logarithmic) limb-darkening coefficients, too. As result a substantially improved solution was obtained, tabulated in Table\,\ref{Tab:lcfoldfit}, and plotted in Fig.\,\ref{C0102698865lcfit}. Our result is in agreement with this spectral classification, as we found ($\log g^*_\mathrm{A}=3.93\pm0.01$) which, substituting the typical mass of a main-sequence, early A-type star, gives a surface gravity about $\log g_\mathrm{A}\approx4.10$ corresponding to this spectral type. 

\begin{figure*}
\includegraphics[width=\columnwidth]{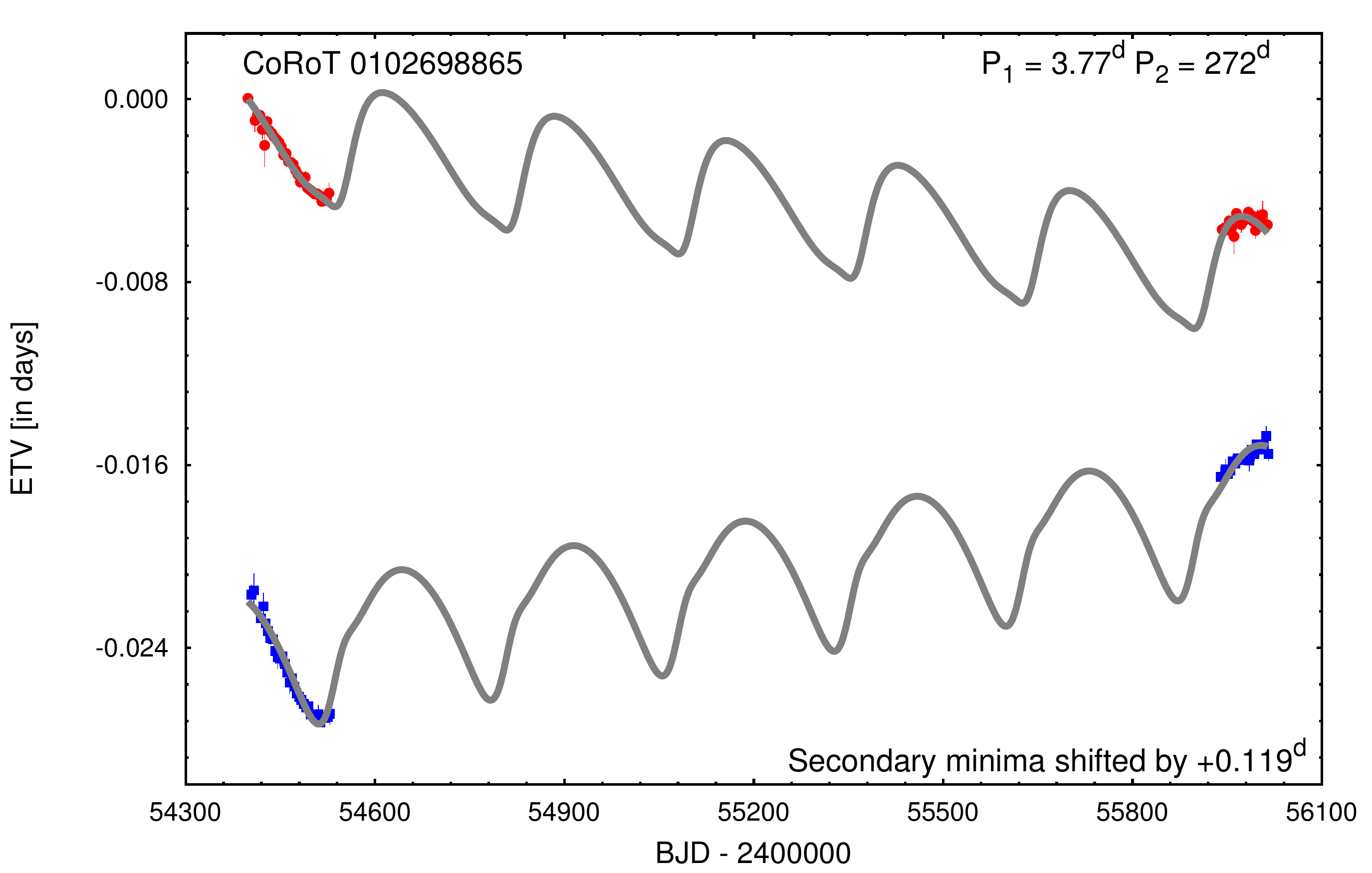}\includegraphics[width=\columnwidth]{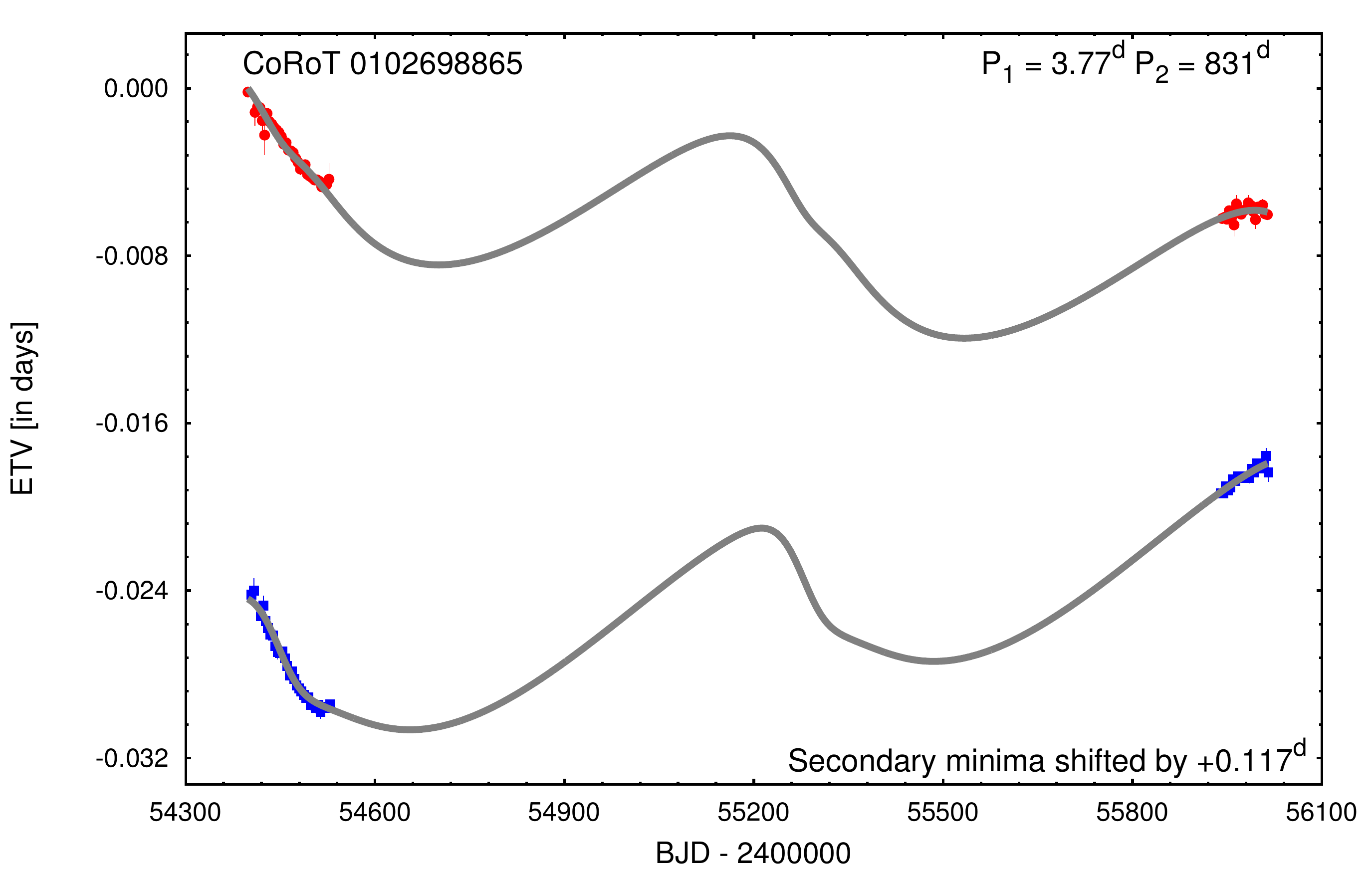}
  \caption{ETVs of primary (red) and secondary (blue) minima of   CoRoT 102698865   together with two different combined LTTE+dyn ETV solutions (grey). In case of the shorter outer period solution (left) the convergence of the two curves reveals relatively rapid apsidal motion. (Note, the secondary curves are upshifted by $\sim0\fd12$ for better visibility.)}
  \label{Fig:C0102698865ETV}
\end{figure*}

Considering the ETVs, the primary and secondary curves clearly converge to each other, which is an evidence of the apsidal motion. Furthermore, the slopes and the curvatures of the curves are very different in the two observing runs. Consequently, the presence of a third star, perturbing the motion of the EB is a reasonable assumption. On the other hand, however, the two segments of the ETVs do not show evident periodicities, which makes the period of the third body, and thus, any quantitative ETV solutions less certain. Therefore, it is not surprising that, instead of a unique solution, we found two similarly acceptable third body configuraions, with substantially different outer periods.\footnote{Strictly speaking, a third set of formal third-body solutions was also found in the outer period range $P_2\sim1100-1300$\,days, but these solutions resulted in astrophysically unrealistic stellar masses and therefore, in what follows, we do not consider them.} We tabulate the results of both solutions in the third and fourth rows of Tables\,\ref{LTTEdyn} and \ref{Tab:EBfullcomb}, and plot them in the two panels of Fig.\,\ref{Fig:C0102698865ETV}. While in the case of the shorter outer period solution the two orbits were found to be practically coplanar, the other solution resulted in a higher mutual inclination $i_\mathrm{m}=39\pm3\degr$. The angle between the ascending nodes of the two orbits was found to be $\Delta\Omega=\Omega_2-\Omega_1=-24\pm7\degr$, which results in an observable inclination of $i_2=58\pm10\degr$ for the outer orbit. This solution predicts an inclination variation of $\Delta i_1\approx0\fdg2$ for the 1617-day long interval between the first and the last CoRoT data points, which, due to the total eclipses, remains below the limit of the observable eclipse depth variations. Furthermore, the obtained masses for the binary members in both solutions are in good agreement with the results of the light curve analysis. Comparing the obtained amount of third light $l_3=0.15\pm0.01$ with the mass of the third component ($m_\mathrm{C}=1.1\pm0.1\,\mathrm{M}_{\sun}$ and $1.9\pm0.3\,\mathrm{M}_{\sun}$ for the shorter and longer outer period solutions, respectively), one can see, that for the second solution this value is in perfect agreement with the expected contribution of a main-sequence tertiary with the given mass, while for the first case the third star might be an evolved object or, some additional sources should also be assumed.

\begin{figure}
\includegraphics[width=\columnwidth]{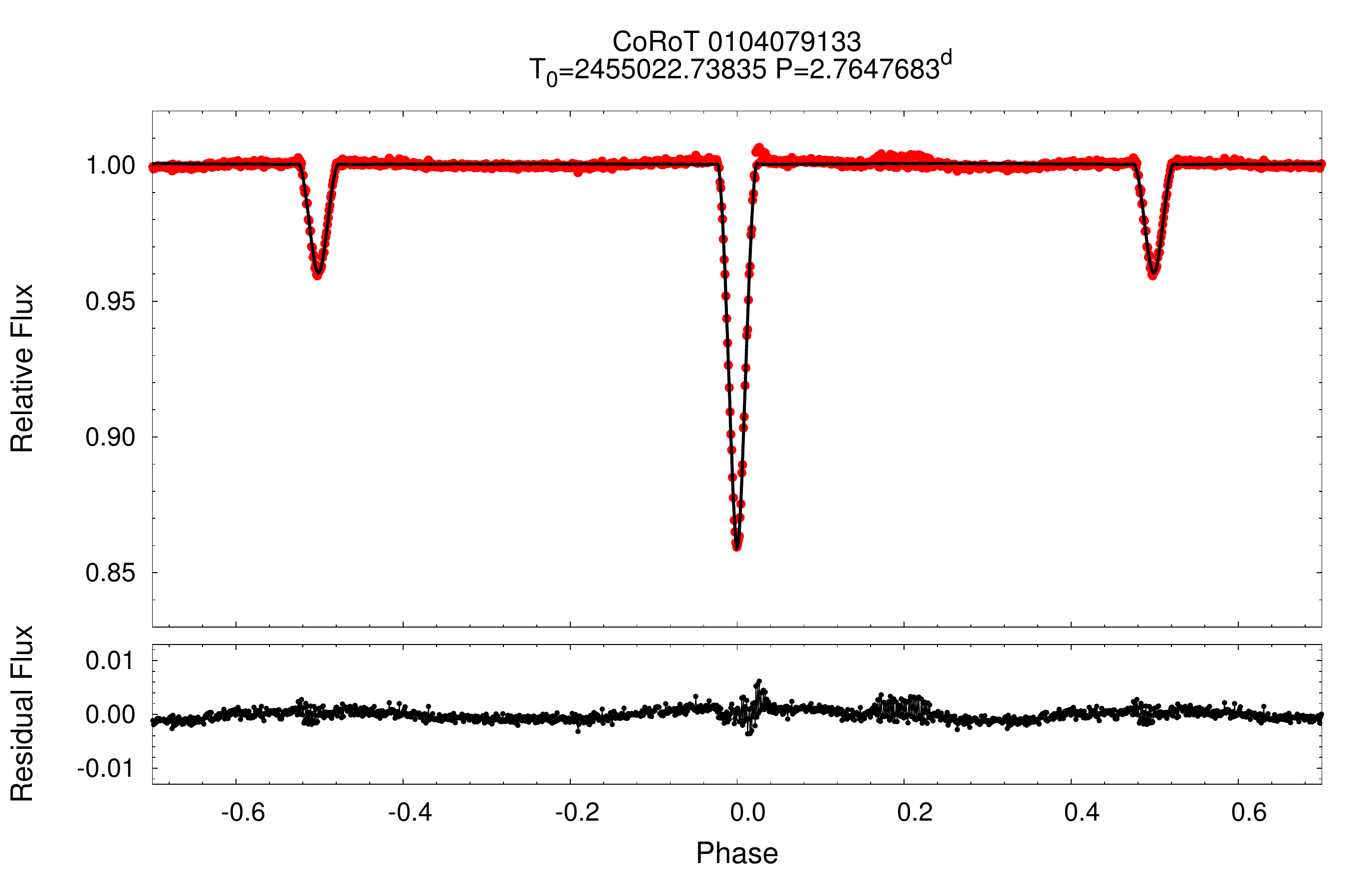}
  \caption{Folded, binned, averaged light curve of CoRoT 104079133 (red) together with the synthesized light curve solution (black) and the residual curve (below).}
  \label{C0104079133lcfit}
\end{figure}

\begin{figure*}
\includegraphics[width=\columnwidth]{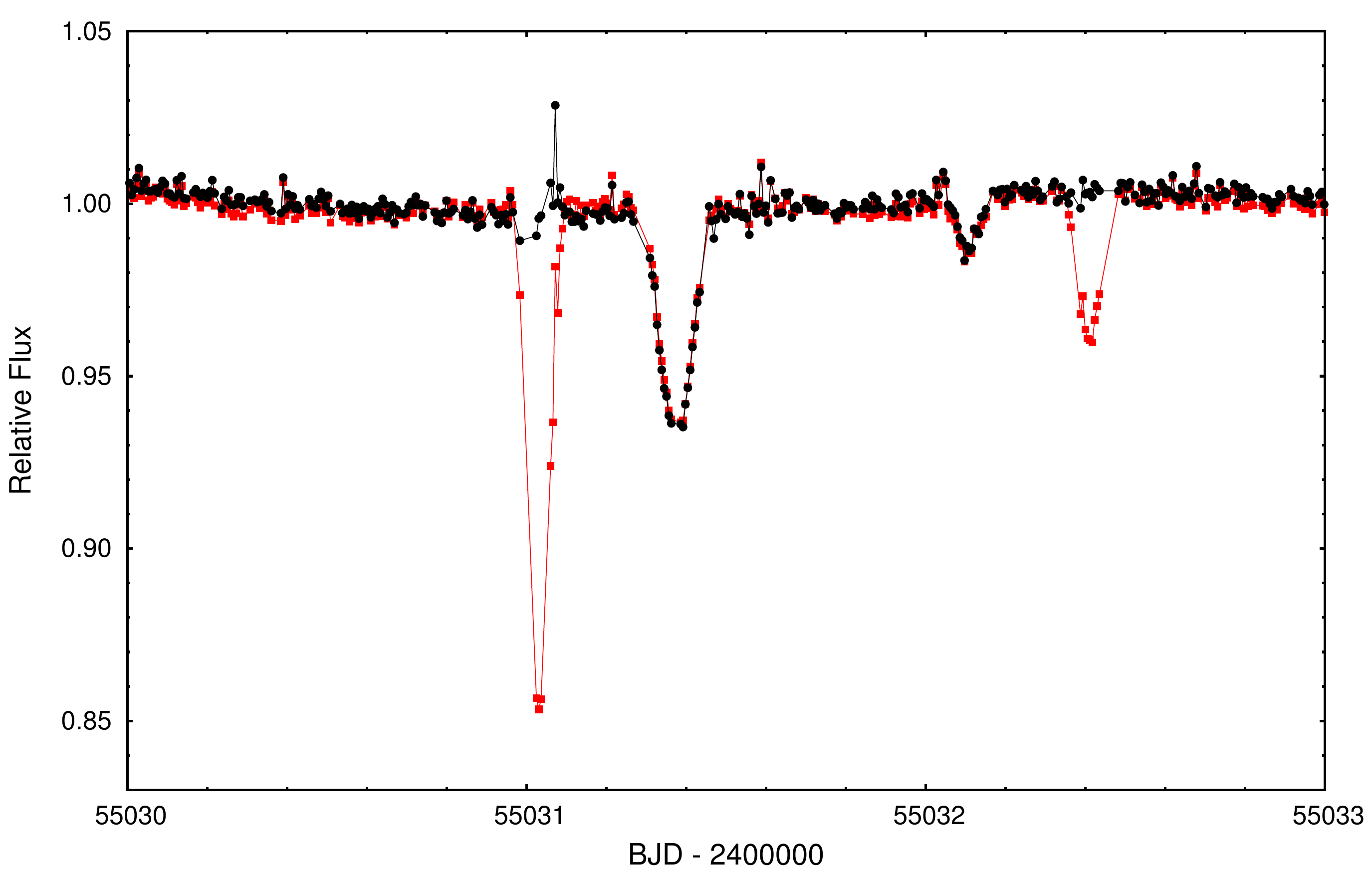}\includegraphics[width=\columnwidth]{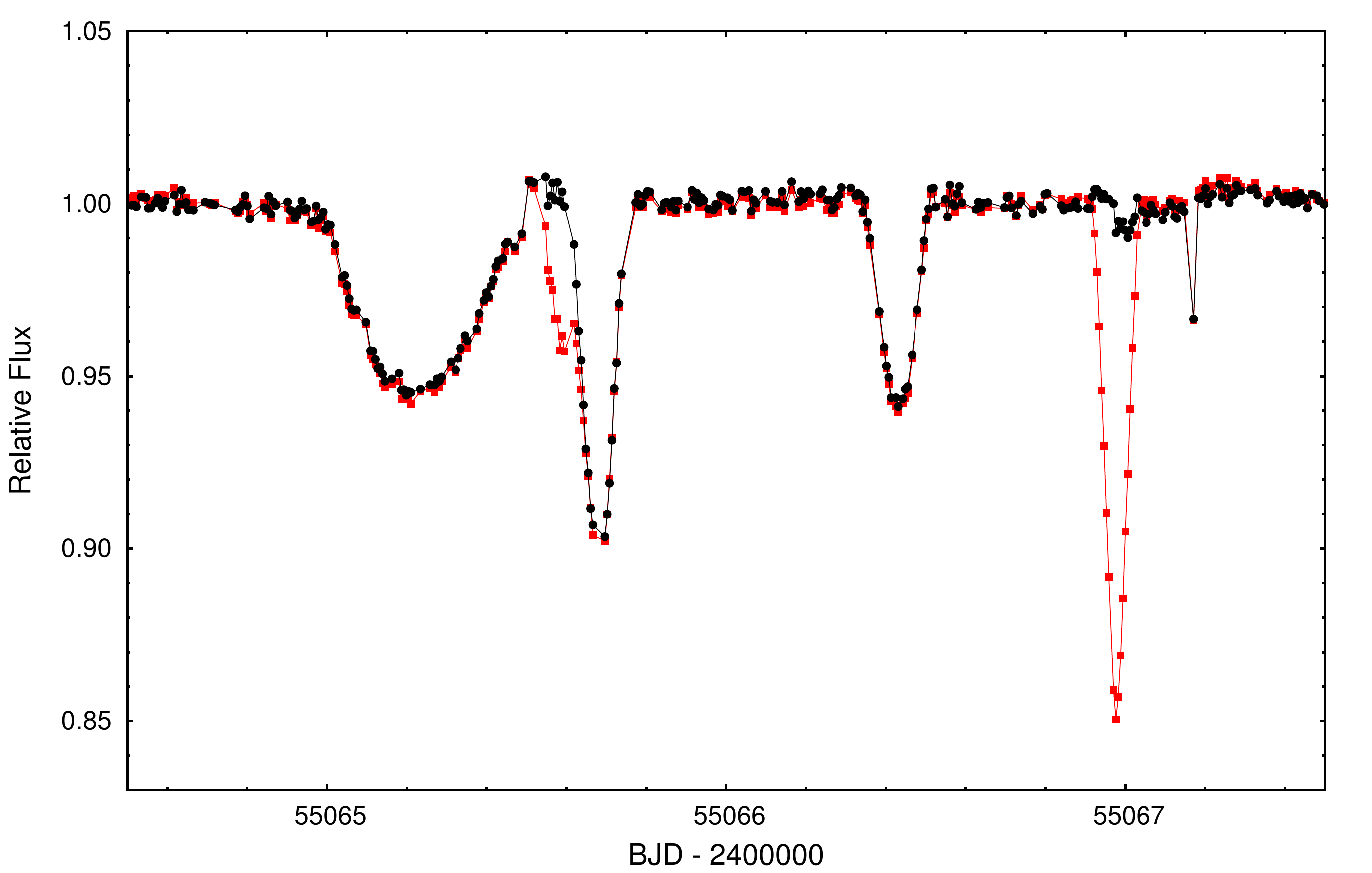}
  \caption{The extra eclipsing events on the observed, detrended light curve of CoRoT 104079133 (red). The black curve represents the residual light curve obtained after the interpolated removal of the phase-folded, averaged light curve from the observed, detrended curve. The complex characteristics of the extra eclipses make it certain that CoRoT 104079133 is (at least) a triply eclipsing hierarchical triple system.}
  \label{Fig:C0104079133E3}
\end{figure*}

{\it CoRoT 104079133} is another marginally eccentric detached EB with a period of $P_1\sim2\fd7$ and moderately differing eclipse depths (see Fig.\,\ref{C0104079133lcfit}), observed during the run LRc04. The most exciting features of the light curve are the two groups of extraneous eclipses with various shapes about BJDs\,2\,455\,031-32 and  2\,455\,065-66 (see Fig.\,\ref{Fig:C0104079133E3}) which makes it very likely that  CoRoT\,104079133  belongs to the small group of triply eclipsing hierarchical triple systems. According to our light curve solution (Table\,\ref{Tab:lcfoldfit}) the extra flux dominates ($l_3\approx72$\%) the CoRoT observations. (The outer contamination rate, given in ExoDat, is about $\sim 2$\%.) Therefore, we may expect again, that the spectral informations given in ExoDat does not refer to the EB members, but to the source of the extra flux.

\begin{figure}
\includegraphics[width=\columnwidth]{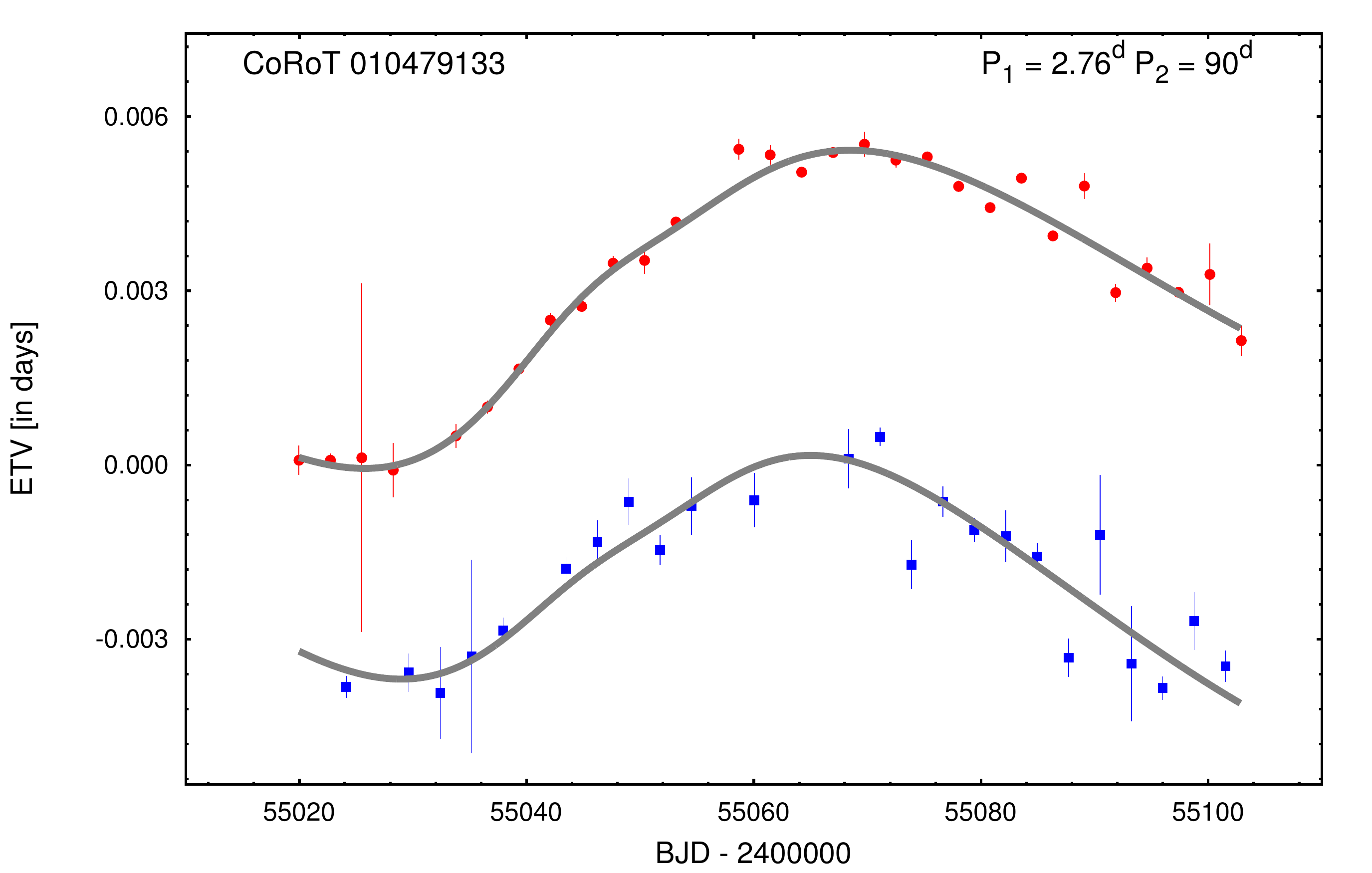}
  \caption{Eclipse timing diagrams of primary (red) and secondary (blue) eclipses of CoRoT 104079133 system together with the accepted LTTE+dynamical ETV solution (grey).}
  \label{Fig:C0104079133ETV}
\end{figure}

We were looking again for a combined LTTE+dynamical ETV solution. Due to the marginal eccentricity of the inner EB we decided to take into account both the primary and secondary ETVs despite the fact that the latter data had significantly larger uncertainties. Naturally, this also implies the inclusion of the apsidal motion terms into our analysis. Our results (see Fig.\,\ref{Fig:C0104079133ETV}) are tabulated in the fifth rows of Tables\,\ref{LTTEdyn} and \ref{Tab:EBfullcomb}. The most important finding is that the moments of the inferior and superior conjunctions of the EB and the third component relative to the Earth ($t_\mathrm{inf}=55\,033\pm1$\,RBJD and $t_\mathrm{sup}=55\,067\pm1$\,RBJD) are in very good agreement with the locations of the extra eclipses in the light curve. This makes it very likely that the ETV and the outer eclipses are caused by the same object.

From the locations of the extra eclipses relative to the two kinds of conjunctions points, one can make a few simple, qualitative statements on the geometry of the extra eclipses. Thus, given that the first set of the extra eclipses (consisting of two individual fadings, see the left panel of Fig.\,\ref{Fig:C0104079133E3}) occurred around the inferior conjunction, i.e. when the third object was located between the Earth and the EB, it shows the tertiary component eclipsing the members of the inner binary. As the first event did happen after a primary eclipse of the EB, in the case of (almost) coplanar orbits with prograde revolutions (i.e. $i_\mathrm{m}=0\fdg9\pm0\fdg6$), the primary component was eclipsed first, while during the second, shallower fading, which occurred before the forthcoming secondary eclipse, the secondary star was eclipsed.

A more complex structure of three extraneous events of the second group of the extra eclipses can be seen in the right panel of Fig.\,\ref{Fig:C0104079133E3}. These events were observed close to the superior conjunction. Therefore, here the EB members eclipsed the third star. The first, long-duration event occurred just before a secondary eclipse, therefore, here the secondary component should have been the eclipser. The eclipse duration was necessarily longer, because the secondary, in its revolution around the primary of the EB was moving in these moments in almost the opposite direction to the revolution in the outer orbit. Then, just after the mid-time of the secondary eclipse, the primary component also eclipsed the tertiary. Due to the similarly directed revolution of the primary on both the inner and outer orbits at those moments, this event was the shortest. Finally, just after the quadrature position of the inner EB, the secondary component eclipsed again the tertiary star. A more detailed, quantitative light curve analysis of this triple star, capable of resolving the ambiguity of prograde vs. retrograde revolution, is planned in a future work.

Considering the masses obtained, our combined analysis resulted in similar masses for the primary and tertiary components ($m_\mathrm{A}=0.95\pm0.16\,\mathrm{M}_{\sun}$ and $m_\mathrm{C}=0.81\pm0.20\,\mathrm{M}_{\sun}$, respectively), and a less massive secondary star ($m_\mathrm{B}=0.32\pm0.06\,\mathrm{M}_{\sun}$). While the masses of both the primary and the tertiary are in agreement with the spectral class G5V given in the ExoDat catalog, there is a slight discrepancy with the high amount of the third light ($l_3=0.726\pm0.004$). This fact emphasizes again the importance of a further, more detailed analysis.

\begin{figure}
\includegraphics[width=\columnwidth]{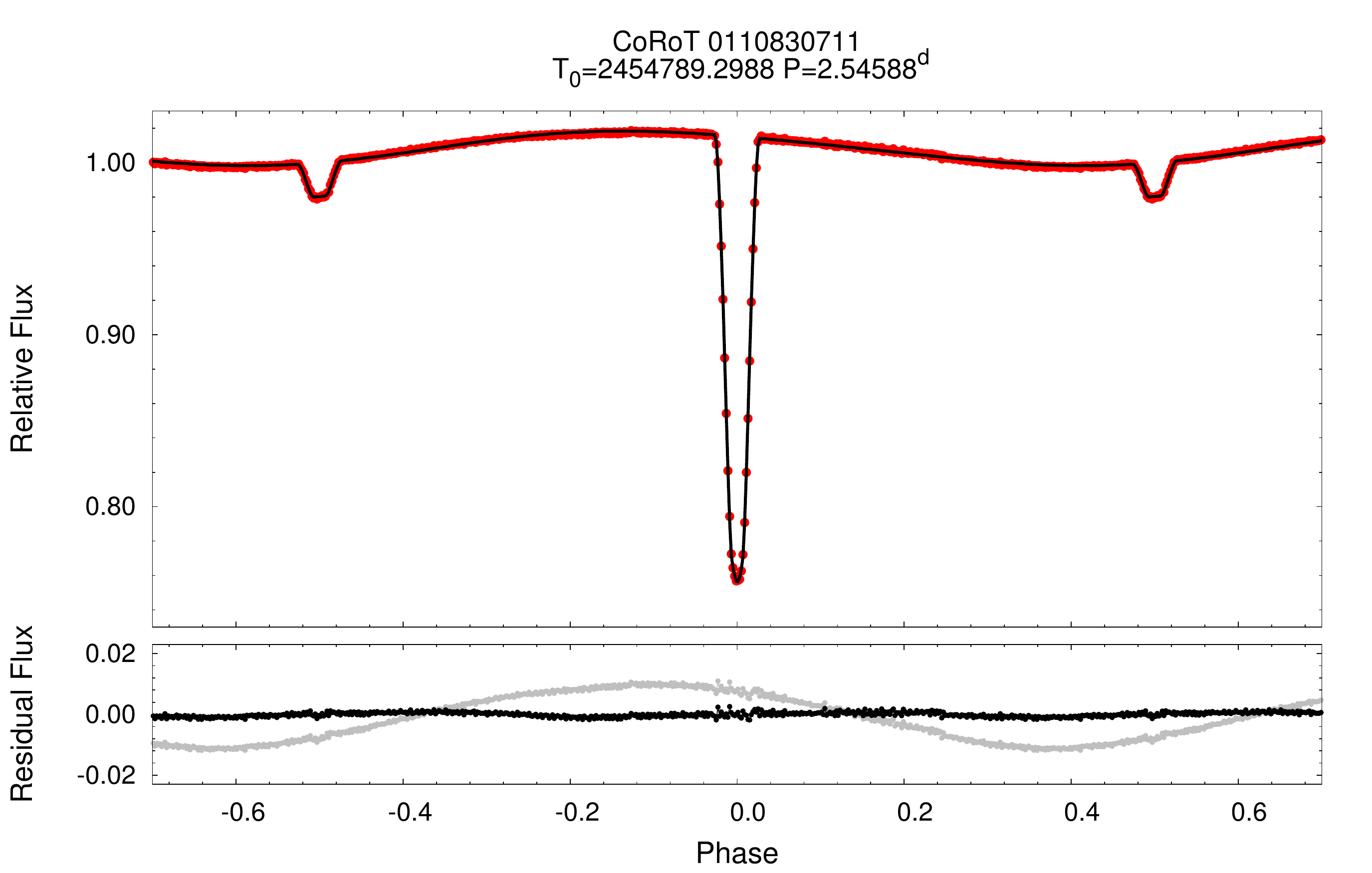}
  \caption{Folded, binned, averaged light curve of CoRoT 110830711 (red) together with the synthesized light curve solution (black). Note, the out-of-eclipse modulation was modelled mathematically with an extra sinusoidal term (see text for details). In the lower panel the residual curves are shown with and without the extra sinusoidal term (black and grey curves, respectively).}
  \label{Fig:C0110830711lcfit}
\end{figure}

{\it CoRoT 110830711} was observed during run LRa02. The Algol-type light curve of this $P\sim2\fd55$-binary shows relatively deep ($\sim 25\%$) primary transits and shallow ($\sim2\%$) secondary occultations. The out-of-eclipse sections exhibit strong quasi-sinusoidal modulations with an amplitude similar to the depths of the secondary eclipses. These modulations remain clearly visible in the folded, binned averaged light curve (see Fig.\,\ref{Fig:C0110830711lcfit}), suggesting rotational origin with a synchronized primary stellar spin rate. Not being our primary interest, for the light curve analysis they were simply modeled mathematically as an extra flux component of the form $\Delta\phi=a\cos(2\upi/P\cdot t)+b\sin(2\upi/P\cdot t)$, where the coefficients $a$ and $b$ were determined with a linear least-squares fitting for each trial set (of light curve parameters) during the MCMC search. The resulting parameters are tabulated in the last column of Table\,\ref{Tab:lcfoldfit}, while the synthesized light curve (and the residuals, with and without the extra trigonometric terms) are plotted in Fig.\,\ref{Fig:C0110830711lcfit}. As it can be seen, the extra flux is almost negligible in this case ($l_3\approx3-7\%$), therefore one can assume that the spectral classification ($F5V$) given in ExoDat refers to the primary component of the EB. Note that our first solutions with freely adjusted mass ratio resulted in unrealistically low mass ratios of $q_1\sim0.002-0.008$. But, as we have emphasized previously, the photometric mass ratio is known to be an ill-determined quantity for detached systems. Therefore we resorted to another kind of constraining the mass ratio, with the combination and inversion of the zero age main-sequence mass--luminosity and mass--radii relations of \citet{toutetal96}, in the same way as Sect.\,7 of \citet{rappaportetal17} did. (To do this, the effective temperature of the primary was set to $T_\mathrm{eff1}=7120$\,K, conforming to its spectral type.) Then, comparing the minima of the $\chi^2$ values of the freely adjusted and constrained-q chains, the difference was about $1.4\%$, while all the values of the other parameters remained within the uncertainties given in Table\,\ref{Tab:lcfoldfit}. Hence we conclude that the extreme mass ratio found in our first MCMC analysis is probably false, and the binary most probably consists of two normal main-sequence stars, but with quite different masses.

\begin{figure}
\includegraphics[width=\columnwidth]{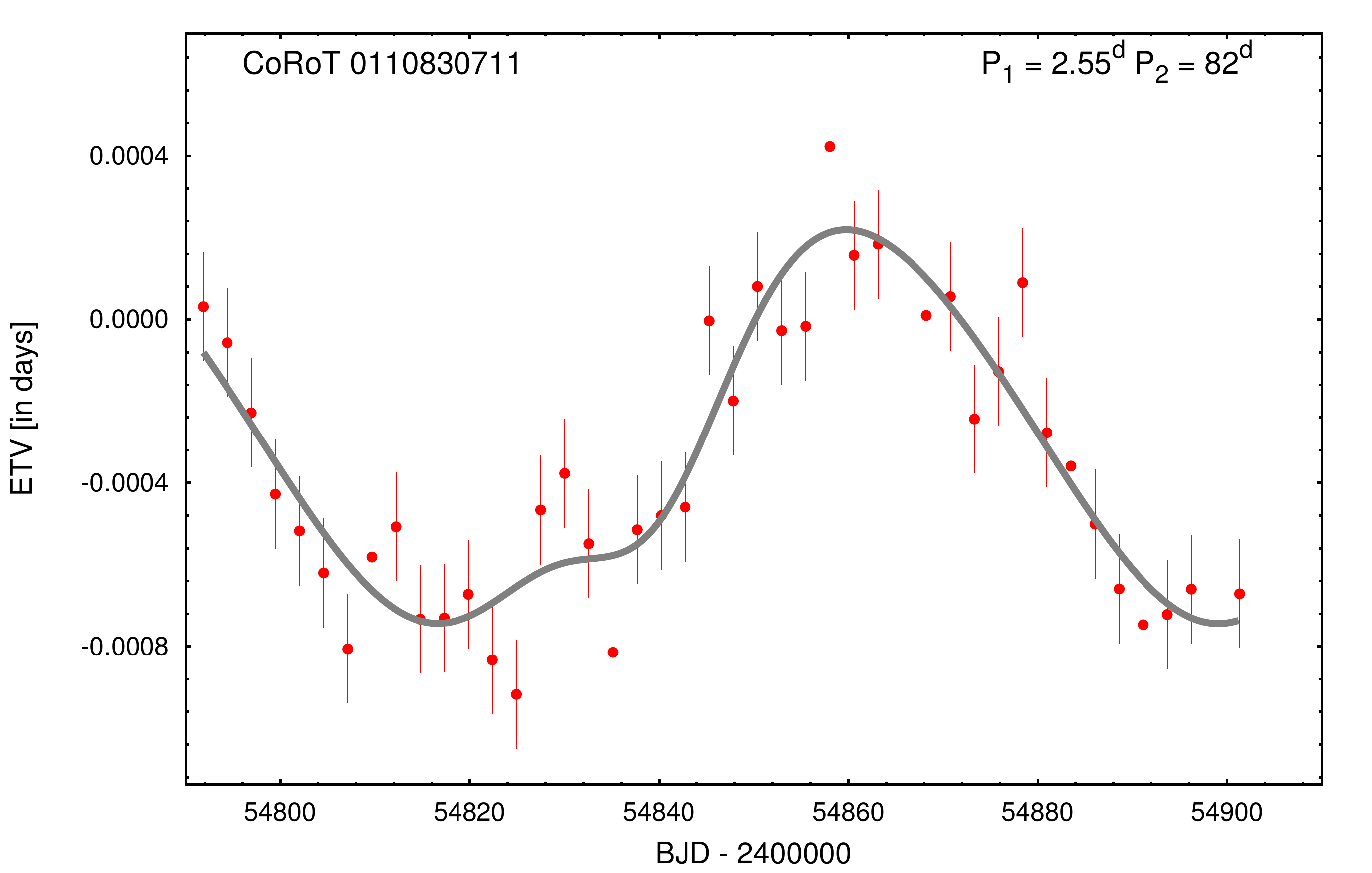}
  \caption{ETV of the primary minima in CoRoT 110830711  and the combined LTTE+dynamical solution curve.}
  \label{Fig:110830711ETV}
\end{figure}

Turning to the ETV solution, we used only the ETV curve of the primary eclipses, and omitted the secondary ETV curve obtained with a substantially larger scatter from the shallow secondary eclipses. Our combined LTTE+dynamical solution, which has the shortest outer period ($P_2=82\pm2$\,d) in our sample, is tabulated in the last row of Table\,\ref{LTTEdyn} and plotted in Fig.\,\ref{Fig:110830711ETV}. According to our results, the two orbits are slightly misaligned ($i_\mathrm{m}=4\fdg9^{+5\fdg2}_{-1\fdg3}$). Therefore one can expect a precession of the EB's orbital plane with an amplitude of $\sim4-10\degr$ on a timescale of $\sim20-40$ years \citep[see, e.g.][for a detailed discussion on the orbital precession induced by a third star on misaligned orbit]{Borkovits2015}. Considering the individual stellar masses deduced from the combination of the light curve and ETV solutions (last row in Table\,\ref{Tab:EBfullcomb}), the mass of the primary ($m_\mathrm{A}=1.25\pm0.24\,\mathrm{M}_{\sun}$) within its $1\sigma$ uncertainty, is in agreement with the expected mass of an $F5V$ star. The secondary and tertiary components were found to have similar masses ($m_\mathrm{B}=0.61\pm0.12\,\mathrm{M}_{\sun}$ and $m_\mathrm{C}=0.62\pm0.13\,\mathrm{M}_{\sun}$, respectively). The expected light contribution of such a less massive star is also in good agreement with the small amount of the third light ($l_3=0.06^{+0.01}_{-0.03}$).

\subsection{Systems with extra eclipse(s), but without detectable ETVs.}
\label{Subsect:others}

\begin{figure}
 \includegraphics[width=\columnwidth]{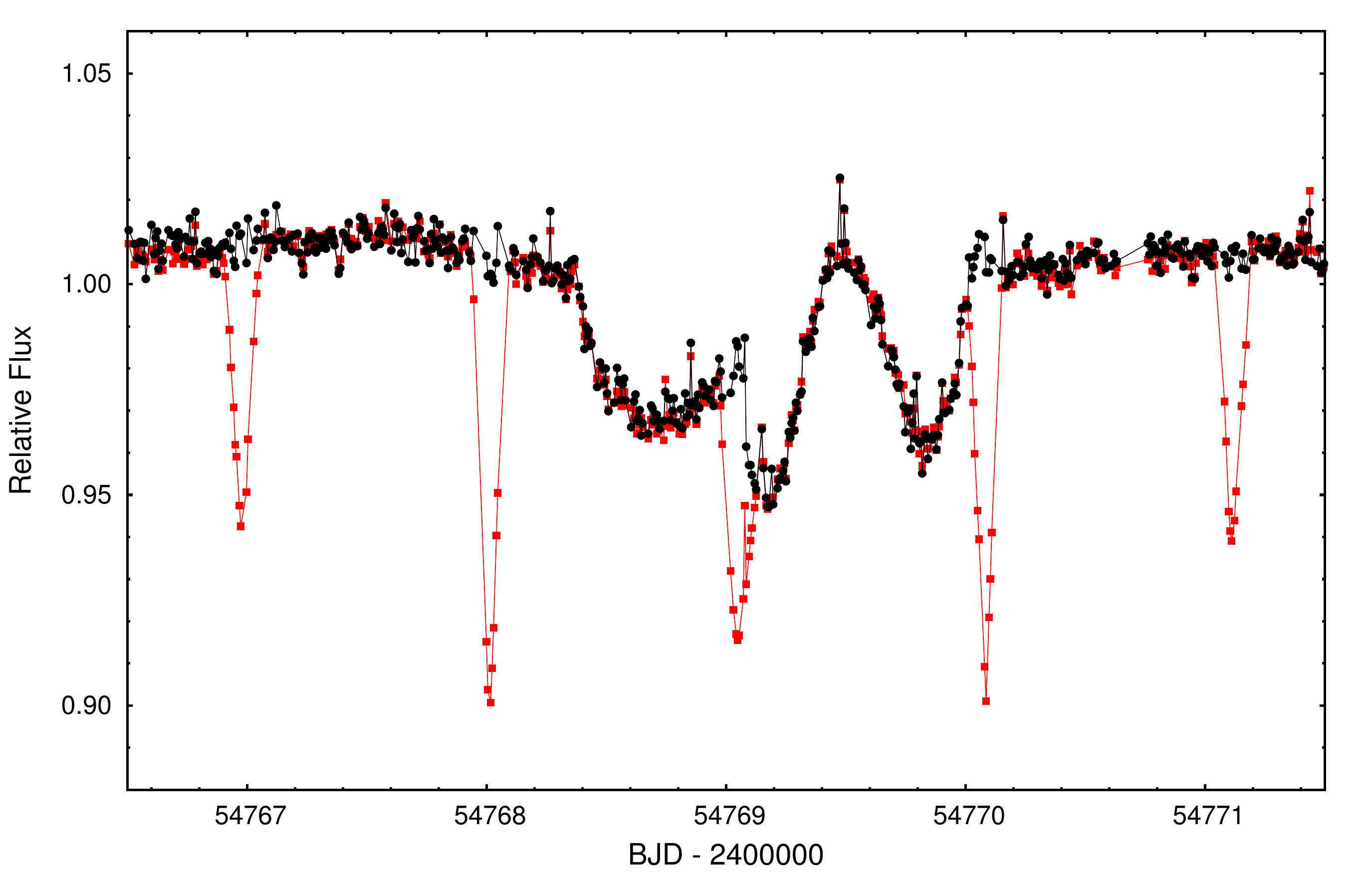}
  \caption{The extra eclipsing events on the observed, detrended light curve of CoRoT 221664856 (red). The black curve represents the residual light curve obtained after the interpolated removal of the phase-folded, averaged light curve from the observed, detrended curve. The complex characteristics of the extra eclipses make it certain that CoRoT 221664856 is at least a triply eclipsing hierarchical triple system.}
  \label{Fig:C0221664856E3}
\end{figure}

We identified some additional CoRoT EBs where extra eclipsing event(s) can be found in the light curve, but do not show detectable third-body signals in their ETV. Among them the most promising hierarchical triple candidate is the SRa02 target CoRoT\,221664856. In this case the complex characteristics of the three extraneous eclipsing events at BJD\,2\,454\,768--2\,454\,770 (Fig.\,\ref{Fig:C0221664856E3}) clearly reveal the triply eclipsing hierarchical triple nature of this system. Unfortunately, the short ($\sim33$-day-long) dataset does not make it possible to get any reasonable ETV solution. Therefore, ground based photometric follow-up observations of this system in the future would be exceptionally worthy. Note, however, that the spectral classification of G2I given in the ExoDat site cannot refer to any of the stars of the inner pair, as it is not possible for such large supergiant stars to form a well-detached $\sim2\fd06$-day-period close binary with any other star. Therefore, if the given luminosity class was valid, the supergiant star should be the tertiary component. 

\begin{figure*}
 \includegraphics[width=\columnwidth]{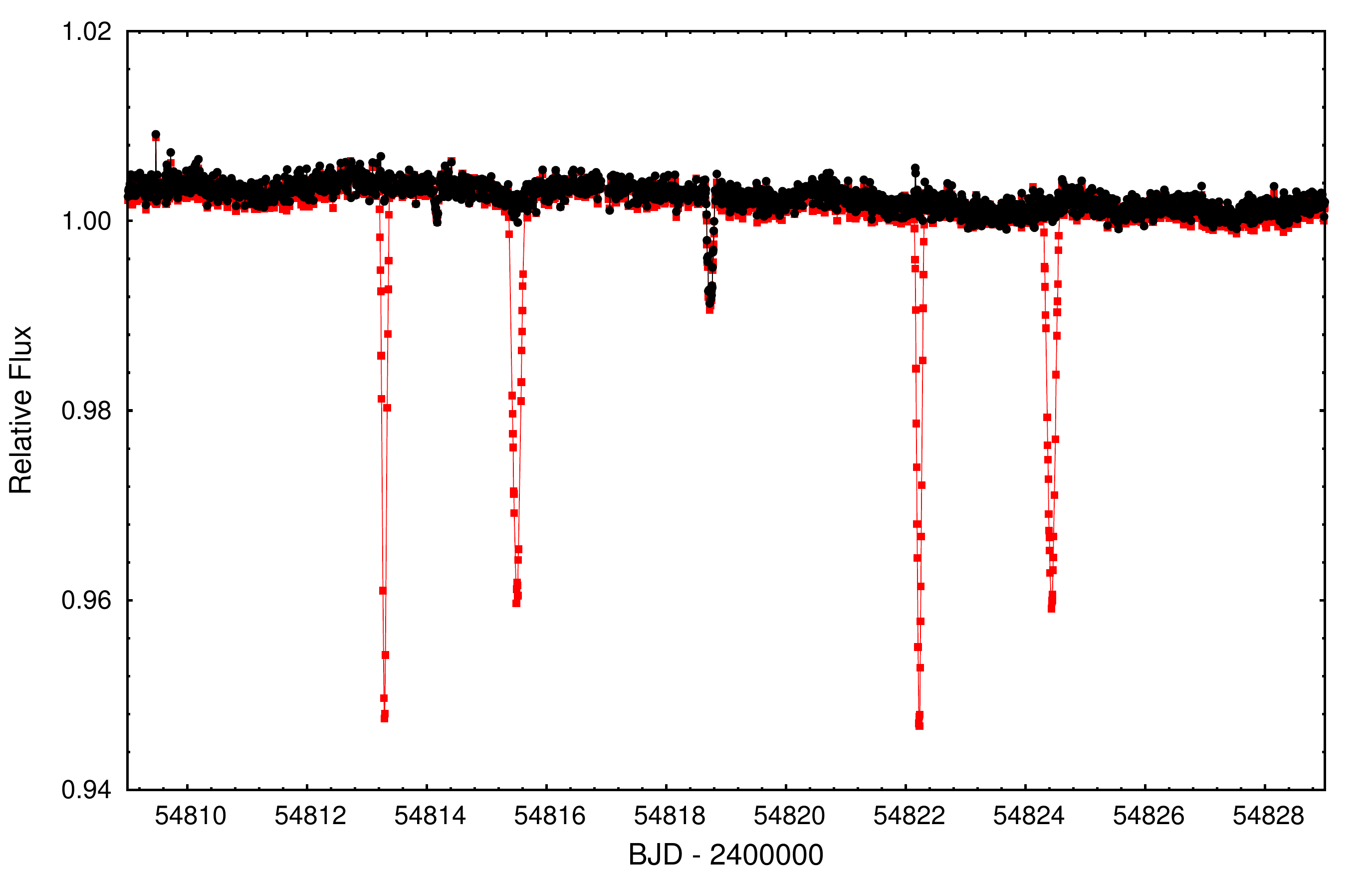}\includegraphics[width=\columnwidth]{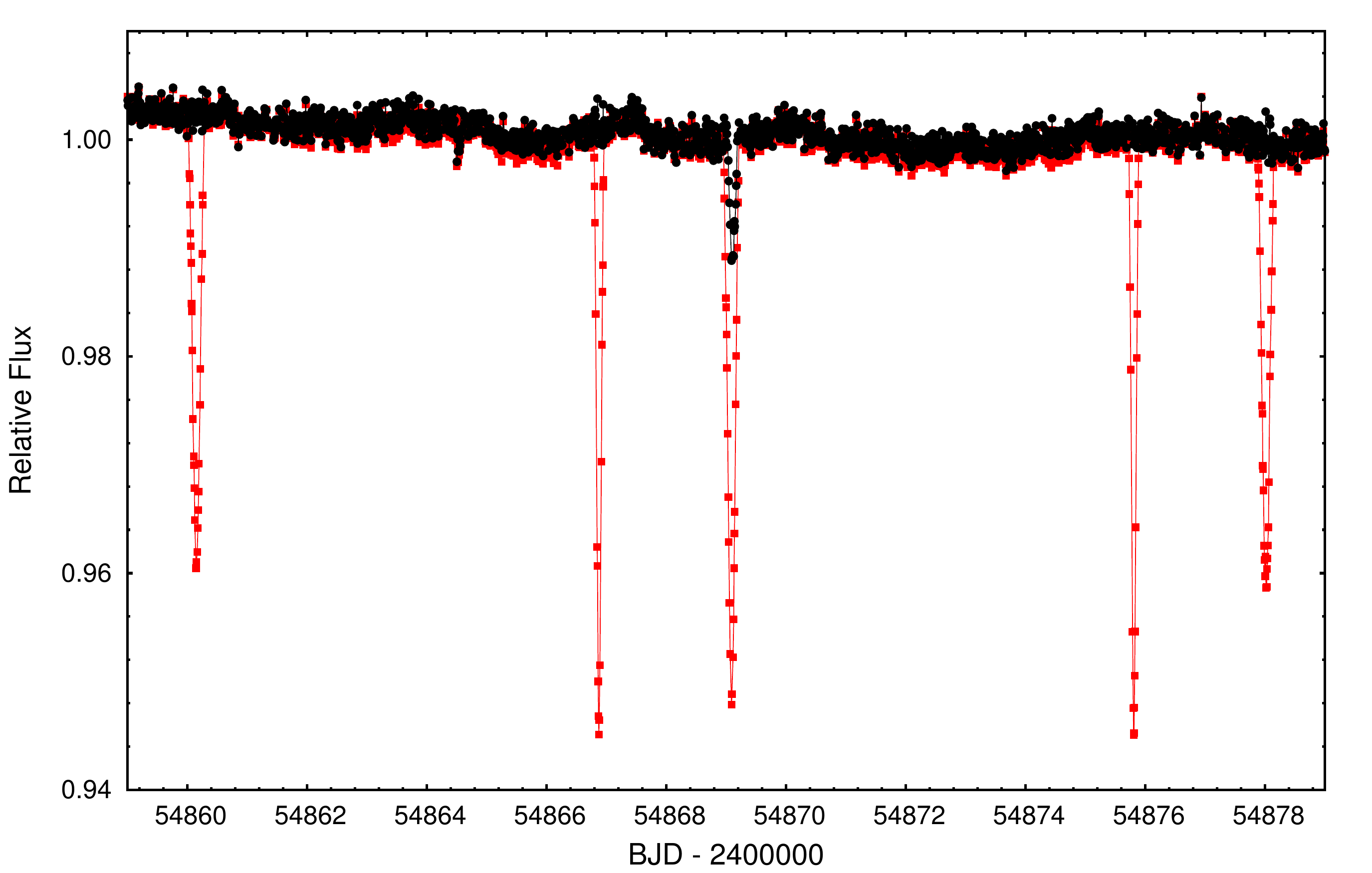}
  \caption{The extra eclipsing events on the observed, detrended light curve of CoRoT 110829335 (red). The black curve represents the residual light curve obtained after the interpolated removal of the phase-folded, averaged light curve from the observed, detrended curve. The two sets of a shallow fading (secondary eclipse of binary B) followed regularly by a somewhat deeper other fading (primary eclipse in binary B) reveal the two-EB blended nature of the light curve.}
  \label{Fig:C0110829335E3}
\end{figure*}

\begin{figure*}
\includegraphics[width=\columnwidth]{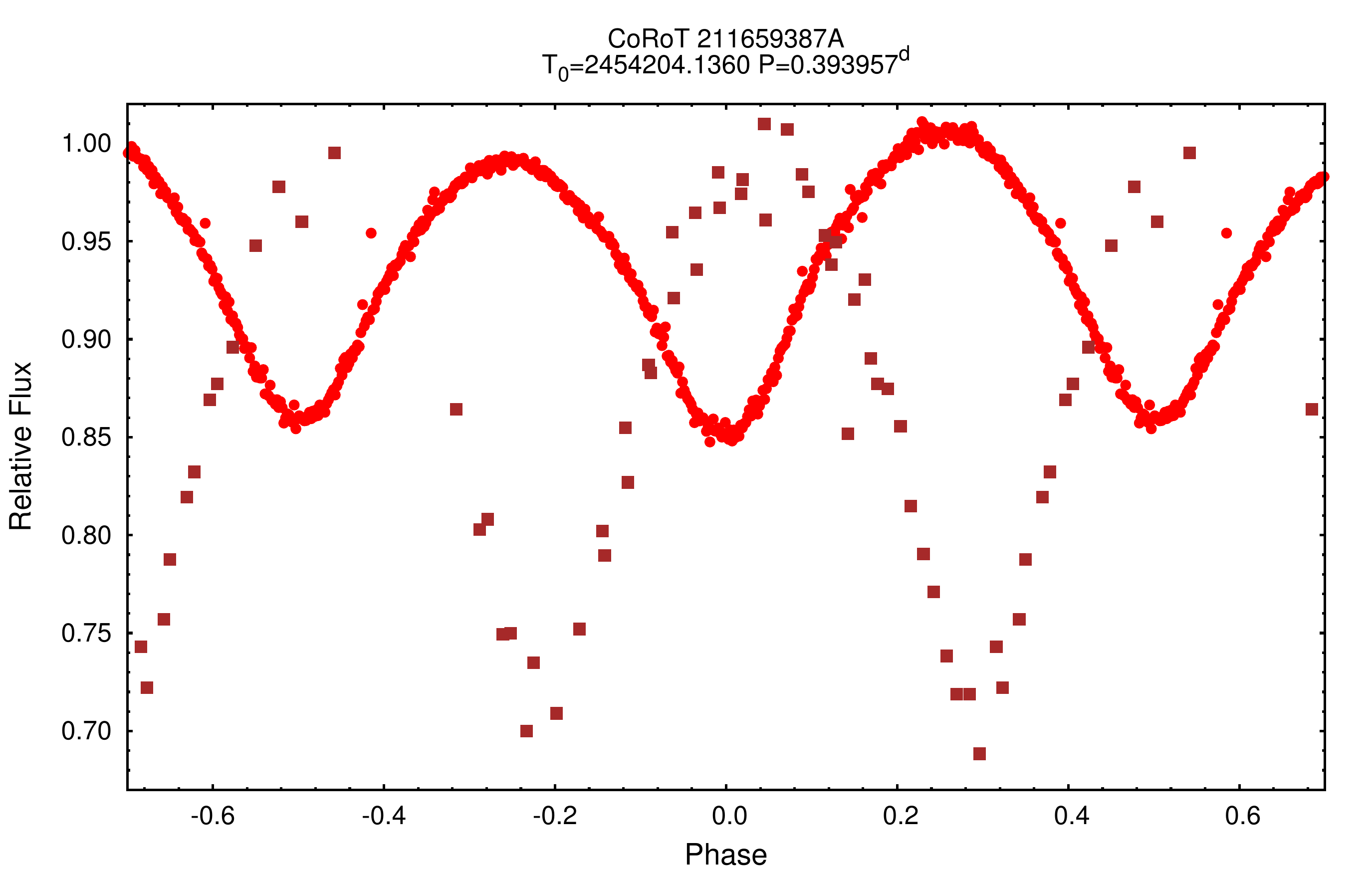}\includegraphics[width=\columnwidth]{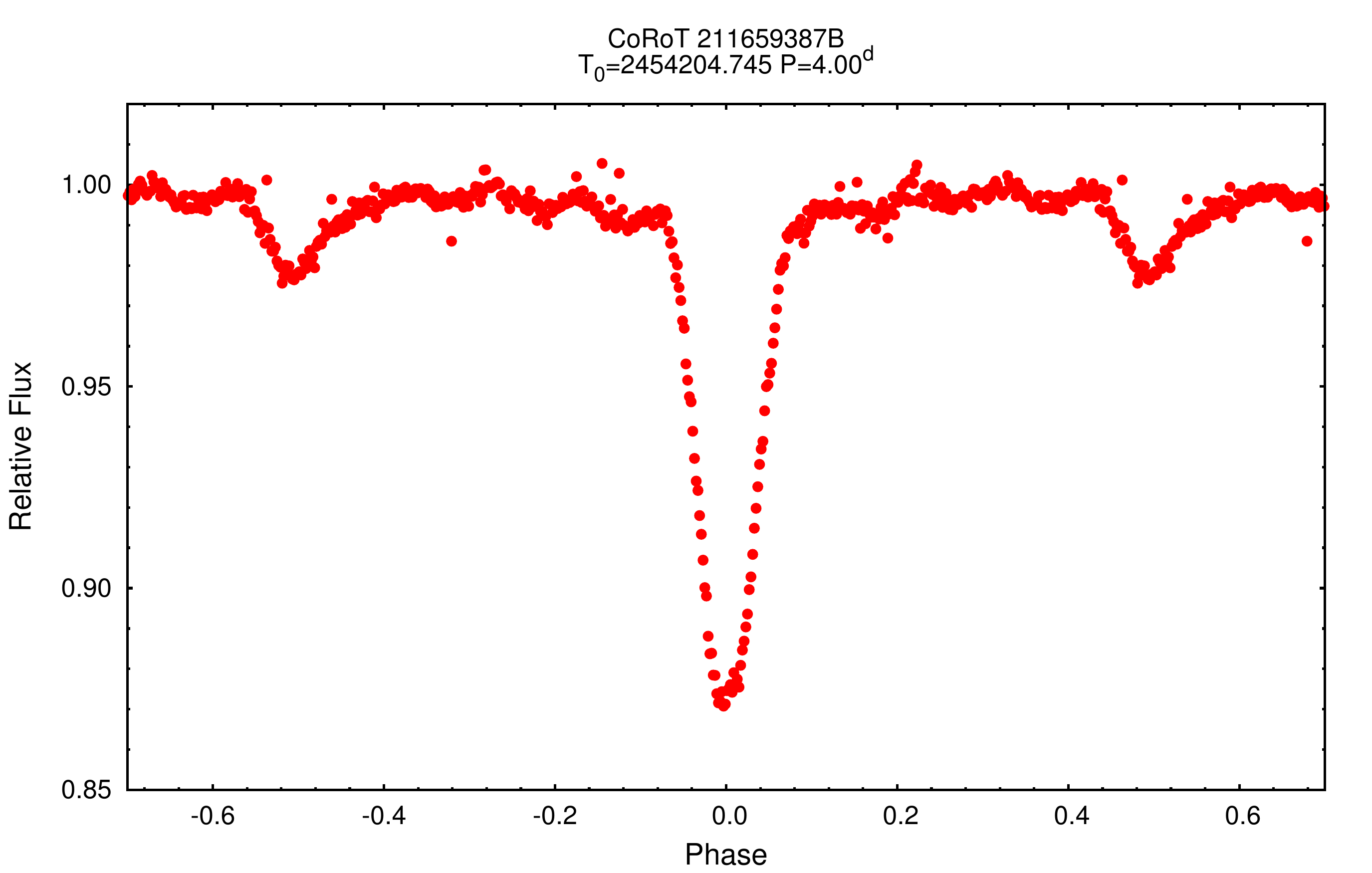}
  \caption{The disentangled, folded, binned, averaged light curves of the two EBs blended in the light curve of the CoRoT target id. 211659387. Note the period ratio of the two EBs are very close to $1:10$. This seems to be an incidental fact. In the left panel we plotted also (with brown boxes) the folded light curve formed from our ( $I$-band) ground-based follow up observations on four nights at August 2015 at Piszk\'estet\H o Observatory, Hungary. As it can be seen, the orbital phases has shifted by almost a quarter of the eclipsing period by the time of our observations, which cannot be explained with the uncertainty of the calculated eclipsing period, but imply some period variations (being either physical or apparent, incidental or continuous) since the epoch of the CoRoT measurements.}
  \label{Fig:C0211659387AB}
\end{figure*}

\begin{figure*}
\includegraphics[width=\columnwidth]{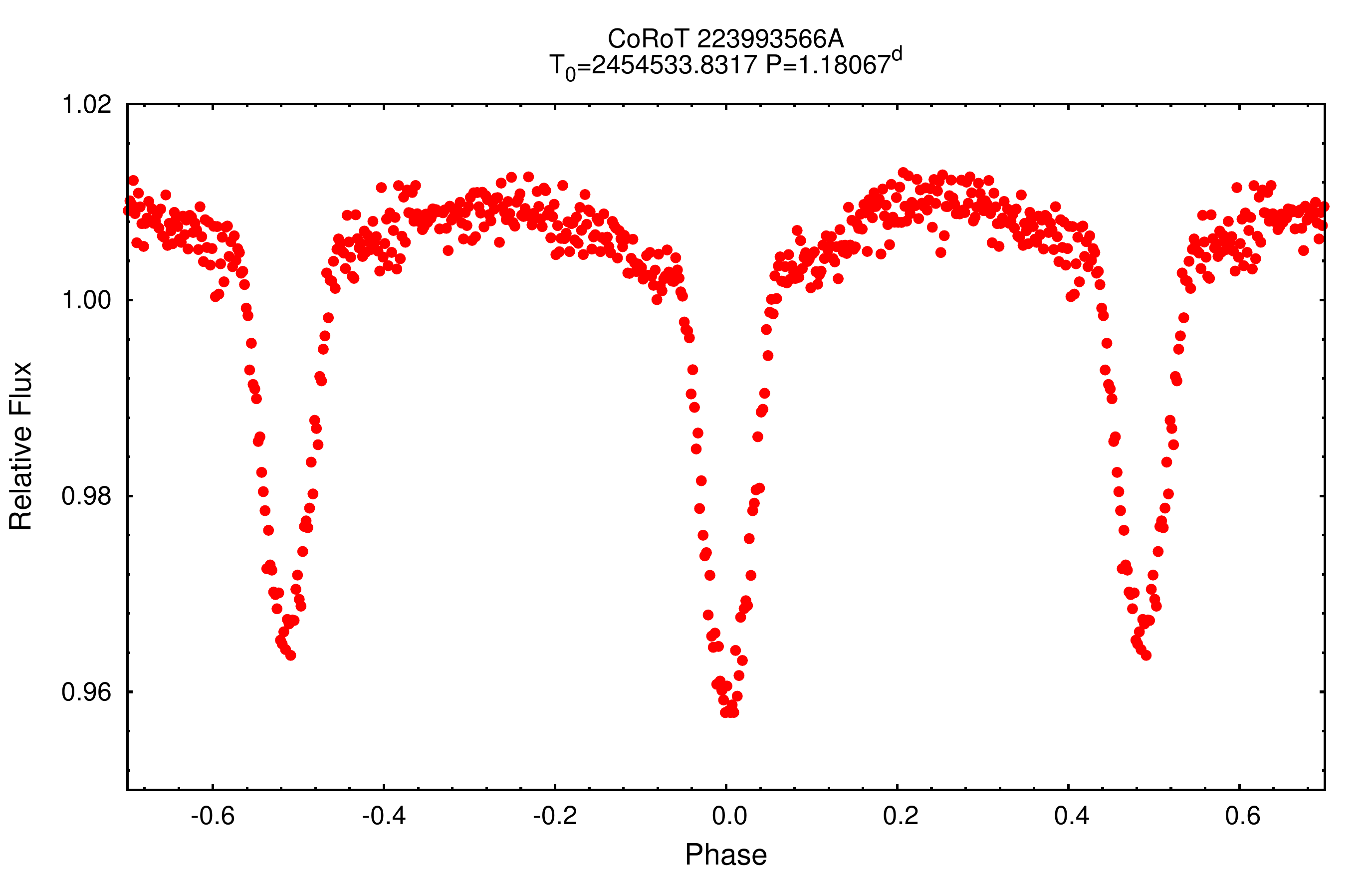}\includegraphics[width=\columnwidth]{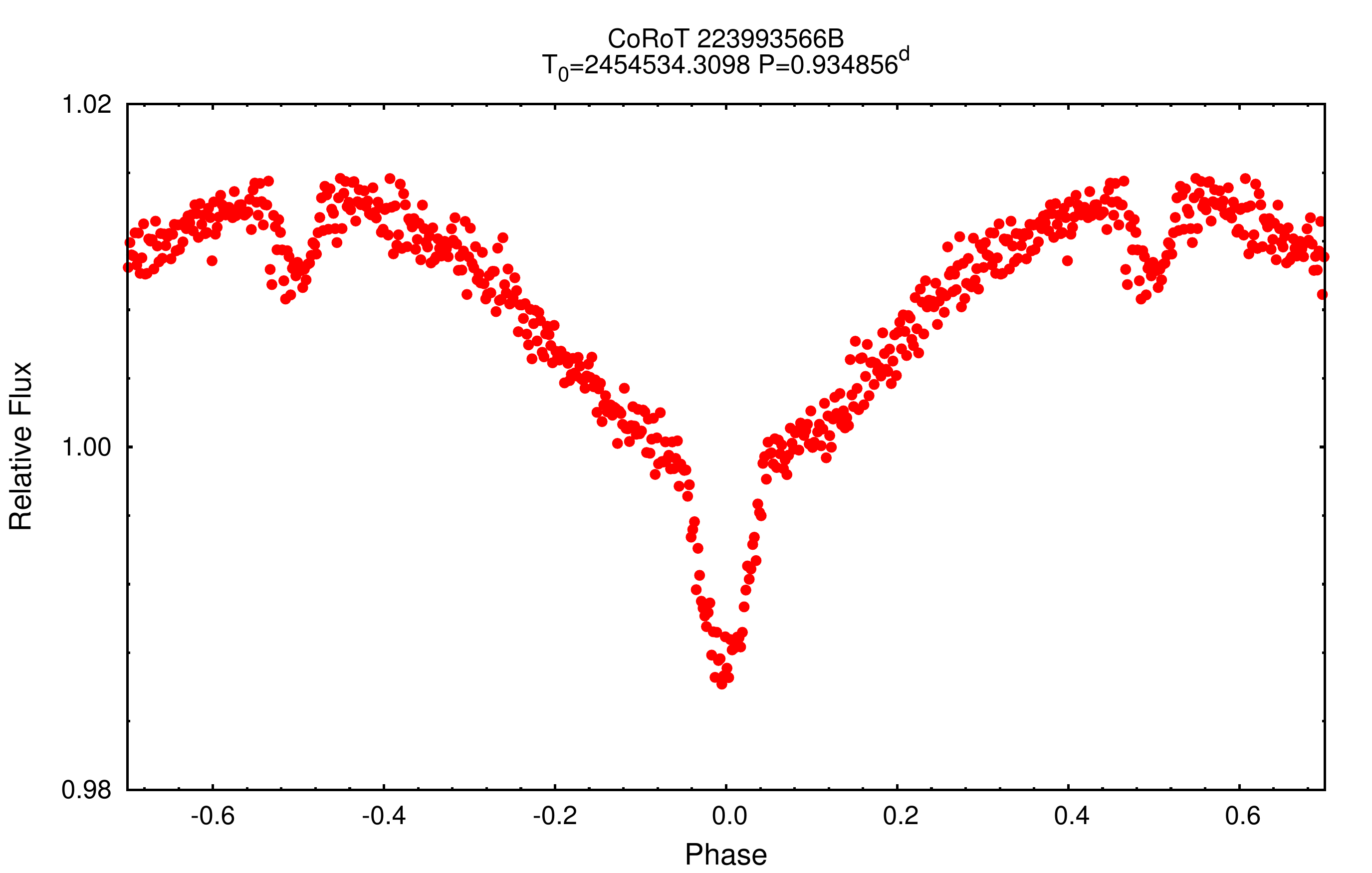}
  \caption{The disentangled, folded, binned, averaged light curves of the two EBs blended in the light curve of the CoRoT target id. 223993566. Note the remarkable reflection/irradiation effect in the second binary.}
  \label{Fig:C0223993566AB}
\end{figure*}

\begin{figure*}
\includegraphics[width=\columnwidth]{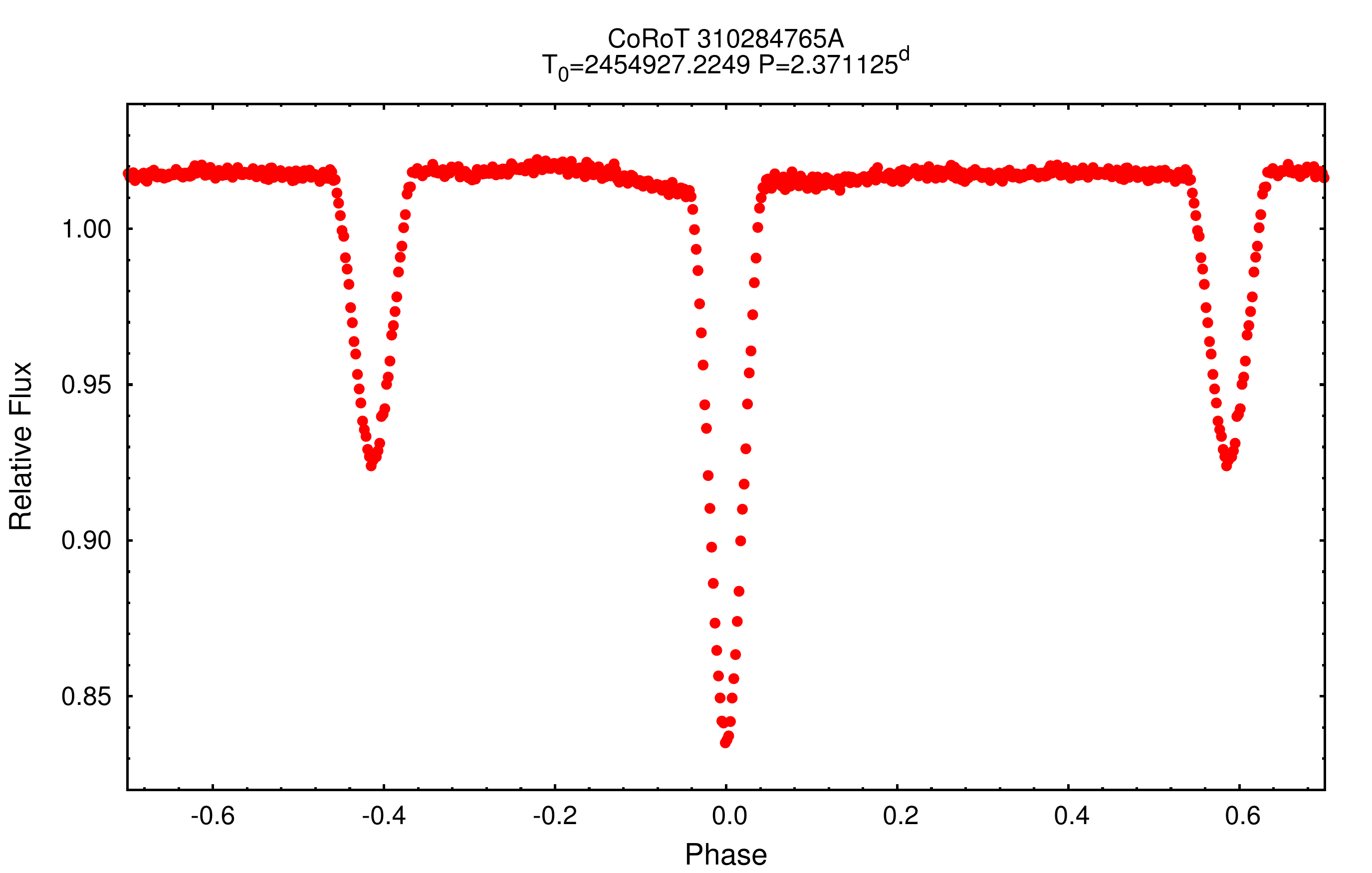}\includegraphics[width=\columnwidth]{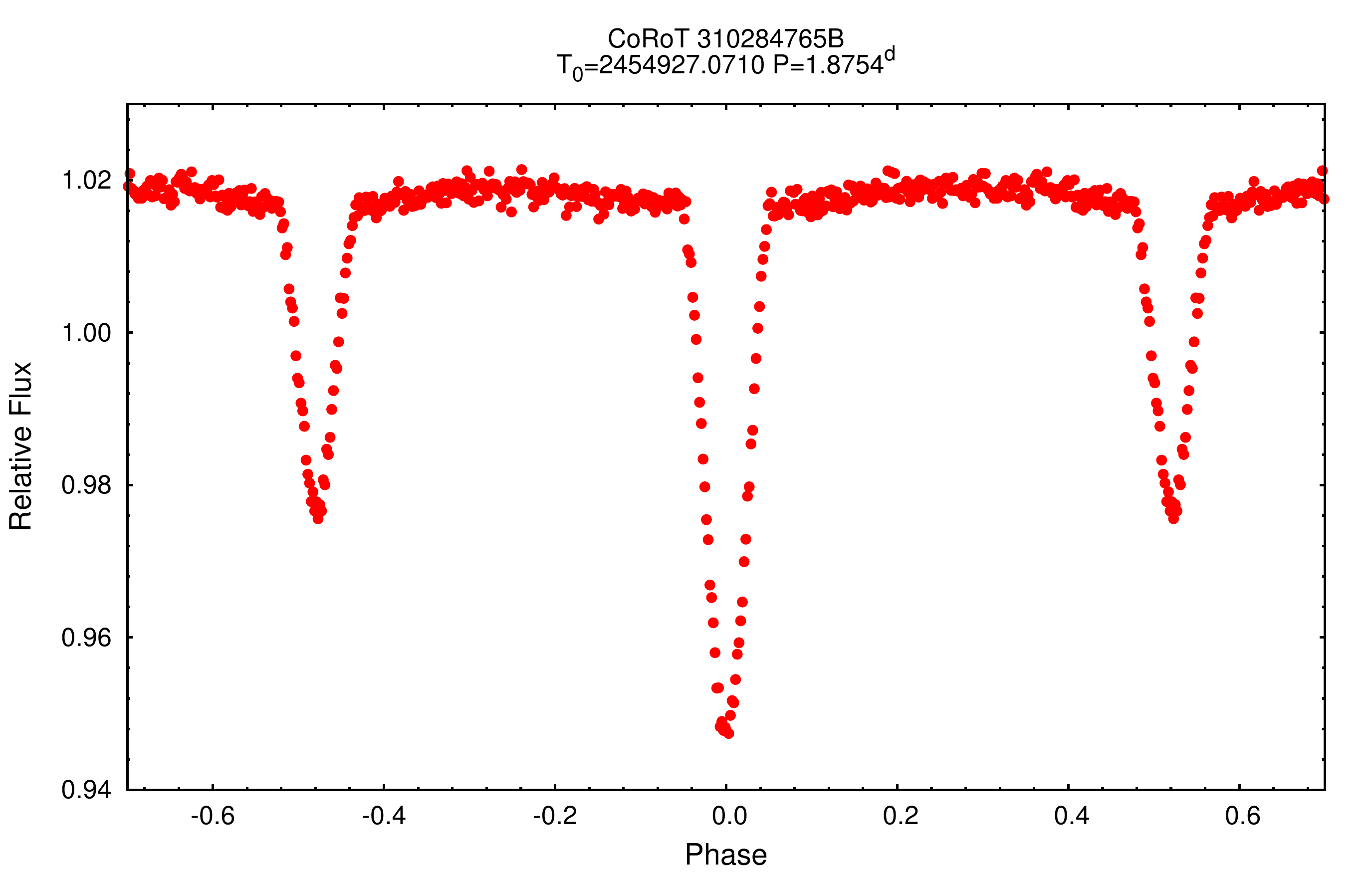}
  \caption{The disentangled, folded, binned, averaged light curves of the two EBs blended in the light curve of the CoRoT target id. 310284765. Note, both EBs have slightly displaced secondary minima, indicating eccentric orbits.}
  \label{Fig:C0310284765AB}
\end{figure*}

We have also identified four new blended EBs (i.e. mixed light curves of two EBs without any detectable interactions between them) in the CoRoT fields. (Note, blended light curves of CoRoT\,211625668 -- \citealt{eriksonetal12} and  CoRoT\,310266512  -- \citealt{fernandezchou15} were reported previously.) For these systems the light curves can be easily disentangled into pairs of separate EBs, and the ETVs do not exhibit any short-term interactions. Therefore we cannot decide whether these systems are hierarchic 2+2 multiples with long outer periods, or unbounded EBs seen in the same direction. We list these systems in Table\,\ref{Tab:blendephem} and plot their light curves in Figs.\,\ref{Fig:C0110829335E3}--\ref{Fig:C0310284765AB}. 

Interestingly, at least five of these eight EBs have eccentric orbits. Amongst them, the binary CoRoT 110829335B, one of the longest period EBs in the whole CoRoT sample ($P_\mathrm{B}\approx50\fd31$), has extremely displaced secondary minima ($\phi_\mathrm{IIB}\approx0\fp91$), therefore, its eccentricity should be $e_\mathrm{B}\gtrsim0.71$.\footnote{For the extremely high eccentricity, we calculated $e_\mathrm{min}$ by the use of the {\em complete, analytical form} of the time displacement, i.e. Eq.\,(\ref{Eq:Delta_apse}), instead of its frequently used first order (in eccentricity) approximation related simply to $e\cos\omega$.} Note, the eccentricity of its ($P_\mathrm{A}\approx8\fd93$-day-period) blended mate, i.e. CoRoT 110829335A should also exceed $e_\mathrm{A,min}\approx0.41$.

\begin{figure}
 \includegraphics[width=\columnwidth]{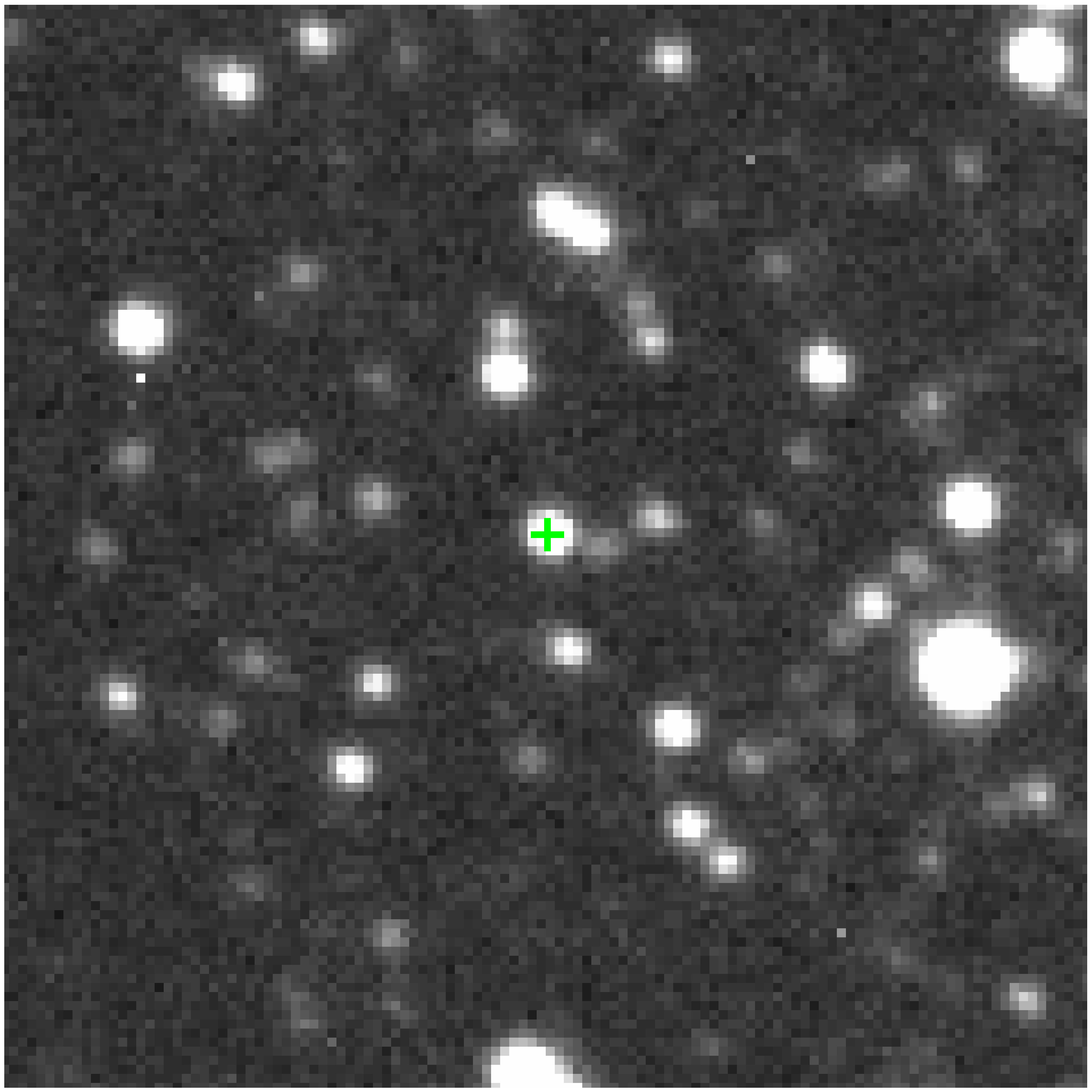}
  \caption{A narrow 2'x2'-section of the field of view around the CoRoT target id. 211659387. The photo was taken with the 90/60cm Schmidt telescope of Konkoly Observatory. The source of the overcontact EB-light curve is matched with green cross. No brightness variations exceeding a $3\sigma$-level were found for the other closest objects which contaminated probably the CoRoT measurements.}
  \label{Fig:C0211659387FoV}
\end{figure}

The light curve of the SRa01 target CoRoT\,211659387 is formed by the blend of a $P_\mathrm{A}\approx0\fd39$-day-period overcontact and a $P_\mathrm{A}\approx4\fd00$-day-period detached EB\footnote{Interestingly, \citet{eriksonetal12} give ephemeris for this latter, detached binary in their Table\,10 without mentioning the $0\fd39$-day overcontact component.} (Fig.\,\ref{Fig:C0211659387AB}). As one can see, the period ratio is almost $\sim1:10$. According to the ExoDat site, however, the contamination ratio of this source is about 55\%. We made some $VRI$-band photometric follow-up observations of this interesting blended source on the nights of 21/22, 23/24, 24/25 and 26/27 August, 2015 with the 90/60cm Schmidt telescope located on the Piszk\'es-tet\H o Mountain Station of the Konkoly Observatory. Our observations cover the full phase of the short period overcontact component (denoted with green cross in Fig.\,\ref{Fig:C0211659387FoV}). We plot also the phase-folded $I$-band light curve in the left panel of Fig.\,\ref{Fig:C0211659387AB}. As one can see, the 2015 light curve folded with the ephemeris determined from the CoRoT measurements (obtained in 2007) shows significant shift in phase which cannot be explained by the uncertainty of the period determination, but imply real variation(s) in the eclipsing period (which might be either physical or apparent). On the other hand, unfortunately, we were not able to observe any light curve variations (practically, eclipses) coming from the longer period binary component. Therefore, further observations are urgently needed.

Another new, interesting blended system is CoRoT\,223993566 which was observed during both the SRa01 and SRa05 runs. Therefore the full length of the data window is almost 4 years, which made it possible to detect evidence of apsidal motion (i.e. slight convergence of the primary and secondary ETV curves) in the $P_\mathrm{A}\approx1\fd18$-day-period eccentric binary A. The other, shorter period ($P_\mathrm{B}\approx0\fd93$) EB in this blended system is likely to have circular orbit, and exhibits a remarkable reflection/irradiation effect (see the right panel of Fig.\,\ref{Fig:C0223993566AB}).
 
Finally, the composite light curve of CoRoT\,310284765 exhibits the mixture of the light curves of two short-period, slightly eccentric Algols (Fig.\,\ref{Fig:C0310284765AB}).

\begin{table}
\caption{Orbital epheremides of the newly identified blended CoRoT EBs.}
\label{Tab:blendephem}
\begin{tabular}{llll}
\hline
CoRoT id & $T_0$ & $P$ & Remark \\
         & (RBJD)& (d) & \\
\hline
110829335 & 54795.4332 & 8.9304 & $\phi_\mathrm{II}=0\fp247$ \\
          & 54818.7400 &50.3075 & $\phi_\mathrm{II}=0\fp909$ \\
211659387 & 54204.1360 &0.393957& \\
          & 54204.7450 & 4.00   & \\
223993566 & 54533.8317 &1.18067 & $\phi_\mathrm{II}=0\fp483^a$\\
          & 54534.3098 &0.934856& strong reflection\\
310284765 & 54927.2249 &2.371125& $\phi_\mathrm{II}=0\fp587$ \\
          & 54927.0710 & 1.8754 & $\phi_\mathrm{II}=0\fp522$ \\
\hline
\end{tabular}
\end{table}

\section{Discussion}
\label{Discussion}

\subsection{Comparison with compact {\em Kepler}-triples}

\begin{figure}
\includegraphics[width=\columnwidth]{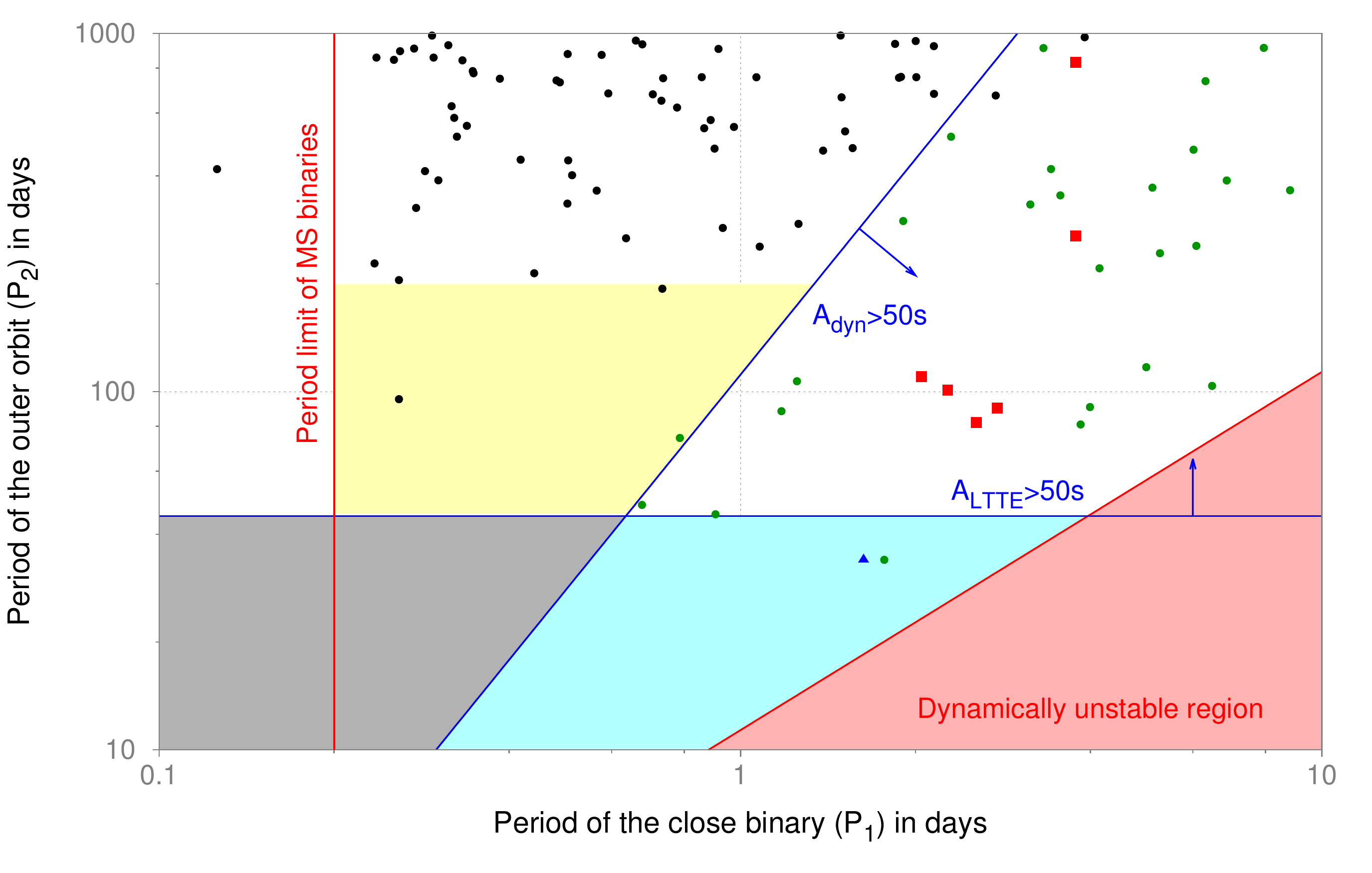}
  \caption{The location of the five triple star candidates (red boxes) on the $P_1$ vs $P_2$ plane. (The two alternative third-body solutions for CoRoT\,102698865 are plotted separately.) For a comparison we plotted those short-period {\em Kepler}-triple system candidates for which the inner and outer periods are $P_1\leq10\,$d and $P_2\leq1000\,$d. Following the work of \citet{Borkovits2016}, the pure LTTE systems are marked with black circles, while triples with combined LTTE+dynamical ETV solution are plotted with green. Furthermore, the first triple system discovered during the K2 mission (HD\,144548   -- \citealp{alonsoetal15}) is also plotted (blue triangle). The blue lines shows the borders of the domains where the amplitudes of the LTTE and dynamical terms may exceed $\sim50\,$sec, which can be regarded as a limit for an unambiguous detection. These limits were calculated for a hypothetical triple system of three, equally solar mass stars, with a typical outer eccentricity of $e_2=0.35$, and quite arbitrarily, $i_2=60\degr$ and $\omega_2\pm90\degr$. The shaded areas have the following meanings: (i) grey: in this region no LTTE or dynamical perturbations are detectable at all via ETV analysis; (ii) cyan: no LTTE can be detected, though dynamical effect may be significant and, therefore, certainly detectable; (iii) yellow: the ``desert'' of close (but clearly LTTE-detectable) third companions around short period EBs (mostly overcontact systems); red: dynamically unstable region, in the sense of the stability criterium of \citet{mardlingaarseth01}. (See text for further details.)}
  \label{Fig:P1vsP2KepCor}
\end{figure}

Despite the fact that the group of the tight hierarchical triple star candidates presented above in the CoRoT sample are not nearly as numerous as in the {\em Kepler}-sample, and the quantitative results obtained above are naturally far less certain than in the latter case, we conclude our study with some qualitative comparison with the findings of \citet{Borkovits2016} on the {\em Kepler}-sample. For this we plotted in Fig.\,\ref{Fig:P1vsP2KepCor} the six possible configurations found for our five triple candidates on the $P_1-P_2$ plane together with the {\em Kepler}-triples having, according to the results of \citet{Borkovits2016}, inner and outer periods $P_1\leq10\,$d and $P_2\leq1000$\,d, respectively. As it can be seen (shaded yellow region in Fig.\,\ref{Fig:P1vsP2KepCor}), similar to the {\em Kepler}-sample, we did not find any short outer period triple amongst the shortest period EBs, which practically means the lack of tight third stellar components revolving around overcontact systems. The absence of such systems from the {\em Kepler}-sample was first noticed by \citet{conroyetal14}. Our results emphasize again that this effect should have an astrophysical (more probably evolutionary) origin.

Turning to our five candidate systems, four of them have inner periods between 2 and 3 days. The sample of \citet{Borkovits2016} contains 17 triple candidates having inner periods in the same domain. The shortest outer period in the {\em Kepler}-sample is $P_2=515$\,d. There are, however, 7 triple candidates amongst the $P_2<2$\,d inner period systems in the {\em Kepler}-sample, the outer periods of which remain $P_2<110$\,d. The lack of systems with similar outer periods in the same inner period domain suggests that our results should be taken with a grain of salt. It is possible that the analysed ETVs cover only smaller portions of the outer orbits instead of almost a full cycle and, therefore, the orbital solutions might be misinterpreted, and the true outer period might be substantially longer. On the other hand, taking into account the low overall population of the $P_2<2-300$\,d region itself, the absence of such short outer period systems in the $2\lesssim P_1\lesssim4$\,d inner period regime from the {\em Kepler} data could be a purely statistical fluctuation. Consequently, the observed distribution difference does not necessarily question the validity of our solutions.

\subsection{Prospects of ground-based follow-up observations}
\label{Subsect:follow-up}

It is evident that further observations are needed to clarify (or refute) our findings. Therefore, in what follows, we briefly discuss the possibilities of future, ground-based follow-up observations. First we consider the spectroscopic measurements. Our candidates are relatively faint, but they would still be available with several instruments equipped to relatively large (3+\,m aperture) telescopes. For two of the three third-light dominated systems (CoRoT\,100805120 and CoRoT\,101290947) we can expect to detect only the lines of the more distant tertiary components (which are most probably red giants). Nevertheless, the determination of the parameters of the outer orbits (including the spectroscopic mass functions) from radial velocity measurements would allow us to lift the high-degree degeneracy between the LTTE and dynamical contributions of the ETV solution (see the discussion in \citealp{Rappaport2013}) and, therefore, would enable us to calculate an accurate dynamical model, including a reliable dynamical mass determination. For the remaining third-light dominated system, the triply eclipsing  CoRoT\,104079133, one can expect to detect both the lines of the primary component of the inner EB, and the tertiary star. Because of the chance of being an SB3 system, CoRoT\,102698865 could be the most promising triple candidate, while CoRoT\,11083077 is expected to be an SB1 system. Note, due to the 3-4\,month-long (short) outer periods of all but one of our triple candidates, the spectroscopic outer orbits could be determined during one observing session with the exception of CoRoT\,102698865. Multi-session observations, however, would also be preferred for the systems with shorter outer periods in order to detect the effects of the longer time-scale three-body perturbations on the orbit(s).

As for the possibility of photometric follow-up observations, which are the most common way of obtaining additional eclipsing minima time measurements over time for ETV studies, our systems are exposed to both favourable and very unfavourable circumstances. Although the short period of the majority of our systems (including other systems, discussed in subsection\,\ref{Subsect:others}, too) would be ideal for such observations, the combination of the very small eclipse depths (at least in the third-flux dominated systems) with the low amplitude ETVs may pose too great a challenge for the earth-based measurements. For example in the case of CoRoT\,100805120 and CoRoT\,101290947 a millimagnitude photometric accuracy of each individual measurement would be required for a satisfactory signal to noise ratio. On the other hand, the situation is more mixed for the case of CoRoT\,110830711: although its primary minima with amplitudes exceeding $0\fm2$ magnitudes can be observed with satisfactory photometric accuracy even with smaller telescopes and in average sky conditions, the full amplitude of the ETV curve is only about $0\fd001$-day, which requires an accuracy of some $10^{-4}$\,days for each time of minimum determination. In this respect the two most promising targets would be CoRoT\,102698865 and  CoRoT\,104079133, having both relatively deep primary eclipses and larger ETV amplitudes at the same time. Furthermore, in the first of them the apsidal motion could also be well followed with ground-based minima observations, while in the latter one the detection of probable future outer eclipses offers a further exciting possibility. This latter statement also holds for CoRoT\,221664856. However, one should keep in mind that the outer eclipse events may last as long as one or two days, meaning that a successful observation of such events would require international campaigns in the future, similar to the one organized by \citet{conroyetal15} for the observation of the forecasted outer eclipse of KIC\,02835289.

\section{Summary and conclusions}\label{summary}

In this paper we reported the results of our search for close, third stellar companions to eclipsing binaries observed with CoRoT spacecraft via ETV, as well as some auxiliary light curve analyses. Despite the short length of the data series, we were able to find third-body solutions (with combination of light-travel time effect and third-body perturbations) for ETV curves of five, relatively short-period Algol-systems, namely the CoRoT ids. 100805120, 101290947, 102698865, 104079133, and 110830711. The periods of the outer orbits were found to be between 82 and 831 days. For one of them, CoRoT\,102698865, we obtained two alternative solutions with outer periods of 272 and 831 days, resectively. Apsidal motion (most probably of dynamical origin) was also detected for three eccentric systems. For three of the five systems the light curve is dominated by the extra (third) flux, suggesting that the spectral information available in the literature for these systems refer to the source of the extra flux rather than the EB itself. By combining the results of the light curve and ETV analyses we were able to calculate in a dynamical manner the individual masses of all the three components and the physical dimensions of the inner EB's stellar components as well. These results, though with relatively higher uncertainties, are consistent with both the available spectral information and the amounts of extra lights deduced from the ligh curve solutions. Our results support that CoRoT\,100805120 (and perhaps CoRoT\,101290947) join the still small group of compact hierarchical triple stars with red giants as their most massive component.

We have identified two EBs exhibiting extraneous eclipses with complex structures. These certain triply eclipsing triple systems are CoRoTs\,104079133 and 221664856. For the first system extra eclipses both around the inferior and superior conjunctions were observed, and we were also able to obtain ETV solution (see the previous paragraph). For the second system the short dataset covering only 1 month was insufficient to provide any meaningful ETV solution.

We have also reported four new composite light curves of blended EBs. Five of the eight blended EBs revolve on eccentric orbits and one of them, CoRoT\,110829335B, was found to be extremely eccentric with $e\geq0.71$, while the sixth blended EB, CoRoT\,223993566B, exhibits remarkable reflection/irradiation effect.

Finally, we discuss briefly the reliability of our ETV solutions. Their fundamental weakness is that they do not satisfy the most natural criterion of a trustworthy three-body interpretation of an ETV curve, which states that the observations should cover at least two outer orbital periods \citep[e.g.][]{conroyetal14,Borkovits2016}. On the other hand, the solutions for these triple candidates fulfill the first three criteria of \citet{Frieboes-Conde1973}. Note, however that, with the extra information obtainable from a combined LTTE + dynamical solution in our hand, we can slightly reformulate and strengthen the original criteria, listed in the Introduction, at some points. Therefore, in our case we can state that we were able to model the timing data of the selected EBs with combined LTTE+dynamical ETV three-body models (i.e. criterion 1), fitting simultaneously the primary and secondary curves (2). From the ETV solutions we derived the total masses ($m_\mathrm{AB}$) of the inner EBs and the masses ($m_\mathrm{C}$) of the third components, which were found to be consistent with the amounts of the third lights ($l_3$), obtained from the auxiliary light curve analyses (3). (Note that the last criterion of \citet{Frieboes-Conde1973} cannot be applied on our systems due to the lack of radial velocity observations.) Finally, in the case of the triply eclipsing system CoRoT\,104079133 we have also found that the extra eclipses were occurred in the vicinity of the inferior and superior conjunction points of the outer orbit of the ETV solution, which makes it very likely that the source of the ETV signal is identical with the outer eclipsing component, strenghtening our confidence regarding the reliability of our ETV solution based third-body model. 

Therefore, we may conclude that, despite the short data lengths compared to the periods of the detected outer orbits, our solutions were found to be physically consistent and, therefore, the third-body hypotheses seem to be well established. Further observations, however, are necessary to confirm and refine, or refute, our results.

\section*{Acknowledgements}
This project has partly been supported by the HAS Wigner RCP - GPU-Lab and the Hungarian National Research, Development and Innovation Office, NKFIH-OTKA grants K-113117 and K-115709. This research has made use of the ExoDat Database, operated at LAM-OAMP, Marseille, France, on behalf of the CoRoT/Exoplanet program. This research has made use of data collected by the CoRoT mission. The research has also made use of the VizieR catalogue access tool, CDS, Strasbourg, France. The original description of the VizieR service was published by \citet{vizier}. The authors are grateful to the referee, K. Conroy, for his valuable comments and suggestions which helped us to substantially improve the quality of the paper, and to G. Kutrov\'atz and I. B. B\'\i r\'o for the linguistic corrections.

\bibliographystyle{mnras}
\bibliography{Hajduetalrev3full}

\begin{thebibliography}{}
\makeatletter
\relax
\def\mn@urlcharsother{\let\do\@makeother \do\$\do\&\do\#\do\^\do\_\do\%\do\~}
\def\mn@doi{\begingroup\mn@urlcharsother \@ifnextchar [ {\mn@doi@}
  {\mn@doi@[]}}
\def\mn@doi@[#1]#2{\def\@tempa{#1}\ifx\@tempa\@empty \href
  {http://dx.doi.org/#2} {doi:#2}\else \href {http://dx.doi.org/#2} {#1}\fi
  \endgroup}
\def\mn@eprint#1#2{\mn@eprint@#1:#2::\@nil}
\def\mn@eprint@arXiv#1{\href {http://arxiv.org/abs/#1} {{\tt arXiv:#1}}}
\def\mn@eprint@dblp#1{\href {http://dblp.uni-trier.de/rec/bibtex/#1.xml}
  {dblp:#1}}
\def\mn@eprint@#1:#2:#3:#4\@nil{\def\@tempa {#1}\def\@tempb {#2}\def\@tempc
  {#3}\ifx \@tempc \@empty \let \@tempc \@tempb \let \@tempb \@tempa \fi \ifx
  \@tempb \@empty \def\@tempb {arXiv}\fi \@ifundefined
  {mn@eprint@\@tempb}{\@tempb:\@tempc}{\expandafter \expandafter \csname
  mn@eprint@\@tempb\endcsname \expandafter{\@tempc}}}

\bibitem[\protect\citeauthoryear{{Alonso}, {Deeg}, {Hoyer}, {Lodieu}, {Palle}
  \& {Sanchis-Ojeda}}{{Alonso} et~al.}{2015}]{alonsoetal15}
{Alonso} R.,  {Deeg} H.~J.,  {Hoyer} S.,  {Lodieu} N.,  {Palle} E.,
  {Sanchis-Ojeda} R.,  2015, \mn@doi [\aap] {10.1051/0004-6361/201527109},
  \href {http://esoads.eso.org/abs/2015A%26A...584L...8A} {584, L8}

\bibitem[\protect\citeauthoryear{{Auvergne} et~al.,}{{Auvergne}
  et~al.}{2009}]{auvergneetal09}
{Auvergne} M.,  et~al., 2009, \mn@doi [\aap] {10.1051/0004-6361/200810860},
  \href {http://esoads.eso.org/abs/2009A%26A...506..411A} {506, 411}

\bibitem[\protect\citeauthoryear{{Baglin} et~al.,}{{Baglin}
  et~al.}{2006}]{2006cosp...36.3749B}
{Baglin} A.,  et~al., 2006, in 36th COSPAR Scientific Assembly.

\bibitem[\protect\citeauthoryear{{Balaji}, {Croll}, {Levine}  \&
  {Rappaport}}{{Balaji} et~al.}{2015}]{balajietal15}
{Balaji} B.,  {Croll} B.,  {Levine} A.~M.,   {Rappaport} S.,  2015, \mn@doi
  [\mnras] {10.1093/mnras/stv031}, \href
  {http://esoads.eso.org/abs/2015MNRAS.448..429B} {448, 429}

\bibitem[\protect\citeauthoryear{{Baudin}, {Maceroni}  \& {Alencar}}{{Baudin}
  et~al.}{2016}]{baudinetal16}
{Baudin} F.,  {Maceroni} C.,   {Alencar} S.~H.~P.,  2016, {IV.3 The wealth of
  stellar variability}.
p.~209, \mn@doi{10.1051/978-2-7598-1876-1.c043}

\bibitem[\protect\citeauthoryear{{Borkovits}, {{\'E}rdi}, {Forg{\'a}cs-Dajka}
  \& {Kov{\'a}cs}}{{Borkovits} et~al.}{2003}]{Borkovits2003}
{Borkovits} T.,  {{\'E}rdi} B.,  {Forg{\'a}cs-Dajka} E.,   {Kov{\'a}cs} T.,
  2003, \mn@doi [\aap] {10.1051/0004-6361:20021688}, \href
  {http://adsabs.harvard.edu/abs/2003A%26A...398.1091B} {398, 1091}

\bibitem[\protect\citeauthoryear{{Borkovits}, {Csizmadia}, {Forg{\'a}cs-Dajka}
  \& {Heged{\"u}s}}{{Borkovits} et~al.}{2011}]{Borkovits2011}
{Borkovits} T.,  {Csizmadia} S.,  {Forg{\'a}cs-Dajka} E.,   {Heged{\"u}s} T.,
  2011, \mn@doi [\aap] {10.1051/0004-6361/201015867}, \href
  {http://adsabs.harvard.edu/abs/2011A%26A...528A..53B} {528, A53}

\bibitem[\protect\citeauthoryear{{Borkovits} et~al.,}{{Borkovits}
  et~al.}{2013}]{borkovitsetal13}
{Borkovits} T.,  et~al., 2013, \mn@doi [\mnras] {10.1093/mnras/sts146}, \href
  {http://esoads.eso.org/abs/2013MNRAS.428.1656B} {428, 1656}

\bibitem[\protect\citeauthoryear{{Borkovits} et~al.,}{{Borkovits}
  et~al.}{2014}]{borkovitsetal14}
{Borkovits} T.,  et~al., 2014, \mn@doi [\mnras] {10.1093/mnras/stu1379}, \href
  {http://adsabs.harvard.edu/abs/2014MNRAS.443.3068B} {443, 3068}

\bibitem[\protect\citeauthoryear{{Borkovits}, {Rappaport}, {Hajdu}  \&
  {Sztakovics}}{{Borkovits} et~al.}{2015}]{Borkovits2015}
{Borkovits} T.,  {Rappaport} S.,  {Hajdu} T.,   {Sztakovics} J.,  2015, \mn@doi
  [\mnras] {10.1093/mnras/stv015}, \href
  {http://adsabs.harvard.edu/abs/2015MNRAS.448..946B} {448, 946}

\bibitem[\protect\citeauthoryear{{Borkovits}, {Hajdu}, {Sztakovics},
  {Rappaport}, {Levine}, {B{\'{\i}}r{\'o}}  \& {Klagyivik}}{{Borkovits}
  et~al.}{2016}]{Borkovits2016}
{Borkovits} T.,  {Hajdu} T.,  {Sztakovics} J.,  {Rappaport} S.,  {Levine} A.,
  {B{\'{\i}}r{\'o}} I.~B.,   {Klagyivik} P.,  2016, \mn@doi [\mnras]
  {10.1093/mnras/stv2530}, \href
  {http://adsabs.harvard.edu/abs/2016MNRAS.455.4136B} {455, 4136}

\bibitem[\protect\citeauthoryear{{Borucki} et~al.,}{{Borucki}
  et~al.}{2010}]{boruckietal10}
{Borucki} W.~J.,  et~al., 2010, \mn@doi [Science] {10.1126/science.1185402},
  \href {http://esoads.eso.org/abs/2010Sci...327..977B} {327, 977}

\bibitem[\protect\citeauthoryear{{Cabrera} et~al.,}{{Cabrera}
  et~al.}{2009}]{cabreraetal09}
{Cabrera} J.,  et~al., 2009, \mn@doi [\aap] {10.1051/0004-6361/200912684},
  \href {http://adsabs.harvard.edu/abs/2009A%26A...506..501C} {506, 501}

\bibitem[\protect\citeauthoryear{{Chaintreuil}, {Deru}, {Baudin}, {Ferrigno},
  {Grolleau}  \& {Romagnan}}{{Chaintreuil} et~al.}{2016}]{chaintreuiletal16}
{Chaintreuil} S.,  {Deru} A.,  {Baudin} F.,  {Ferrigno} A.,  {Grolleau} E.,
  {Romagnan} R.,  2016, {II.4 The ''ready to use'' CoRoT data}.
p.~61, \mn@doi{10.1051/978-2-7598-1876-1.c024}

\bibitem[\protect\citeauthoryear{{Chandler}}{{Chandler}}{1888}]{Chandler1888}
{Chandler} S.~C.,  1888, Bulletin Astronomique, Serie I, \href
  {http://adsabs.harvard.edu/abs/1888BuAsI...5R.499.} {5, 499}

\bibitem[\protect\citeauthoryear{{Conroy}, {Pr{\v s}a}, {Stassun}, {Orosz},
  {Fabrycky}  \& {Welsh}}{{Conroy} et~al.}{2014}]{conroyetal14}
{Conroy} K.~E.,  {Pr{\v s}a} A.,  {Stassun} K.~G.,  {Orosz} J.~A.,  {Fabrycky}
  D.~C.,   {Welsh} W.~F.,  2014, \mn@doi [\aj] {10.1088/0004-6256/147/2/45},
  \href {http://esoads.eso.org/abs/2014AJ....147...45C} {147, 45}

\bibitem[\protect\citeauthoryear{{Conroy}, {Prsa}, {Stassun}  \&
  {Orosz}}{{Conroy} et~al.}{2015}]{conroyetal15}
{Conroy} K.,  {Prsa} A.,  {Stassun} K.,   {Orosz} J.,  2015, Information
  Bulletin on Variable Stars, \href
  {http://esoads.eso.org/abs/2015IBVS.6138....1C} {6138}

\bibitem[\protect\citeauthoryear{{Cowling}}{{Cowling}}{1938}]{cowling38}
{Cowling} T.~G.,  1938, \mn@doi [\mnras] {10.1093/mnras/98.9.734}, \href
  {http://esoads.eso.org/abs/1938MNRAS..98..734C} {98, 734}

\bibitem[\protect\citeauthoryear{{Deleuil} et~al.,}{{Deleuil}
  et~al.}{2009}]{ExoDAT}
{Deleuil} M.,  et~al., 2009, \mn@doi [\aj] {10.1088/0004-6256/138/2/649}, \href
  {http://adsabs.harvard.edu/abs/2009AJ....138..649D} {138, 649}

\bibitem[\protect\citeauthoryear{{Derekas} et~al.,}{{Derekas}
  et~al.}{2011}]{Derekas2011}
{Derekas} A.,  et~al., 2011, \mn@doi [Science] {10.1126/science.1201762}, \href
  {http://adsabs.harvard.edu/abs/2011Sci...332..216D} {332, 216}

\bibitem[\protect\citeauthoryear{{Erikson} et~al.,}{{Erikson}
  et~al.}{2012}]{eriksonetal12}
{Erikson} A.,  et~al., 2012, \mn@doi [\aap] {10.1051/0004-6361/201116934},
  \href {http://adsabs.harvard.edu/abs/2012A%26A...539A..14E} {539, A14}

\bibitem[\protect\citeauthoryear{{Fabrycky} \& {Tremaine}}{{Fabrycky} \&
  {Tremaine}}{2007}]{fabryckytremaine07}
{Fabrycky} D.,  {Tremaine} S.,  2007, \mn@doi [\apj] {10.1086/521702}, \href
  {http://esoads.eso.org/abs/2007ApJ...669.1298F} {669, 1298}

\bibitem[\protect\citeauthoryear{{Fern{\'a}ndez Fern{\'a}ndez} \&
  {Chou}}{{Fern{\'a}ndez Fern{\'a}ndez} \& {Chou}}{2015}]{fernandezchou15}
{Fern{\'a}ndez Fern{\'a}ndez} J.,  {Chou} D.-Y.,  2015, \mn@doi [\pasp]
  {10.1086/681626}, \href {http://esoads.eso.org/abs/2015PASP..127..421F} {127,
  421}

\bibitem[\protect\citeauthoryear{{Frieboes-Conde} \&
  {Herczeg}}{{Frieboes-Conde} \& {Herczeg}}{1973}]{Frieboes-Conde1973}
{Frieboes-Conde} H.,  {Herczeg} T.,  1973, \aaps, \href
  {http://adsabs.harvard.edu/abs/1973A%26AS...12....1F} {12, 1}

\bibitem[\protect\citeauthoryear{{Gaulme}, {McKeever}, {Rawls}, {Jackiewicz},
  {Mosser}  \& {Guzik}}{{Gaulme} et~al.}{2013}]{gaulmeetal13}
{Gaulme} P.,  {McKeever} J.,  {Rawls} M.~L.,  {Jackiewicz} J.,  {Mosser} B.,
  {Guzik} J.~A.,  2013, \mn@doi [\apj] {10.1088/0004-637X/767/1/82}, \href
  {http://esoads.eso.org/abs/2013ApJ...767...82G} {767, 82}

\bibitem[\protect\citeauthoryear{{Gies}, {Matson}, {Guo}, {Lester}, {Orosz}  \&
  {Peters}}{{Gies} et~al.}{2015}]{giesetal15}
{Gies} D.~R.,  {Matson} R.~A.,  {Guo} Z.,  {Lester} K.~V.,  {Orosz} J.~A.,
  {Peters} G.~J.,  2015, \mn@doi [\aj] {10.1088/0004-6256/150/6/178}, \href
  {http://esoads.eso.org/abs/2015AJ....150..178G} {150, 178}

\bibitem[\protect\citeauthoryear{{Gimenez} \& {Garcia-Pelayo}}{{Gimenez} \&
  {Garcia-Pelayo}}{1983}]{gimenezgarcia83}
{Gimenez} A.,  {Garcia-Pelayo} J.~M.,  1983, \mn@doi [\apss]
  {10.1007/BF00653602}, \href {http://esoads.eso.org/abs/1983Ap%26SS..92..203G}
  {92, 203}

\bibitem[\protect\citeauthoryear{{Hambleton} et~al.,}{{Hambleton}
  et~al.}{2013}]{hambletonetal13}
{Hambleton} K.~M.,  et~al., 2013, \mn@doi [\mnras] {10.1093/mnras/stt886},
  \href {http://esoads.eso.org/abs/2013MNRAS.434..925H} {434, 925}

\bibitem[\protect\citeauthoryear{{Irwin}}{{Irwin}}{1952}]{Irwin1952}
{Irwin} J.~B.,  1952, \mn@doi [\apj] {10.1086/145604}, \href
  {http://adsabs.harvard.edu/abs/1952ApJ...116..211I} {116, 211}

\bibitem[\protect\citeauthoryear{{Irwin}}{{Irwin}}{1959}]{irwin59}
{Irwin} J.~B.,  1959, \mn@doi [\aj] {10.1086/107913}, \href
  {http://esoads.eso.org/abs/1959AJ.....64..149I} {64, 149}

\bibitem[\protect\citeauthoryear{{Kirk} et~al.,}{{Kirk}
  et~al.}{2016}]{kirketal16}
{Kirk} B.,  et~al., 2016, \mn@doi [\aj] {10.3847/0004-6256/151/3/68}, \href
  {http://esoads.eso.org/abs/2016AJ....151...68K} {151, 68}

\bibitem[\protect\citeauthoryear{{Kiseleva}, {Eggleton}  \&
  {Mikkola}}{{Kiseleva} et~al.}{1998}]{kiselevaetal98}
{Kiseleva} L.~G.,  {Eggleton} P.~P.,   {Mikkola} S.,  1998, \mn@doi [\mnras]
  {10.1046/j.1365-8711.1998.01903.x}, \href
  {http://esoads.eso.org/abs/1998MNRAS.300..292K} {300, 292}

\bibitem[\protect\citeauthoryear{{Klinglesmith} \& {Sobieski}}{{Klinglesmith}
  \& {Sobieski}}{1970}]{klinglesmithsobieski70}
{Klinglesmith} D.~A.,  {Sobieski} S.,  1970, \mn@doi [\aj] {10.1086/110960},
  \href {http://adsabs.harvard.edu/abs/1970AJ.....75..175K} {75, 175}

\bibitem[\protect\citeauthoryear{{Kumar}, {Reddy}  \& {Lambert}}{{Kumar}
  et~al.}{2011}]{kumaretal11}
{Kumar} Y.~B.,  {Reddy} B.~E.,   {Lambert} D.~L.,  2011, \mn@doi [\apjl]
  {10.1088/2041-8205/730/1/L12}, \href
  {http://adsabs.harvard.edu/abs/2011ApJ...730L..12K} {730, L12}

\bibitem[\protect\citeauthoryear{{Mardling} \& {Aarseth}}{{Mardling} \&
  {Aarseth}}{2001}]{mardlingaarseth01}
{Mardling} R.~A.,  {Aarseth} S.~J.,  2001, \mn@doi [\mnras]
  {10.1046/j.1365-8711.2001.03974.x}, \href
  {http://adsabs.harvard.edu/abs/2001MNRAS.321..398M} {321, 398}

\bibitem[\protect\citeauthoryear{{Mayer}}{{Mayer}}{1990}]{Mayer1990}
{Mayer} P.,  1990, Bulletin of the Astronomical Institutes of Czechoslovakia,
  \href {http://adsabs.harvard.edu/abs/1990BAICz..41..231M} {41, 231}

\bibitem[\protect\citeauthoryear{{Naoz} \& {Fabrycky}}{{Naoz} \&
  {Fabrycky}}{2014}]{naozfabrycky14}
{Naoz} S.,  {Fabrycky} D.~C.,  2014, \mn@doi [\apj]
  {10.1088/0004-637X/793/2/137}, \href
  {http://esoads.eso.org/abs/2014ApJ...793..137N} {793, 137}

\bibitem[\protect\citeauthoryear{{Naoz}, {Fragos}, {Geller}, {Stephan}  \&
  {Rasio}}{{Naoz} et~al.}{2016}]{naozetal16}
{Naoz} S.,  {Fragos} T.,  {Geller} A.,  {Stephan} A.~P.,   {Rasio} F.~A.,
  2016, \mn@doi [\apjl] {10.3847/2041-8205/822/2/L24}, \href
  {http://esoads.eso.org/abs/2016ApJ...822L..24N} {822, L24}

\bibitem[\protect\citeauthoryear{{Ochsenbein}, {Bauer}  \&
  {Marcout}}{{Ochsenbein} et~al.}{2000}]{vizier}
{Ochsenbein} F.,  {Bauer} P.,   {Marcout} J.,  2000, \mn@doi [\aaps]
  {10.1051/aas:2000169}, \href
  {http://adsabs.harvard.edu/abs/2000A%26AS..143...23O} {143, 23}

\bibitem[\protect\citeauthoryear{{Perets} \& {Fabrycky}}{{Perets} \&
  {Fabrycky}}{2009}]{peretzfabrycky09}
{Perets} H.~B.,  {Fabrycky} D.~C.,  2009, \mn@doi [\apj]
  {10.1088/0004-637X/697/2/1048}, \href
  {http://esoads.eso.org/abs/2009ApJ...697.1048P} {697, 1048}

\bibitem[\protect\citeauthoryear{{Pr{\v s}a} \& {Zwitter}}{{Pr{\v s}a} \&
  {Zwitter}}{2005}]{prsazwitter05}
{Pr{\v s}a} A.,  {Zwitter} T.,  2005, \mn@doi [\apj] {10.1086/430591}, \href
  {http://esoads.eso.org/abs/2005ApJ...628..426P} {628, 426}

\bibitem[\protect\citeauthoryear{{Rappaport}, {Deck}, {Levine}, {Borkovits},
  {Carter}, {El Mellah}, {Sanchis-Ojeda}  \& {Kalomeni}}{{Rappaport}
  et~al.}{2013}]{Rappaport2013}
{Rappaport} S.,  {Deck} K.,  {Levine} A.,  {Borkovits} T.,  {Carter} J.,  {El
  Mellah} I.,  {Sanchis-Ojeda} R.,   {Kalomeni} B.,  2013, \mn@doi [\apj]
  {10.1088/0004-637X/768/1/33}, \href
  {http://adsabs.harvard.edu/abs/2013ApJ...768...33R} {768, 33}

\bibitem[\protect\citeauthoryear{{Rappaport} et~al.,}{{Rappaport}
  et~al.}{2017}]{rappaportetal17}
{Rappaport} S.,  et~al., 2017, \mn@doi [\mnras] {10.1093/mnras/stx143}, \href
  {http://esoads.eso.org/abs/2017MNRAS.467.2160R} {467, 2160}

\bibitem[\protect\citeauthoryear{{Rouan}, {Baglin}, {Copet}, {Schneider},
  {Barge}, {Deleuil}, {Vuillemin}  \& {L{\'e}ger}}{{Rouan}
  et~al.}{1998}]{1998EM&P...81...79R}
{Rouan} D.,  {Baglin} A.,  {Copet} E.,  {Schneider} J.,  {Barge} P.,  {Deleuil}
  M.,  {Vuillemin} A.,   {L{\'e}ger} A.,  1998, Earth Moon and Planets, \href
  {http://adsabs.harvard.edu/abs/1998EM%26P...81...79R} {81, 79}

\bibitem[\protect\citeauthoryear{{Sarro} et~al.,}{{Sarro}
  et~al.}{2013}]{sarroetal13}
{Sarro} L.~M.,  et~al., 2013, \mn@doi [\aap] {10.1051/0004-6361/201220184},
  \href {http://adsabs.harvard.edu/abs/2013A%26A...550A.120S} {550, A120}

\bibitem[\protect\citeauthoryear{{Shappee} \& {Thompson}}{{Shappee} \&
  {Thompson}}{2013}]{shappeethompson13}
{Shappee} B.~J.,  {Thompson} T.~A.,  2013, \mn@doi [\apj]
  {10.1088/0004-637X/766/1/64}, \href
  {http://esoads.eso.org/abs/2013ApJ...766...64S} {766, 64}

\bibitem[\protect\citeauthoryear{{Soderhjelm}}{{Soderhjelm}}{1975}]{Soderhjelm1975}
{Soderhjelm} S.,  1975, \aap, \href
  {http://adsabs.harvard.edu/abs/1975A%26A....42..229S} {42, 229}

\bibitem[\protect\citeauthoryear{{Southworth}, {Zucker}, {Maxted}  \&
  {Smalley}}{{Southworth} et~al.}{2004}]{southworthetal04}
{Southworth} J.,  {Zucker} S.,  {Maxted} P.~F.~L.,   {Smalley} B.,  2004,
  \mn@doi [\mnras] {10.1111/j.1365-2966.2004.08389.x}, \href
  {http://esoads.eso.org/abs/2004MNRAS.355..986S} {355, 986}

\bibitem[\protect\citeauthoryear{{Southworth} et~al.,}{{Southworth}
  et~al.}{2011}]{southworthetal11}
{Southworth} J.,  et~al., 2011, \mn@doi [\mnras]
  {10.1111/j.1365-2966.2011.18559.x}, \href
  {http://esoads.eso.org/abs/2011MNRAS.414.2413S} {414, 2413}

\bibitem[\protect\citeauthoryear{{Sterken}}{{Sterken}}{2005}]{sterken05}
{Sterken} C.,  2005, in {Sterken} C.,  ed.,  Astronomical Society of the
  Pacific Conference Series Vol. 335, The Light-Time Effect in Astrophysics:
  Causes and cures of the O-C diagram. p.~3

\bibitem[\protect\citeauthoryear{{Sterne}}{{Sterne}}{1939}]{sterne39}
{Sterne} T.~E.,  1939, \mn@doi [\mnras] {10.1093/mnras/99.5.451}, \href
  {http://esoads.eso.org/abs/1939MNRAS..99..451S} {99, 451}

\bibitem[\protect\citeauthoryear{{Tauris} \& {van den Heuvel}}{{Tauris} \& {van
  den Heuvel}}{2014}]{taurisvandenheuvel14}
{Tauris} T.~M.,  {van den Heuvel} E.~P.~J.,  2014, \mn@doi [\apjl]
  {10.1088/2041-8205/781/1/L13}, \href
  {http://esoads.eso.org/abs/2014ApJ...781L..13T} {781, L13}

\bibitem[\protect\citeauthoryear{{Tokovinin}}{{Tokovinin}}{2004}]{tokovinin04}
{Tokovinin} A.,  2004, in {Allen} C.,  {Scarfe} C.,  eds,  Revista Mexicana de
  Astronomia y Astrofisica Conference Series Vol. 21, Revista Mexicana de
  Astronomia y Astrofisica Conference Series. pp 7--14

\bibitem[\protect\citeauthoryear{{Tokovinin}}{{Tokovinin}}{2014}]{tokovinin14}
{Tokovinin} A.,  2014, \mn@doi [\aj] {10.1088/0004-6256/147/4/87}, \href
  {http://esoads.eso.org/abs/2014AJ....147...87T} {147, 87}

\bibitem[\protect\citeauthoryear{{Tokovinin}, {Thomas}, {Sterzik}  \&
  {Udry}}{{Tokovinin} et~al.}{2006}]{Tokovininetal06}
{Tokovinin} A.,  {Thomas} S.,  {Sterzik} M.,   {Udry} S.,  2006, \mn@doi [\aap]
  {10.1051/0004-6361:20054427}, \href
  {http://esoads.eso.org/abs/2006A%26A...450..681T} {450, 681}

\bibitem[\protect\citeauthoryear{{Tout}, {Pols}, {Eggleton}  \& {Han}}{{Tout}
  et~al.}{1996}]{toutetal96}
{Tout} C.~A.,  {Pols} O.~R.,  {Eggleton} P.~P.,   {Han} Z.,  1996, \mn@doi
  [\mnras] {10.1093/mnras/281.1.257}, \href
  {http://esoads.eso.org/abs/1996MNRAS.281..257T} {281, 257}

\bibitem[\protect\citeauthoryear{{Tran}, {Levine}, {Rappaport}, {Borkovits},
  {Csizmadia}  \& {Kalomeni}}{{Tran} et~al.}{2013}]{Tran2013}
{Tran} K.,  {Levine} A.,  {Rappaport} S.,  {Borkovits} T.,  {Csizmadia} S.,
  {Kalomeni} B.,  2013, \mn@doi [\apj] {10.1088/0004-637X/774/1/81}, \href
  {http://adsabs.harvard.edu/abs/2013ApJ...774...81T} {774, 81}

\bibitem[\protect\citeauthoryear{{Woltjer}}{{Woltjer}}{1922}]{Woltjer1922}
{Woltjer} Jr. J.,  1922, \bain, \href
  {http://adsabs.harvard.edu/abs/1922BAN.....1...93W} {1, 93}

\makeatother
\end{thebibliography}

\appendix
\onecolumn
\section{Tables of times of minima for the five analysed systems}
\label{App:ToM}

In this Appendix we tabulate (in Tables\,\ref{Tab:CoRoT_0100805120_ToM}\,--\,\ref{Tab:CoRoT_0110830711_ToM}) the individual minima times of the primary and secondary eclipses for the five EBs analysed in Sect.\,\ref{Subsect:ETVsolution}.

\begin{table*}
\caption{Times of minima of CoRoT 100805120}
 \label{Tab:CoRoT_0100805120_ToM}
\begin{center}
\begin{tabular}{@{}crlcrlcrl}
\hline
Time & Cycle  & std. dev. & Time & Cycle  & std. dev. & Time & Cycle  & std. dev. \\ 
(RBJD)& no. & \multicolumn{1}{c}{$(d)$} & (RBJD) & no. & \multicolumn{1}{c}{$(d)$} & (RBJD) & no. &   \multicolumn{1}{c}{$(d)$} \\ 
\hline
54237.146244 &   -0.5 & 0.000484 & 54284.855460 &   20.5 & 0.000417 & 54332.559937 &   41.5 & 0.000316 \\ 
54238.276918 &    0.0 & 0.000200 & 54285.985888 &   21.0 & 0.000148 & 54333.689369 &   42.0 & 0.000150 \\ 
54239.416462 &    0.5 & 0.000627 & 54287.131227 &   21.5 & 0.000354 & 54334.833181 &   42.5 & 0.000529 \\ 
54240.547533 &    1.0 & 0.000296 & 54288.258995 &   22.0 & 0.000137 & 54335.962548 &   43.0 & 0.000141 \\ 
54241.683430 &    1.5 & 0.000777 & 54289.396303 &   22.5 & 0.000361 & 54337.099641 &   43.5 & 0.000453 \\ 
54242.821438 &    2.0 & 0.000131 & 54290.531240 &   23.0 & 0.000128 & 54338.236396 &   44.0 & 0.000148 \\ 
54243.956181 &    2.5 & 0.001031 & 54291.670284 &   23.5 & 0.000625 & 54339.377322 &   44.5 & 0.000446 \\ 
54245.093709 &    3.0 & 0.000139 & 54292.803150 &   24.0 & 0.000152 & 54340.506463 &   45.0 & 0.000128 \\ 
54246.231633 &    3.5 & 0.000384 & 54293.938262 &   24.5 & 0.000493 & 54341.655812 &   45.5 & 0.000500 \\ 
54247.363295 &    4.0 & 0.000138 & 54295.073629 &   25.0 & 0.000134 & 54342.775583 &   46.0 & 0.000159 \\ 
54248.512360 &    4.5 & 0.000550 & 54296.212573 &   25.5 & 0.000342 & 54343.916132 &   46.5 & 0.000366 \\ 
54249.636145 &    5.0 & 0.000135 & 54297.347735 &   26.0 & 0.000150 & 54345.048339 &   47.0 & 0.000141 \\ 
54250.769848 &    5.5 & 0.000343 & 54298.483956 &   26.5 & 0.000333 & 54346.192355 &   47.5 & 0.000420 \\ 
54251.908411 &    6.0 & 0.000130 & 54299.616685 &   27.0 & 0.000129 & 54347.320875 &   48.0 & 0.000136 \\ 
54253.045220 &    6.5 & 0.000376 & 54300.756720 &   27.5 & 0.000392 & 54348.460530 &   48.5 & 0.000357 \\ 
54254.180943 &    7.0 & 0.000122 & 54301.888663 &   28.0 & 0.000143 & 54349.592363 &   49.0 & 0.000162 \\ 
54255.313813 &    7.5 & 0.000296 & 54303.028314 &   28.5 & 0.000404 & 54350.723338 &   49.5 & 0.000332 \\ 
54256.453357 &    8.0 & 0.000137 & 54304.162229 &   29.0 & 0.000133 & 54351.863418 &   50.0 & 0.000144 \\ 
54257.588317 &    8.5 & 0.000400 & 54305.300347 &   29.5 & 0.000362 & 54353.003473 &   50.5 & 0.000463 \\ 
54258.722736 &    9.0 & 0.000159 & 54306.435202 &   30.0 & 0.000141 & 54354.135536 &   51.0 & 0.000131 \\ 
54259.858996 &    9.5 & 0.000330 & 54307.570767 &   30.5 & 0.000366 & 54355.274980 &   51.5 & 0.000355 \\ 
54260.996471 &   10.0 & 0.000141 & 54308.704172 &   31.0 & 0.000131 & 54356.409038 &   52.0 & 0.000126 \\ 
54262.133187 &   10.5 & 0.000364 & 54309.841034 &   31.5 & 0.000417 & 54357.547080 &   52.5 & 0.000420 \\ 
54263.266710 &   11.0 & 0.000151 & 54310.976296 &   32.0 & 0.000140 & 54358.678231 &   53.0 & 0.000145 \\ 
54264.400351 &   11.5 & 0.000312 & 54312.116524 &   32.5 & 0.000379 & 54359.820410 &   53.5 & 0.000458 \\ 
54265.538578 &   12.0 & 0.000171 & 54313.247733 &   33.0 & 0.000143 & 54360.952306 &   54.0 & 0.000122 \\ 
54266.678577 &   12.5 & 0.000425 & 54314.390893 &   33.5 & 0.000535 & 54362.088349 &   54.5 & 0.000368 \\ 
54267.811294 &   13.0 & 0.000183 & 54315.520056 &   34.0 & 0.000135 & 54363.223458 &   55.0 & 0.000164 \\ 
54268.955192 &   13.5 & 0.000353 & 54316.660742 &   34.5 & 0.000378 & 54364.360909 &   55.5 & 0.000364 \\ 
54270.080501 &   14.0 & 0.000148 & 54317.790585 &   35.0 & 0.000142 & 54365.498407 &   56.0 & 0.000156 \\ 
54271.221441 &   14.5 & 0.000450 & 54318.927932 &   35.5 & 0.000388 & 54366.637514 &   56.5 & 0.000369 \\ 
54272.353532 &   15.0 & 0.000148 & 54320.064655 &   36.0 & 0.000148 & 54367.766874 &   57.0 & 0.000155 \\ 
54273.489057 &   15.5 & 0.000363 & 54321.203593 &   36.5 & 0.000417 & 54368.904260 &   57.5 & 0.000456 \\ 
54274.624308 &   16.0 & 0.000144 & 54322.335402 &   37.0 & 0.000139 & 54370.041899 &   58.0 & 0.000141 \\ 
54275.764066 &   16.5 & 0.000353 & 54323.467666 &   37.5 & 0.000462 & 54371.179952 &   58.5 & 0.000482 \\ 
54276.897710 &   17.0 & 0.000150 & 54324.605754 &   38.0 & 0.000125 & 54372.312244 &   59.0 & 0.000153 \\ 
54278.038013 &   17.5 & 0.000325 & 54325.741717 &   38.5 & 0.000393 & 54373.453391 &   59.5 & 0.000432 \\ 
54279.172369 &   18.0 & 0.000141 & 54326.877578 &   39.0 & 0.000139 & 54374.583291 &   60.0 & 0.000128 \\ 
54280.310154 &   18.5 & 0.000381 & 54328.016675 &   39.5 & 0.000349 & 54375.722690 &   60.5 & 0.000418 \\ 
54281.442787 &   19.0 & 0.000139 & 54329.150308 &   40.0 & 0.000136 & 54376.856843 &   61.0 & 0.000140 \\ 
54282.581336 &   19.5 & 0.000399 & 54330.284630 &   40.5 & 0.000480 & 54377.990420 &   61.5 & 0.000387 \\ 
54283.714030 &   20.0 & 0.000129 & 54331.418343 &   41.0 & 0.000146 &&& \\ 
\hline
\end{tabular}
\end{center}
\end{table*}

\begin{table*}
\caption{Times of minima of CoRoT 101290947}
 \label{Tab:CoRoT_0101290947_ToM}
\begin{center}
\begin{tabular}{@{}crlcrlcrl}
\hline
Time & Cycle  & std. dev. & Time & Cycle  & std. dev. & Time & Cycle  & std. dev. \\ 
(RBJD)& no. & \multicolumn{1}{c}{$(d)$} & (RBJD) & no. & \multicolumn{1}{c}{$(d)$} & (RBJD) & no. &   \multicolumn{1}{c}{$(d)$} \\ 
\hline
54237.664345 &    0.0 & 0.000136 & 54285.819733 &   23.5 & 0.000100 & 54332.937739 &   46.5 & 0.000102 \\ 
54238.691992 &    0.5 & 0.000153 & 54286.843417 &   24.0 & 0.000111 & 54333.960702 &   47.0 & 0.000098 \\ 
54239.712795 &    1.0 & 0.000195 & 54287.867778 &   24.5 & 0.000104 & 54334.986409 &   47.5 & 0.000125 \\ 
54241.764243 &    2.0 & 0.000171 & 54288.892845 &   25.0 & 0.000095 & 54336.009693 &   48.0 & 0.000101 \\ 
54242.789330 &    2.5 & 0.000214 & 54289.916369 &   25.5 & 0.000185 & 54337.033980 &   48.5 & 0.000156 \\ 
54243.809338 &    3.0 & 0.000237 & 54290.940862 &   26.0 & 0.000065 & 54338.058765 &   49.0 & 0.000128 \\ 
54244.839577 &    3.5 & 0.000401 & 54291.965491 &   26.5 & 0.000108 & 54339.082601 &   49.5 & 0.000135 \\ 
54245.860059 &    4.0 & 0.000128 & 54292.989790 &   27.0 & 0.000096 & 54340.107498 &   50.0 & 0.000098 \\ 
54246.884222 &    4.5 & 0.000195 & 54294.014191 &   27.5 & 0.000068 & 54341.131670 &   50.5 & 0.000166 \\ 
54247.911246 &    5.0 & 0.000280 & 54295.037985 &   28.0 & 0.000101 & 54342.155977 &   51.0 & 0.000073 \\ 
54248.934356 &    5.5 & 0.000280 & 54296.062574 &   28.5 & 0.000124 & 54343.180189 &   51.5 & 0.000118 \\ 
54249.958775 &    6.0 & 0.000139 & 54297.087141 &   29.0 & 0.000090 & 54344.204196 &   52.0 & 0.000106 \\ 
54250.984407 &    6.5 & 0.000177 & 54298.111220 &   29.5 & 0.000152 & 54345.228730 &   52.5 & 0.000132 \\ 
54252.007605 &    7.0 & 0.000147 & 54299.135127 &   30.0 & 0.000131 & 54346.253046 &   53.0 & 0.000066 \\ 
54253.034024 &    7.5 & 0.000139 & 54300.160342 &   30.5 & 0.000228 & 54347.277621 &   53.5 & 0.000090 \\ 
54254.057975 &    8.0 & 0.000230 & 54301.183593 &   31.0 & 0.000142 & 54348.302294 &   54.0 & 0.000086 \\ 
54255.084080 &    8.5 & 0.000144 & 54302.208552 &   31.5 & 0.000093 & 54349.326276 &   54.5 & 0.000142 \\ 
54256.106618 &    9.0 & 0.000169 & 54303.232914 &   32.0 & 0.000143 & 54350.350159 &   55.0 & 0.000087 \\ 
54257.132911 &    9.5 & 0.000288 & 54304.257115 &   32.5 & 0.000114 & 54351.374618 &   55.5 & 0.000118 \\ 
54258.156915 &   10.0 & 0.000258 & 54305.281609 &   33.0 & 0.000104 & 54352.398490 &   56.0 & 0.000109 \\ 
54259.182723 &   10.5 & 0.000135 & 54306.305817 &   33.5 & 0.000127 & 54353.423787 &   56.5 & 0.000079 \\ 
54260.206253 &   11.0 & 0.000155 & 54307.330000 &   34.0 & 0.000085 & 54354.447537 &   57.0 & 0.000142 \\ 
54261.230979 &   11.5 & 0.000221 & 54308.354612 &   34.5 & 0.000172 & 54355.472712 &   57.5 & 0.000220 \\ 
54262.254950 &   12.0 & 0.000293 & 54309.377971 &   35.0 & 0.000123 & 54356.496855 &   58.0 & 0.000130 \\ 
54263.280628 &   12.5 & 0.000106 & 54310.403464 &   35.5 & 0.000161 & 54357.521976 &   58.5 & 0.000106 \\ 
54264.304673 &   13.0 & 0.000109 & 54311.427224 &   36.0 & 0.000121 & 54358.545687 &   59.0 & 0.000100 \\ 
54265.329571 &   13.5 & 0.000118 & 54312.450731 &   36.5 & 0.000144 & 54359.570417 &   59.5 & 0.000088 \\ 
54266.353985 &   14.0 & 0.000098 & 54313.475550 &   37.0 & 0.000094 & 54360.595315 &   60.0 & 0.000120 \\ 
54267.378598 &   14.5 & 0.000116 & 54314.500992 &   37.5 & 0.000142 & 54361.619942 &   60.5 & 0.000155 \\ 
54268.402493 &   15.0 & 0.000136 & 54315.524518 &   38.0 & 0.000093 & 54362.644580 &   61.0 & 0.000136 \\ 
54269.428514 &   15.5 & 0.000149 & 54316.549024 &   38.5 & 0.000205 & 54363.669371 &   61.5 & 0.000095 \\ 
54270.452001 &   16.0 & 0.000123 & 54317.572424 &   39.0 & 0.000082 & 54364.694482 &   62.0 & 0.000099 \\ 
54271.477196 &   16.5 & 0.000092 & 54318.597368 &   39.5 & 0.000118 & 54365.719097 &   62.5 & 0.000102 \\ 
54272.501680 &   17.0 & 0.000170 & 54319.621211 &   40.0 & 0.000187 & 54366.742601 &   63.0 & 0.000086 \\ 
54273.525393 &   17.5 & 0.000159 & 54320.645228 &   40.5 & 0.000071 & 54367.768854 &   63.5 & 0.000092 \\ 
54274.549739 &   18.0 & 0.000094 & 54321.670242 &   41.0 & 0.000121 & 54368.793323 &   64.0 & 0.000133 \\ 
54275.574562 &   18.5 & 0.000093 & 54322.694492 &   41.5 & 0.000094 & 54369.817711 &   64.5 & 0.000123 \\ 
54276.598997 &   19.0 & 0.000101 & 54323.718548 &   42.0 & 0.000111 & 54370.842836 &   65.0 & 0.000131 \\ 
54277.624756 &   19.5 & 0.000115 & 54324.742314 &   42.5 & 0.000104 & 54371.867285 &   65.5 & 0.000169 \\ 
54278.647709 &   20.0 & 0.000069 & 54325.766584 &   43.0 & 0.000122 & 54372.891139 &   66.0 & 0.000287 \\ 
54279.672781 &   20.5 & 0.000110 & 54326.791675 &   43.5 & 0.000078 & 54373.916472 &   66.5 & 0.000087 \\ 
54280.696598 &   21.0 & 0.000079 & 54327.815501 &   44.0 & 0.000091 & 54374.941461 &   67.0 & 0.000085 \\ 
54281.721854 &   21.5 & 0.000087 & 54328.839829 &   44.5 & 0.000078 & 54375.965606 &   67.5 & 0.000069 \\ 
54282.745718 &   22.0 & 0.000089 & 54329.864088 &   45.0 & 0.000130 & 54376.990141 &   68.0 & 0.000126 \\ 
54283.769996 &   22.5 & 0.000108 & 54330.888834 &   45.5 & 0.000106 & 54378.015104 &   68.5 & 0.000073 \\ 
54284.794378 &   23.0 & 0.000095 & 54331.912850 &   46.0 & 0.000123 &&& \\ 
\hline
\end{tabular}
\end{center}
\end{table*}

\begin{table*}
\caption{Times of minima of CoRoT 102698865}
 \label{Tab:CoRoT_0102698865_ToM}
\begin{center}
\begin{tabular}{@{}crlcrlcrl}
\hline
Time & Cycle  & std. dev. & Time & Cycle  & std. dev. & Time & Cycle  & std. dev. \\ 
(RBJD)& no. & \multicolumn{1}{c}{$(d)$} & (RBJD) & no. & \multicolumn{1}{c}{$(d)$} & (RBJD) & no. &   \multicolumn{1}{c}{$(d)$} \\ 
\hline
54398.608096 &    0.0 & 0.000094 & 54472.047495 &   19.5 & 0.000051 & 55949.537829 &  411.0 & 0.000074 \\ 
54400.352100 &    0.5 & 0.000088 & 54474.076262 &   20.0 & 0.000052 & 55951.294555 &  411.5 & 0.000104 \\ 
54404.127336 &    1.5 & 0.000107 & 54475.820765 &   20.5 & 0.000040 & 55953.311804 &  412.0 & 0.000063 \\ 
54407.901090 &    2.5 & 0.000057 & 54477.849641 &   21.0 & 0.000050 & 55955.068274 &  412.5 & 0.000059 \\ 
54409.927825 &    3.0 & 0.000100 & 54479.594185 &   21.5 & 0.000042 & 55957.085055 &  413.0 & 0.000063 \\ 
54411.674986 &    3.5 & 0.000088 & 54481.622876 &   22.0 & 0.000082 & 55958.842228 &  413.5 & 0.000062 \\ 
54413.701646 &    4.0 & 0.000083 & 54483.367611 &   22.5 & 0.000045 & 55960.858257 &  414.0 & 0.000058 \\ 
54415.448825 &    4.5 & 0.000082 & 54485.396617 &   23.0 & 0.000045 & 55962.615712 &  414.5 & 0.000071 \\ 
54417.475185 &    5.0 & 0.000072 & 54487.141003 &   23.5 & 0.000044 & 55964.632828 &  415.0 & 0.000081 \\ 
54419.220562 &    5.5 & 0.000064 & 54489.170217 &   24.0 & 0.000067 & 55966.389493 &  415.5 & 0.000060 \\ 
54421.248132 &    6.0 & 0.000063 & 54490.914422 &   24.5 & 0.000055 & 55968.406117 &  416.0 & 0.000050 \\ 
54422.994651 &    6.5 & 0.000160 & 54492.943320 &   25.0 & 0.000053 & 55970.163013 &  416.5 & 0.000058 \\ 
54425.021002 &    7.0 & 0.000145 & 54494.688024 &   25.5 & 0.000061 & 55972.179471 &  417.0 & 0.000079 \\ 
54426.767475 &    7.5 & 0.000104 & 54496.716803 &   26.0 & 0.000058 & 55973.936551 &  417.5 & 0.000060 \\ 
54428.795605 &    8.0 & 0.000059 & 54498.461234 &   26.5 & 0.000053 & 55975.953188 &  418.0 & 0.000064 \\ 
54430.540713 &    8.5 & 0.000060 & 54500.490291 &   27.0 & 0.000062 & 55977.710169 &  418.5 & 0.000067 \\ 
54432.568783 &    9.0 & 0.000063 & 54502.234783 &   27.5 & 0.000058 & 55979.726835 &  419.0 & 0.000070 \\ 
54434.313957 &    9.5 & 0.000043 & 54504.263756 &   28.0 & 0.000064 & 55981.483771 &  419.5 & 0.000073 \\ 
54436.342227 &   10.0 & 0.000053 & 54506.008214 &   28.5 & 0.000045 & 55983.500711 &  420.0 & 0.000069 \\ 
54438.087481 &   10.5 & 0.000051 & 54508.037328 &   29.0 & 0.000061 & 55985.257240 &  420.5 & 0.000121 \\ 
54440.115633 &   11.0 & 0.000064 & 54509.781938 &   29.5 & 0.000054 & 55987.274157 &  421.0 & 0.000069 \\ 
54441.860534 &   11.5 & 0.000050 & 54511.810821 &   30.0 & 0.000046 & 55989.031257 &  421.5 & 0.000055 \\ 
54443.889083 &   12.0 & 0.000055 & 54513.555162 &   30.5 & 0.000059 & 55991.047439 &  422.0 & 0.000072 \\ 
54445.633825 &   12.5 & 0.000042 & 54515.584129 &   31.0 & 0.000057 & 55992.804643 &  422.5 & 0.000085 \\ 
54447.662532 &   13.0 & 0.000055 & 54517.328917 &   31.5 & 0.000037 & 55994.820608 &  423.0 & 0.000143 \\ 
54449.407340 &   13.5 & 0.000043 & 54519.357722 &   32.0 & 0.000050 & 55996.578636 &  423.5 & 0.000090 \\ 
54451.435896 &   14.0 & 0.000049 & 54521.102481 &   32.5 & 0.000040 & 55998.594774 &  424.0 & 0.000076 \\ 
54453.180993 &   14.5 & 0.000049 & 54523.131364 &   33.0 & 0.000064 & 56000.351969 &  424.5 & 0.000056 \\ 
54455.209106 &   15.0 & 0.000060 & 54524.876093 &   33.5 & 0.000059 & 56002.369950 &  425.0 & 0.000268 \\ 
54456.954225 &   15.5 & 0.000047 & 54526.905187 &   34.0 & 0.000070 & 56004.125615 &  425.5 & 0.000081 \\ 
54458.982738 &   16.0 & 0.000039 & 54528.649795 &   34.5 & 0.000058 & 56006.142000 &  426.0 & 0.000074 \\ 
54460.727426 &   16.5 & 0.000057 & 55939.973721 &  408.5 & 0.000065 & 56007.899256 &  426.5 & 0.000062 \\ 
54462.755950 &   17.0 & 0.000051 & 55941.990731 &  409.0 & 0.000071 & 56009.915150 &  427.0 & 0.000060 \\ 
54464.500540 &   17.5 & 0.000086 & 55943.747292 &  409.5 & 0.000054 & 56011.673261 &  427.5 & 0.000070 \\ 
54466.529478 &   18.0 & 0.000053 & 55945.764353 &  410.0 & 0.000066 & 56013.688681 &  428.0 & 0.000060 \\ 
54468.274289 &   18.5 & 0.000047 & 55947.521194 &  410.5 & 0.000068 & 56015.446049 &  428.5 & 0.000070 \\ 
54470.302961 &   19.0 & 0.000063 &&&&& \\ 
\hline
\end{tabular}
\end{center}
\end{table*}

\begin{table*}
\caption{Times of minima of CoRoT 104079133}
 \label{Tab:CoRoT_0104079133_ToM}
\begin{center}
\begin{tabular}{@{}crlcrlcrl}
\hline
Time & Cycle  & std. dev. & Time & Cycle  & std. dev. & Time & Cycle  & std. dev. \\ 
(RBJD)& no. & \multicolumn{1}{c}{$(d)$} & (RBJD) & no. & \multicolumn{1}{c}{$(d)$} & (RBJD) & no. &   \multicolumn{1}{c}{$(d)$} \\ 
\hline
55019.972986 &   -1.0 & 0.000155 & 55047.622616 &    9.0 & 0.000108 & 55076.647065 &   19.5 & 0.000282 \\ 
55021.350336 &   -0.5 & 0.000383 & 55049.000823 &    9.5 & 0.000430 & 55078.034802 &   20.0 & 0.000075 \\ 
55022.737608 &    0.0 & 0.000940 & 55050.387289 &   10.0 & 0.000168 & 55079.411200 &   20.5 & 0.000254 \\ 
55024.116019 &    0.5 & 0.000175 & 55051.764613 &   10.5 & 0.000209 & 55080.799060 &   21.0 & 0.000112 \\ 
55025.502275 &    1.0 & 0.002930 & 55053.152572 &   11.0 & 0.000055 & 55082.175713 &   21.5 & 0.000394 \\ 
55026.879109 &    1.5 & 0.000308 & 55054.529997 &   11.5 & 0.000479 & 55083.564192 &   22.0 & 0.000054 \\ 
55028.266685 &    2.0 & 0.000448 & 55055.916590 &   12.0 & 0.000349 & 55084.939992 &   22.5 & 0.000307 \\ 
55029.645513 &    2.5 & 0.000378 & 55057.291556 &   12.5 & 0.000263 & 55086.327821 &   23.0 & 0.000081 \\ 
55031.033565 &    3.0 & 0.001328 & 55058.683073 &   13.0 & 0.000207 & 55087.702872 &   23.5 & 0.000314 \\ 
55032.409788 &    3.5 & 0.000643 & 55060.059344 &   13.5 & 0.000436 & 55089.093304 &   24.0 & 0.000194 \\ 
55033.796523 &    4.0 & 0.000200 & 55061.447601 &   14.0 & 0.000114 & 55090.469613 &   24.5 & 0.001141 \\ 
55035.175040 &    4.5 & 0.001587 & 55062.826905 &   14.5 & 0.000246 & 55091.856092 &   25.0 & 0.000203 \\ 
55036.561645 &    5.0 & 0.000078 & 55064.211927 &   15.0 & 0.000087 & 55093.232017 &   25.5 & 0.000460 \\ 
55037.940110 &    5.5 & 0.000229 & 55066.976889 &   16.0 & 0.000050 & 55094.621138 &   26.0 & 0.000199 \\ 
55039.326924 &    6.0 & 0.000076 & 55068.353929 &   16.5 & 0.000519 & 55095.996220 &   26.5 & 0.000211 \\ 
55040.706802 &    6.5 & 0.000557 & 55069.741656 &   17.0 & 0.000092 & 55097.385346 &   27.0 & 0.000057 \\ 
55042.092389 &    7.0 & 0.000146 & 55071.118930 &   17.5 & 0.000114 & 55098.761998 &   27.5 & 0.000446 \\ 
55043.470419 &    7.5 & 0.000220 & 55072.506007 &   18.0 & 0.000103 & 55100.150278 &   28.0 & 0.000452 \\ 
55044.857249 &    8.0 & 0.000052 & 55073.881356 &   18.5 & 0.000406 & 55101.525844 &   28.5 & 0.000239 \\ 
55046.235511 &    8.5 & 0.000316 & 55075.270684 &   19.0 & 0.000100 & 55102.913762 &   29.0 & 0.000252 \\ 
\hline
\end{tabular}
\end{center}
\end{table*}

\begin{table*}
\caption{Times of minima of CoRoT 110830711}
 \label{Tab:CoRoT_0110830711_ToM}
\begin{center}
\begin{tabular}{@{}crlcrlcrl}
\hline
Time & Cycle  & std. dev. & Time & Cycle  & std. dev. & Time & Cycle  & std. dev. \\ 
(RBJD)& no. & \multicolumn{1}{c}{$(d)$} & (RBJD) & no. & \multicolumn{1}{c}{$(d)$} & (RBJD) & no. &   \multicolumn{1}{c}{$(d)$} \\ 
\hline
54788.023078 &   -0.5 & 0.000402 & 54824.940971 &   14.0 & 0.000050 & 54863.130202 &   29.0 & 0.000059 \\ 
54789.296425 &    0.0 & 0.000236 & 54826.214821 &   14.5 & 0.000241 & 54864.401442 &   29.5 & 0.000133 \\ 
54790.572760 &    0.5 & 0.000168 & 54827.487298 &   15.0 & 0.000056 & 54866.948855 &   30.5 & 0.000479 \\ 
54791.845541 &    1.0 & 0.000065 & 54828.760033 &   15.5 & 0.000118 & 54869.494468 &   31.5 & 0.000117 \\ 
54793.119517 &    1.5 & 0.000180 & 54830.033262 &   16.0 & 0.000068 & 54870.767700 &   32.0 & 0.000065 \\ 
54794.391327 &    2.0 & 0.000079 & 54832.578973 &   17.0 & 0.000075 & 54872.040394 &   32.5 & 0.000283 \\ 
54795.664033 &    2.5 & 0.000199 & 54833.850810 &   17.5 & 0.000126 & 54873.313283 &   33.0 & 0.000079 \\ 
54796.937032 &    3.0 & 0.000056 & 54835.124576 &   18.0 & 0.000065 & 54874.587249 &   33.5 & 0.000160 \\ 
54798.210657 &    3.5 & 0.000306 & 54836.392527 &   18.5 & 0.001064 & 54875.859268 &   34.0 & 0.000051 \\ 
54799.482707 &    4.0 & 0.000068 & 54837.670751 &   19.0 & 0.000077 & 54877.132981 &   34.5 & 0.000129 \\ 
54800.756760 &    4.5 & 0.000121 & 54838.944440 &   19.5 & 0.000209 & 54878.405360 &   35.0 & 0.000081 \\ 
54802.028493 &    5.0 & 0.000062 & 54840.216662 &   20.0 & 0.000072 & 54879.678157 &   35.5 & 0.000122 \\ 
54803.301066 &    5.5 & 0.000254 & 54841.487854 &   20.5 & 0.000294 & 54880.950873 &   36.0 & 0.000070 \\ 
54804.574266 &    6.0 & 0.000080 & 54842.762557 &   21.0 & 0.000088 & 54882.225031 &   36.5 & 0.000240 \\ 
54805.849405 &    6.5 & 0.000282 & 54844.035997 &   21.5 & 0.000109 & 54883.496662 &   37.0 & 0.000078 \\ 
54807.119961 &    7.0 & 0.000058 & 54845.308887 &   22.0 & 0.000063 & 54884.769767 &   37.5 & 0.000187 \\ 
54808.394553 &    7.5 & 0.000218 & 54846.581572 &   22.5 & 0.000236 & 54886.042394 &   38.0 & 0.000063 \\ 
54809.666055 &    8.0 & 0.000058 & 54847.854569 &   23.0 & 0.000088 & 54887.315450 &   38.5 & 0.000149 \\ 
54810.939690 &    8.5 & 0.000253 & 54849.126682 &   23.5 & 0.000128 & 54888.588111 &   39.0 & 0.000062 \\ 
54812.212004 &    9.0 & 0.000073 & 54850.400722 &   24.0 & 0.000066 & 54889.861642 &   39.5 & 0.000257 \\ 
54813.485392 &    9.5 & 0.000177 & 54851.673887 &   24.5 & 0.000159 & 54891.133899 &   40.0 & 0.000076 \\ 
54814.757655 &   10.0 & 0.000051 & 54852.946496 &   25.0 & 0.000069 & 54892.406984 &   40.5 & 0.000223 \\ 
54816.030821 &   10.5 & 0.000239 & 54854.220417 &   25.5 & 0.000252 & 54893.679800 &   41.0 & 0.000089 \\ 
54817.303532 &   11.0 & 0.000055 & 54855.492376 &   26.0 & 0.000093 & 54894.954592 &   41.5 & 0.000256 \\ 
54818.577371 &   11.5 & 0.000150 & 54856.766127 &   26.5 & 0.000085 & 54896.225738 &   42.0 & 0.000074 \\ 
54819.849472 &   12.0 & 0.000072 & 54858.038690 &   27.0 & 0.000074 & 54897.497495 &   42.5 & 0.000157 \\ 
54821.122230 &   12.5 & 0.000291 & 54859.311530 &   27.5 & 0.000093 & 54898.772165 &   43.0 & 0.000082 \\ 
54822.395180 &   13.0 & 0.000067 & 54860.584306 &   28.0 & 0.000069 & 54900.044392 &   43.5 & 0.000184 \\ 
54823.669011 &   13.5 & 0.000101 & 54861.856686 &   28.5 & 0.000163 & 54901.317483 &   44.0 & 0.000081 \\ 
\hline
\end{tabular}
\end{center}
\end{table*}

\end{document}